\documentclass[aps, prx, reprint, superscriptaddress, longbibliography, amsmath, amssymb, floatfix]{revtex4-2}

\usepackage{dcolumn}
\usepackage[colorlinks=true,
            linkcolor=blue,
            citecolor=blue,
            urlcolor=blue]{hyperref}
\usepackage{siunitx}

\usepackage{xcolor}
\usepackage[utf8]{inputenc}
\usepackage[T1]{fontenc}

\usepackage{amsmath}
\usepackage{amssymb}
\usepackage{glossaries}
\usepackage{graphicx}
\usepackage{here}
\usepackage{times}
\usepackage{float}
\usepackage{booktabs}
\usepackage{bm}

\usepackage{multirow}
\usepackage{mathtools}
\usepackage{xspace}

\usepackage{chemformula}
\NewChemBond{quintuple}{
  \foreach \i in {-.7ex,-.35ex,0ex,.35ex,.7ex}{
    \draw[chembond]
      ([yshift=\i]chemformula-bond-start) -- ([yshift=\i]chemformula-bond-end) ;
  }
}
\NewChemCompoundProperty{|}{\bond{quintuple}}

\newcommand{\vb}{$V_{\mathrm{B}}$\xspace}
\newcommand{\vn}{$V_{\mathrm{N}}$\xspace}

\begin{document}

%\title{Insights into Interlayer Bonding in hBN/SiC Heterostructures:{\texorpdfstring{\\}{}} Toward Transition-Metal Single-Atom Isolation}
\title{Stabilisation of hBN/SiC Heterostructures with Vacancies and Transition-Metal Atoms}
\author{Arsalan Hashemi}
\email{arsalan.hashemi@aalto.fi}
\affiliation{MSP Group, Quantum Technology Finland Center of Excellence, Department of Applied Physics, Aalto University, FI-00076 Espoo, Finland}
\author{Nima Ghafari Cherati}
\affiliation{HUN-REN Wigner Research Centre for Physics, PO Box 49, H-1525 Budapest, Hungary}
\affiliation{Department of Atomic Physics, Institute of Physics, Budapest University of Technology and Economics, M\H{u}egyetem rkp. 3, H-1111 Budapest, Hungary}
\author{Sadegh Ghaderzadeh}
\affiliation{School of Chemistry, University of Nottingham, Nottingham NG7 2RD, UK}
\author{Yanzhou Wang}
\affiliation{MSP Group, Quantum Technology Finland Center of Excellence, Department of Applied Physics, Aalto University, FI-00076 Espoo, Finland}
\author{Mahdi Ghorbani-Asl}
\affiliation{Institute of Ion Beam Physics and Materials Research, Helmholtz-Zentrum Dresden-Rossendorf, 01328 Dresden, Germany}
\author{Tapio Ala-Nissila}
\email{tapio.ala-nissila@aalto.fi}
\affiliation{MSP Group, Quantum Technology Finland Center of Excellence, Department of Applied Physics, Aalto University, FI-00076 Espoo, Finland}
\affiliation{Interdisciplinary Centre for Mathematical Modelling and Department of Mathematical Sciences, Loughborough University, Loughborough, Leicestershire LE11 3TU, United Kingdom}

%\date{\nodate}

\begin{abstract}
When two-dimensional atomic layers of different materials are brought into close proximity to form van der Waals (vdW) heterostructures, interactions between adjacent layers significantly influence their physicochemical properties.
These effects seem particularly pronounced when the interface exhibits local order and near-perfect structural alignment, leading to the emergence of Moir\'e patterns.
Using quantum mechanical density functional theory calculations, we propose a prototypical bilayer heterostructure composed of hexagonal boron nitride (hBN) and silicon carbide (SiC), characterized by a lattice mismatch of 18.77\% between their primitive unit cells.
We find that the removal of boron atoms from specific lattice sites can convert the interlayer interaction from weak vdW coupling to robust localized silicon-nitrogen covalent bonding.
Motivated by this, we study the binding of transition-metal adatoms and formulate design guidelines to enhance surface reactivity, thereby enabling the controlled isolation of single-metal atoms.
Our machine-learning-assisted molecular dynamics simulations confirm both dynamical stability and metal anchoring feasibility at finite temperatures.
Our results suggest the hBN/SiC heterostructure as a versatile platform for atomically precise transition-metal functionalization, having potential for next-generation catalytic energy-conversion technologies.
\end{abstract}

\maketitle

%%%%%%%%%%%%%%%%%%%%%%%%%%%%%%%%%%%%%%%%%%%%%%%%%%%%%%%%%%%%%%%%%%%%%%%%%%%
\section{Introduction}
\label{sect:intro}
%parag1
Since the discovery of graphene \cite{geim2013van}, the two-dimensional (2D) allotrope of carbon,
various 2D forms of different materials have been the main focus of novel materials science
research. Currently, a vast collection of materials can exhibit either atomistically flat or
corrugated 2D phases, such as graphene and its various allotropes, silicene (Si),
phosphorene (P), hexagonal boron nitride (hBN), MoS$_2$, NbSe$_2$, and SnTe. They display an
extraordinary variety of correlated magnetic and electronic states \cite{jingyang2021,burch2018}.
The range of physical properties of 2D materials can be greatly extended by integration of dissimilar van der Waals (vdW) 2D materials into heterostructures, which has enabled novel electronic and optoelectronic functionalities~\cite{Qi_2023,CastellanosGomez2022,geim2013van}.
Examples include (i) switchable charge-carrier injection for diodes~\cite{cheng2018high},
(ii) reconfigurable memory operation for transistors~\cite{sun2022reconfigurable}, and
(iii) tunneling field-effect transistors exhibiting sub-thermionic switching~\cite{liu2016van}.
When 2D materials are vertically stacked~\cite{mounet2018two,yang2022two,liu2021promises},
their intrinsic properties can hybridize,
unique interfacial interactions emerge owing to lattice mismatch-induced strain,
and interlayer charge transfer can occur~\cite{Hybertsen1990_APS,Lu2022_AM,li2024relaxation}.
This process creates a multifunctional platform that may not only enhance the performance of the parent materials but also enable the emergence of new properties that are not intrinsic to the individual materials~\cite{Wang2021_AFM,Lotsch2015_armr}.

%parag2
With its matured fabrication techniques~\cite{futaba2023hexagonal}, wide electronic bandgap~\cite{Kubota2007_sci}, high lattice thermal conductivity~\cite{wang2016superior}, transparency~\cite{xu2022magnetically}, and exceptional mechanical strength~\cite{ding2016mechanical}, hBN has been consistently used in various applications from (opto-)electronics~\cite{Moon2023_AM} to environmental sensing~\cite{LI2022101486} and energy storage/transformation~\cite{Gong2021_ACS}.
The development of layered heterostructures has also driven the combination of hBN with other 2D materials, such as graphene and transition metal dichalcogenides~\cite{ZAHOOR20243, BALTA2024205}.

%parag3
A prototypical example appears when hBN is stacked onto semi-metallic graphene: two structurally similar sheets form a seamless interface~\cite{dean2010_nature,yankowitz2019van,juma2021direct,Ogawa2022,Zihao2019_sci,moore2021nanoscale,Rupini2024_pnas}.
This heterostructure exhibits several emergent properties, including the opening of an electronic bandgap in graphene that effectively transforms it from a semimetal to a semiconductor~\cite{OSullivan2025}, ultrahigh charge carrier mobility in graphene facilitated by hBN phonon contributions~\cite{huang2020ultra}, and enhanced spin-orbit coupling, which extends the spin lifetimes to $1-10$ ns~\cite{PhysRevB.103.075129}.
Importantly, twisting one layer introduces a tunable Moir\'{e} superlattice, adding an additional degree of freedom that renders the heterostructure both more complex and intriguing~\cite{Jingfeng_acsnano_2023,jat2024_nature,ma2025_nature,lu2025_nature}.

%parag4
Inspired by the remarkable properties of hBN/graphene heterostructures, a crucial question arises:
\emph{Which additional 2D materials can be integrated with hBN to advance the development of novel vdW heterostructures?}
The answer to this question depends on both synthetic feasibility and potential application advantages, which can often be challenging to evaluate experimentally, particularly as the material space continues to expand.
To address this challenge, computer-aided modeling serves as a preliminary screening step to narrow down the material space to a few compounds, ensuring a feasible sample size for experimental validation while minimizing complexities~\cite{hafner2006toward,dingreville2016review}.
Several studies have investigated this topic by incorporating molybdenum disulfide (MoS$_2$), tungsten disulfide (WS$_2$), and borophene into hBN monolayers~\cite{Li2017_RSC, LIU2025100687, DUAN2025118129, Zhao2022_APL,doi:10.1021/acs.jpcc.3c07843}.

%parag5
Beyond Moir\'{e} patterns~\cite{PhysRevB.99.075422}, point defects can also exert a dominant influence on the properties of assembled 2D heterostructures~\cite{Rakib2022_JAP,D4CP01673D}.
Such defects are ubiquitous, arising both naturally and through intentional engineering.
In semiconductors, defects can introduce new electronic states within the energy gap, profoundly affecting the material's opto-electronic properties~\cite{PhysRevB.80.155425,doi:10.1021/ja400637n,PhysRevLett.109.205502}.
Additionally, defects may enhance the surface activity of solid materials by promoting charge accumulation, particularly when employed as catalysts in electrochemical and sensing devices~\cite{Singh2018_jpcc,rhodes2019disorder,Chen2023_acr}.
Despite their importance, point defects in Moir\'{e} superlattices remain scarcely explored.

%parag6
Point defects in hBN can be broadly categorized into intrinsic and extrinsic types~\cite{doi:10.1021/jp410716q,Zhang2020_JAP,Cholsuk2024_JPCC,Zeng2024_acsami,smart2021intersystem}.
Intrinsic defects primarily originate from atomic vacancies, with boron monovacancy (\vb) being the most prevalent~\cite{langle2024,gong2023coherent,Jin2009_PRL,zhu2017taming}.
The presence of \vb\ plays a crucial role in modifying the electronic properties of hBN, influencing its optical characteristics and charge transport behavior~\cite{D2NR07234C,babar2024low}.
In addition to carbon- and oxygen-related dopants~\cite{PhysRevB.106.014107,C7TA08515J,PhysRevB.87.035404,Liu2019_AOM,asif2021computational}, extrinsic point defects can be engineered by the deposition of metal atoms into either pre-existing or irradiation-induced vacancies~\cite{GHORBANIASL2022259,Bui_small_2023,Kohlrausch_AdSci_2025}, a strategy that enhances surface activity by promoting electron transfer~\cite{Yu_2017_iop,lian2024theoretical,zhong2022first,C3RA23132A,D2NA00843B,Zhang2024_Langmuir}. This capability suggests that vacancy sites can act as active centers for adsorption and functionalization, rendering these defects highly relevant for applications in catalysis~\cite{Kozubek2018_acs,lu2023_nature}, quantum information processing~\cite{gottscholl2021spin,fang2024quantum}, and optoelectronic devices~\cite{fischer2021controlled,white2022electrical}.

%parag7
In the present work, we propose a promising platform for a bilayer heterostructure with tunable electronic properties. To this end, we couple a hBN monolayer with a semiconducting silicon carbide (SiC) monolayer; the resulting bilayer heterostructure is hereafter denoted as hBN/SiC.
Despite their similar bonding nature, the large lattice mismatch induces Moir\'{e} superlattices.  
Unlike hBN, the isolation of monolayer SiC has been explored in only a few experimental studies~\cite{Polley2023_PRL,Da_2024}, motivating a systematic computational study of its stability and interfacial properties.
Meanwhile, the SiC surface has been recognized as a promising substrate for the epitaxial growth of 2D materials~\cite{mokhov2023growth,zhao2024_nature}.
The SiC monolayer adopts a flat honeycomb lattice, whereas in bulk hexagonal SiC, each layer exhibits a buckled structure with Si and C atoms positioned in separate layers.
We focus on the SiC as supported by hBN and modified by defects in hBN and metallic impurity atoms.
%Assuming SiC serves as the substrate, the second question emerges: \emph{How do point defects in hBN impact the functionality of the hBN/SiC system?}

%parag8
Using quantum density functional theory (DFT), we explore hBN/SiC across four fronts:
(i) inherent stability and electronic-structural properties of the pristine system;
(ii) the influence of \vb and \vn defects on lattice geometry and emergent novel physical behavior;
(iii) electronic enhancements via transition metal (TM) implantation; and
(iv) engineering feasibility of robust single‐atom isolation.
We then propose design strategies to translate these insights into targeted applications.

%%%%%%%%%%%%%%%%%%%%%%%%%%%%%%%%%%%%%%%%%%%%%%%%%%%%%%%%%%%%%%%%%%%%%%%%%%%%%%%%%
\section{Computational Details}
\label{sect:computationaldetails}

%parag1
Our spin-polarized DFT calculations were performed with VASP~\cite{kres1}, employing projector augmented-wave (PAW) pseudopotentials~\cite{PAW1994}.
Exchange-correlation effects were treated using the Perdew-Burke-Ernzerhof (PBE) functional~\cite{Perdew1996_PRL}, and, where necessary, the Heyd-Scuseria-Ernzerhof (HSE) hybrid functional~\cite{Heyd2003_JCP,Heyd2004_JCP}.
A plane-wave kinetic-energy cutoff of 500 eV was employed.
Total energy and ionic force convergence thresholds were set to $10^{-6}$ eV and $0.01$ eV/\AA, respectively.
The Brillouin zone of the pristine primitive unit cells of hBN and SiC was sampled using a 9$\times$9$\times$1 Monkhorst-Pack grid, ensuring accurate electronic structure convergence.
To suppress spurious interactions between periodic images, we added a $25$ \AA\ vacuum gap along the out-of-plane direction.

%parag2
Among the available vdW correction schemes (see Table S1 of the Supplemental Material (SM)~\cite{supplemental_material}), the Grimme DFT-D3 method with Becke-Johnson damping~\cite{Grimme2011_JCC} best reproduces the experimental hBN interlayer spacing ($\approx 3.3$ \AA) \cite{Pease1952AnXS,paszkowicz2002lattice}.
This method yielded interlayer distances of $3.26$ \AA\ for bilayer hBN and $3.23$ \AA\ for bilayer SiC, respectively.
The experimentally reported lattice constants for monolayer hBN and SiC are $2.5$ \AA~\cite{KORSAKS2015976} and $3.12$ \AA~\cite{Polley2023_PRL}, respectively, which are in close agreement with our PBE/HSE-calculated values of $2.51$/$2.49$ \AA\ and $3.09$/$3.07$ \AA.
Furthermore, our calculated electronic bandgap of monolayer hBN using the HSE functional is $5.46$ eV, in agreement with experimental values ranging from $5.5$ to $5.9$ eV~\cite{Chahal2022_JPCC,C7TC04300G,Chimene2015_AM}.
This validation confirms the reliability of our computational approach for achieving the objectives of this study.

\begin{figure*}[htb!]
    \centering
    \includegraphics[scale=1.07]{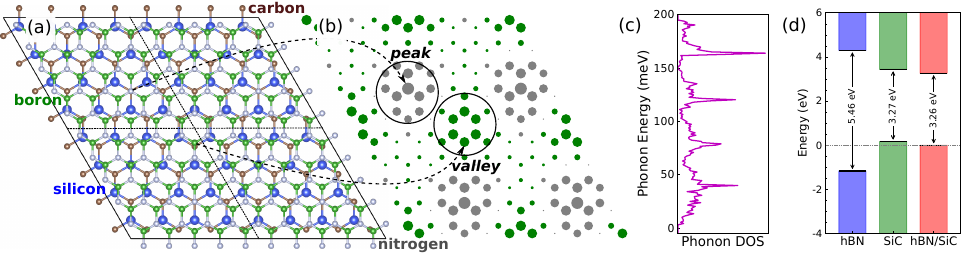}
    \caption{(a) Heterostructure supercell model. Boron, nitrogen, silicon, and carbon are represented in green, gray, blue, and brown, respectively.
    (b) Out-of-plane displacements of hBN layer in response to the lattice mismatch of $1.48$\%.
    The size of each symbol represents the deviation of an atom's $z$-coordinate from the average value.
    Green symbols indicate atoms with $z$-coordinates below the average ($\Delta z^{\rm min} = -0.17$~\AA), while gray symbols represent atoms with $z$-coordinates above the average ($\Delta z^{\rm max} = 0.23$~\AA).
    For the SiC layer $\Delta z^{\rm min}$ and $\Delta z^{\rm max}$ are about $-0.05$ and $0.05$~\AA, which were not depicted.
    (c) Phonon density of states (DOS) of the hBN/SiC system, shown in units of states/meV.
    (d) Energy levels and bandgap for hBN, SiC, and the hBN/SiC heterostructure.
    Energy levels are shifted to valence band maximum (VBM) of the heterostructure system and set as zero.}
    \label{fig:sc-dist}
\end{figure*}

%parag3
To construct the hBN/SiC supercell, the crystalline axes of the hBN and SiC monolayers were aligned, with a 5:4 ratio of their lattice constants.
This alignment corresponds to 5$\times$5$\times$1 and 4$\times$4$\times$1 supercells for hBN and SiC, respectively, resulting in a lattice mismatch of $1.5\%$.
To facilitate defect studies while minimizing fictitious defect interactions, the supercell was duplicated along each in-plane axis (see Fig.~\ref{fig:sc-dist}(a)).
The final supercell consists of $100$ boron (B), $100$ nitrogen (N), $64$ silicon (Si), and $64$ carbon (C) atoms.
For calculations involving the supercell, reciprocal space was only sampled at the $\Gamma$-point.

%parag4
In the supercell employed, removing or adding a single atom produces a defect density of $0.046$ nm$^{-2}$.
Defect formation energies were computed, following the protocol described in Section II of the SM, to evaluate the stability of various charge states of \vb and \vn defects.
Subsequently, binding energy of a TM atom to the surface is defined as:  
\begin{equation}
    E_{\text{bind}} = E_{\text{TM@surface}} - (E_{\text{surface}} + E_{\text{TM}}),
    \label{eq:bindene}
\end{equation}  
where $E_{\text{TM@surface}}$ is the total energy of the optimized complex, while $E_{\text{surface}}$ and $E_{\text{TM}}$ represent the total energies of the \vb-containing defective system (hBN/SiC heterostructure or hBN monolayer) and the isolated TM atom, respectively.
A more negative $E_{\text{bind}}$ indicates a stronger bond to the surface.

%

%parag5
When needed, the lattice vibrations were evaluated using the Phonopy package~\cite{TOGO2015}.
To reach larger system sizes and longer time scales, we carried out machine-learning molecular dynamics (MLMD) simulations using the GPUMD package~\cite{FAN201710}.
The neural evolution potential (NEP) framework~\cite{Ying_2024} with radial and angular cutoffs of $5$ and $3.5$ \AA\ was utilized to represent the potential energy surface.
A detailed explanation of the dataset can be found in Section III of the SM.

%%%%%%%%%%%%%%%%%%%%%%%%%%%%%%%%%%%%%%%%%%%%%%%%%%%%%%%%%%%%%%%%%%%%%%%%%%%%%%%
%

\section{Results}
\label{sect:results}

\subsection{Pristine \texorpdfstring{\MakeLowercase{h}BN/S\MakeLowercase{i}C}{hBN/SiC}}
%\section{Pristine \MakeLowercase{h}BN/S\MakeLowercase{i}C}
\label{sec:pristine-hbn-sic}

%
%parag1
Structurally, hBN resembles graphene, but the \ch{B-N} bonds exhibit partial double-bond character due to $\pi$-electron delocalization between nitrogen's lone $p_{z}$ electrons and the empty $p_{z}$ orbitals of boron.
This broken sublattice symmetry opens a wide bandgap, making hBN an electrical insulator, in contrast to graphene's semi-metallic behavior.
Table S2 compares the fundamental properties of pristine hBN and SiC with those of graphene.
The \ch{B-N} bond length in hBN is $1.45$ \AA, shorter than the sum of the covalent radii of B and N ($1.55$ \AA), and comparable to the \ch{C-C} bond length in graphene ($1.42$ \AA).
SiC exhibits a larger lattice constant and bond length ($1.78$ \AA), consistent with the greater covalent radii of Si and C atoms ($1.84$ \AA).
%hBN ---> 0.84+0.71
%SiC ---> 1.11+0.73

%parag2
Structural symmetry reduces the elastic tensor to only three independent components: $C_{11}$, $C_{12}$, and $C_{66}$.
According to the Born-Huang stability criteria~\cite{Mouhat2014_PRB}, where $C_{11} > C_{12} > 0$ and $C_{66} > 0$), the SiC monolayer and the other systems in Table S2 are elastically stable.
Notably, SiC exhibits a lower Young's modulus than both graphene and hBN, indicating greater mechanical flexibility.
As a result, the SiC monolayer is expected to undergo larger deformation under strain induced by lattice mismatch.

%parag3
The hBN/SiC interface has a binding energy of $-17.9$ meV/\AA$^{2}$, comparable to an hBN bilayer ($-16.9$ meV/\AA$^{2}$) but less in absolute value than that of a SiC bilayer ($-28.7$ meV/\AA$^{2}$).
Note that the magnitudes depend on the choice of vdW correction (see Table S3).
To minimize the initial mismatch of $1.5\%$, the optimized lattice constant of the hBN/SiC supercell undergoes strains of $-0.65\%$ on the hBN layer and $0.86\%$ on the SiC layer, relative to their respective monolayers.
The pronounced mismatch of $18.77\%$ (calculated as $\frac{a_{\rm{SiC}} - a_{\rm{hBN}}}{a_{\rm{SiC}}} \times 100 \%$ where $a$ is the in-plane lattice constant) between unit cells naturally induces a long-period moir\'{e} pattern, resulting in structural non-uniformity across the bilayer (see Fig.~\ref{fig:sc-dist}(b)).

%parag4
The interlayer distance between hBN and SiC sheets varies from $3.21$ to $3.60$ \AA, depending on the local stacking pattern.
While the SiC plane is under tensile strain, it exhibits a slight displacement along the $z$-axis of about $0.05$ \AA; when the stacking resembles AA$^{\prime}$ stacking, where N(C) and Si(B) are positioned over each other, the layers come closer together, forming a valley.
The interlayer distance reaches its maximum when a B atom is positioned at the center of the SiC hexagon, and N atoms are aligned above C atoms, as shown in Fig.~\ref{fig:sc-dist}(b).
The variation in interlayer spacing is primarily attributable to the buckling of the hBN layer.

%parag5
To explore the alternative configuration (different stacking) to the one shown in Fig.~\ref{fig:sc-dist}(a), we swapped the positions of B and N atoms and subsequently assessed the relative stability of the resulting configuration.
In this case, the total energy of the supercell increased slightly (by about $0.13$ eV), which is negligible for $328$ atoms.
In what follows, we focus on the most stable configuration, which is dynamically stable (see Fig.~\ref{fig:sc-dist}(c)).

%parag6
Our MLMD simulations (see Figure S2 and Movie S1) show that the hBN/SiC interface is mechanically robust yet elastically compliant.
Because of weak interlayer coupling, the monolayers slide readily past one another.
The time-averaged interlayer spacing increases by $3\%$ across the entire $25-900$ K temperature range, underscoring interfacial stability.
Meanwhile, the distribution of interlayer distances broadens markedly with temperature, reflecting enhanced out-of-plane rippling of the sheets.

\begin{figure*}[htb!]
    \centering
    \includegraphics[scale=1.10]{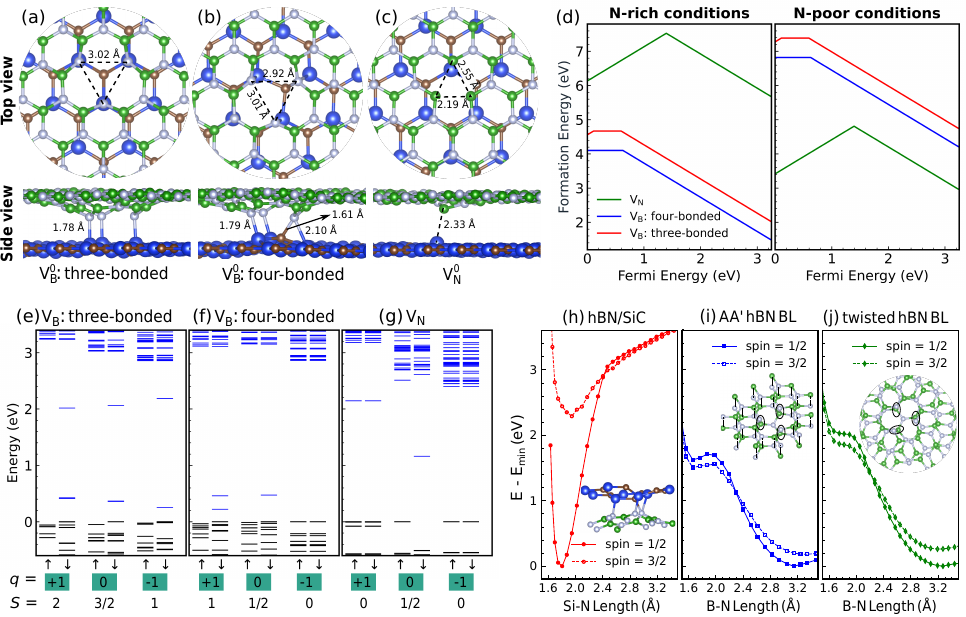}
    \caption{Local defect structures of the (a) three-bonded \vb, (b) four-bonded \vb, and (c) \vn
    are presented along with detailed structural information.
    (d) Formation energy of the defects as a function of the position of the Fermi-level under N-rich and N-poor conditions:
    The VBM is aligned to zero for the sake of simplicity.
    The conduction band minimum (CBM) is at $3.26$ eV.
    For each defect the ($+$), ($0$), and ($-$) charge states may appear.
    The N-rich and N-poor conditions are defined in section II of the SM.
    (e-g) Electronic structure for the ground states of the defects.
    Kohn-Sham (KS) levels are represented in spin-up ($\uparrow$) and spin-down ($\downarrow$) channels.
    The occupied and unoccupied levels are represented in black and blue, respectively, for each charge state ($q$).
    The number of unpaired electrons can be determined from the spin state ($S$).
    For example, $S = 1$ indicates two unpaired electrons.
    (h-j) Energy profile for interlayer distance (bond length) scan of the \vb defect in different systems: the hBN/SiC heterostructure, AA$^{\prime}$ stacking, and the $14^{\circ}$-twisted bilayer.
    Total energies are shifted for simplicity. Both low and high spin states are investigated, and a four-bonded system is considered for the heterostructure.}
    \label{fig:vb}
\end{figure*}
%
%parag7
Bader charge analysis~\cite{Tang_2009} reveals that in a monolayer hBN, $0.72\,e$ is transferred between B and N atoms to form a covalent bond, with B acting as the donor and N as the acceptor.
In a SiC monolayer, Si donates $0.84\,e$ to the bonded C atom, indicating that the \ch{Si-C} bond has a greater ionic character, while the \ch{B-N} bond is more covalent in nature.
In the hBN/SiC system, a total charge transfer of $10^{-3}\,e$ ($\sim 1.859 \times 10^{10}\,e/{\rm{cm}}^{2}$) occurs from the hBN layer to the SiC layer, which is negligible~\cite{Lin2021_AdMat,Back2019_nanoadv,deOliveria2015jcp}.
Thus, there is almost no redistribution of charge across the interface.

%parag8
The electronic bandgap of the heterostructure is $3.26$ eV, which is almost identical to that of pristine SiC (see Fig.~\ref{fig:sc-dist}(d)).
The band alignment analysis indicates that the heterostructure exhibits type I alignment, in which the bandgap of SiC is fully nested within the bandgap of hBN.
As a result, both electrons and holes are confined within the SiC layer.
Nevertheless, this study focuses on hBN, where defect incorporation is experimentally more accessible, owing to the extensive, long-term efforts devoted to this system~\cite{PhysRevLett.109.205502,Vlassiouk2025_arXiv,langle2024}.

%%%%%%%%%%%%%%%%%%%%%%%%%%%%%%%%%%%%%%%%%%%%%%%%%%%%%%%%%%%%%%%%%%%%%%%%
\subsection{Monovacancies \texorpdfstring{\MakeLowercase{in} \MakeLowercase{h}BN/S\MakeLowercase{i}C}{hBN/SiC}}
\label{subsect:VBVN}

%parag1
The response of hBN to vacancy creation is site‑specific: both the identity of the missing atom and its crystallographic position dictate the local geometry, underscoring the intrinsic interfacial heterogeneity.
Energetically, the stability of \vb is strongly influenced by the local environment, particularly the proximity of N atoms to Si atoms and, more prominently, B to C atoms in the adjacent layer.
The introduction of neutral \vb at specific sites in hBN leads to the formation of strong chemical bonds between layers that were formerly bound by dispersive vdW interactions.
This covalent interlayer coupling manifests only at a subset of positions, with the largest effect observed for vacancies located at, or near, regions of quasi-ideal stacking within the heterostructure.

%parag2
To obtain a comprehensive picture of vacancy behaviour, we evaluated $15$ symmetry-inequivalent neutral \vb sites afforded by the chosen supercell.
Sorting them by the number of interlayer covalent bonds formed reveals five families:
(i) Non-bonded (or vdW-bonded):
A \vb positioned almost directly above a Si atom leaves three neighboring N atoms with dangling bonds that point toward the hollow sites of the SiC layer.
These N atoms sit at the centers of the bottom-layer SiC' hexagons.
Lacking suitable bonding partners, the N radicals neither dimerize nor form additional bonds, and the resulting configuration is $5.4$ eV higher in energy than the ground-state defect.
(ii) One-bonded:
Because only one adjacent \ch{Si–N} pair is available, a single \ch{Si–N} bond forms, raising the total energy by $3.1$ eV relative to the ground state configuration.
The resulting defect shows strong magnetic correlations: its lowest manifold is a superposition of eigenstates
(i.e., $|\Psi\rangle = \alpha\,\bigl|S=\tfrac12\bigr\rangle + \beta\,\bigl|S=\tfrac32\bigr\rangle$).
Such spin mixing is typical of highly correlated electronic systems~\cite{Thiering_PRB2016}.  
The two remaining N atoms stay isolated and do not participate in additional bonding.
(iii) Two-bonded:
This configuration forms two \ch{Si–N} bonds, lies $0.42$ eV above the ground state, and carries total spin $S=\frac{3}{2}$.
(iv) Three-bonded:
The defect makes three equivalent \ch{Si–N} bonds of length $1.81$ \AA, which are $0.03$~\AA\ ($\approx1.7\%$) longer than the native \ch{Si–C} bond ($1.78$~\AA).
Its spin is $S=\frac{3}{2}$, and its energy is only $0.35$ eV higher than the ground state (Fig.~\ref{fig:vb}(a)).
(v) Four-bonded:
The energetically preferred structure contains four bonds: three \ch{Si–N} bonds
($1.81$ \AA, $1.81$ \AA, and $2.09$ \AA) and one \ch{C–N} bond ($1.61$ \AA).  It exhibits total spin $S= \frac{1}{2}$ and defines the zero of the relative-energy scale (Fig.~\ref{fig:vb}(b)).
The relative energies quoted above translate directly into differences in formation energy.

%parag3
The lowest-energy configuration (the four-bonded defect, shown in Fig.~\ref{fig:vb}(b)) arises when the absence of a B atom allows two neighboring N atoms to relax toward adjacent Si atoms, while the third N atom settles on the Si-C bond axis between the two sublattices. 
These findings reveal that, at the hBN/SiC interface, the \vb defect nucleates preferentially at sites where the local stacking registry promotes \ch{Si–N} bonding.
The driving force is the donor-acceptor character of the elements involved: Si is a stronger electron donor than C, which makes \ch{Si–N} bonds energetically more favorable~\cite{Rasool2015_AM}.

%A new parag
We performed MLMD simulations across a range of \vb densities.
For multivacancy cases, defects were randomly distributed.
The results are summarized in Fig. S3 and dynamically illustrated in Movie S2 (monovacancy) and Movie S3 (multivacancy).
At the lowest density ($\rho = 0.046~\rm{nm}^{-2}$), introducing a single B vacancy induces layer sliding, aligning Si atoms with opposing N radicals and spontaneously forming an interlayer \ch{Si–N} covalent bond that persists throughout the $5$ ns run.
These bonds pin the hBN monolayer to the SiC layer and suppress further sliding.
At higher vacancy concentrations, competing N radicals generate additional local strain; consequently, the majority form stable bonds, while the rest fluctuate between transiently coordinated and dangling states.

%parag4
A similar analysis for \vn reveals no interlayer chemical bond formation, only a weak tendency for B to remain near Si.
Overall, the total energies of different \vn defects are comparable, varying by at most $0.4$ eV across different sites.
The most stable configuration exhibits a \ch{Si-B} distance of $2.33$ \AA\ (see Fig.~\ref{fig:vb}(c)).
All the \vn defects display a spin state of $\frac{1}{2}$.

%parag5
By comparing the formation energies of defects (see Fig.~\ref{fig:vb}(d)), we find that \vb and \vn are most stable in N-rich and N-poor conditions, respectively.
However, \vn does not exist in a charge-neutral form.
Defect stability in monolayer hBN depends sensitively on the nitrogen chemical potential.
Under N-poor conditions, \vn is energetically preferred over \vb throughout the entire chemical-potential range.
In contrast, under N-rich conditions the two formation energies intersect near VBM + $2.4$ eV, beyond which \vb becomes the lower-energy defect \cite{Linderalv2021_PRB,Weston2018_aps}.
The large band gap of hBN further permits stabilization of a wider spectrum of charge states than is typical in narrower-gap hBN/SiC.

%parag6
Our analysis of the positively charged \vn reveals a significant weakening of the \ch{B-Si} interaction, with both the hBN and SiC monolayers remaining almost flat (i.e., \ch{B-Si} distance $\approx 3.2$ \AA).
For the negatively charged \vn defect, one B atom relaxes toward a neighboring Si, shortening the \ch{Si–B} distance to $2.09$ \AA.  At the same time, the two remaining B atoms approach one another, reaching an inter-boron separation of $2.05$ \AA, indicative of \ch{B-B} dimerisation.
The ($+$|$-$) charge-transition level lies $1.47$ eV above the valence-band maximum (VBM).
As expected, the four-bonded \vb is more stable than the three-bonded form.
The latter is stable in the +$1$ charge state at very low Fermi levels, up to VBM + $0.137$ eV. We focus on the most stable \vb, where a charge transition from ($0$|$-$) occurs at VBM + $0.67$ eV.
Comparison of the optimized geometries shows that only the longer \ch{Si-N} bond length shortens by 2 pm, while the others remain unchanged.

\begin{figure*}[htb!]
    \centering
    \includegraphics[scale=2.06]{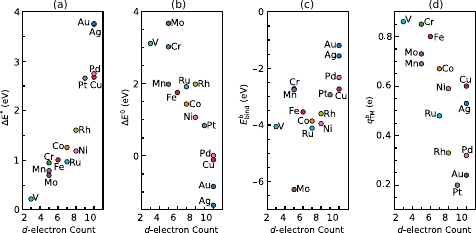}
    \caption{(a) Activation barrier ($\Delta E^{*}$) for converting the bonded hBN/SiC configuration to its non-bonded counterpart.
    (b) Total-energy difference between the bonded and non-bonded minima ($\Delta E^{0}$).
    (c) Binding energy of a single-atom transition metal (TM) to the chemically bonded hBN/SiC interface, $E_{\mathrm{bind}}^{\mathrm{b}}$.
    (d) Net Bader charge transferred from the TM atom to the interface in the bonded configuration, $q_{\mathrm{TM}}^{\mathrm{b}}$.}
    \label{fig:tmbonded-nonbonded}
\end{figure*}
%
%parag7
The electronic structures of the defects in the ground state are shown in Fig.~\ref{fig:vb}(e-g).
Defects introduce states into the electronic gap.
Electron injection into the \emph{three-bonded} \vb defect draws the lower energy defect states downward to the VBM and simultaneously pushes the upper defect level upward toward the conduction-band minimum (CBM); the two lower-lying levels remain degenerate.
In the \emph{four-bonded} \vb, the unoccupied states are near the VBM, and adding electrons moves them closer to the VBM until they merge.
In contrast, \vn exhibits a different electronic configuration, where two empty states in separate channels remain stable in the positively charged state.
Adding extra electrons brings these states closer to the VBM, effectively clearing the gap.

%parag8
For bilayer hBN, the formation of interlayer covalent bonds between the \ch{B-N} columns at the edges of monovacancies has been observed in \textit{highly} negatively charged defects, with bond lengths ranging from $1.61$ to $1.64$ \AA~\cite{PhysRevLett.106.126102}; Nitrogen vacancies do not form interlayer bonds.
To compare the likelihood of interlayer covalent bond formation between hBN/SiC and hBN/hBN in the presence of \vb, we present the configuration coordinate diagram for the heterostructure, AA$'$-stacked hBN, and $14^{\circ}$-rotated bilayer hBN---chosen for its local AA$^{\prime}$ stacking and as the smallest twist angle that remains computationally feasible---in Figs.~\ref{fig:vb}(h)-(j).
In the symmetric hBN/hBN systems, the reaction pathway reduces to a one-dimensional coordinate.
Accordingly, we rigidly displaced the six atoms that form the prospective interlayer bond (three B and three N stacked directly above one another) along their interlayer difference vector, while allowing all other atoms to relax.

%parag9
The hBN/SiC heterostructure minimizes energy without a barrier through \ch{Si-N} bond formation in the low-spin state, as previously discussed.
Our partial charge calculations reveal that $2.03\,e$ ($3.773 \times 10^{13}\,e/{\rm{cm}}^{2}$) is transferred from SiC to hBN upon chemical bond formation.
In contrast, a neutral \vb in hBN/hBN systems does not induce structural changes in the surrounding atoms, with bond formation occurring only as a metastable state at an energy level approximately $1.5$ eV higher.
Additionally, the total energies of the low-spin and high-spin states stay close ($\approx 0.2~\rm{eV}$), with a transition from low-spin to high-spin occurring as the atoms move closer together to form bonds.

%%%%%%%%%%%%%%%%%%%%%%%%%%%%%%%%%%%%%%%%%%%%%%%%%%%%%%%%%%%%%%%%%%%%%%%%%%%%%%%%
\subsection{Transition Metal-Doped \texorpdfstring{\MakeLowercase{h}BN/S\MakeLowercase{i}C}{hBN/SiC}}
\label{sec:tmdoping}

%parag1
Introducing \vb leaves the hBN sheet electron-deficient.
The charge imbalance drives the spontaneous formation of interlayer \ch{Si-N} bonds.
This insight prompts a key question: can a \emph{single} TM atom migrate on the same donor–acceptor pathway to anchor to an isolated \vb site rather than agglomerating into clusters?
If successful, it would provide a rational route for dispersing catalytic TM centers at the single-atom limit.

%parag2
Our calculations show that TM atoms placed at \vb act as local electron reservoirs, compensating for the vacancy's electron deficit while reshaping the interlayer bonding network.
Depending on the dopant, the system may preserve, weaken, or altogether break the newly formed \ch{Si-N} bridges, creating well-defined structural distortions and a rich spectrum of electronic and magnetic states.
A detailed understanding of these TM-decorated point defects, both with and without interlayer bonds, is therefore essential.

%parag3
To quantify the stability of the interlayer covalent bond and the TM anchoring site, we map the potential-energy profile along the structural transformation pathway.
To this end, for each TM species, we could fully relax both the bonded and non-bonded hBN/SiC configurations, verifying that the resulting geometries were true local minima separated by barriers that were not crossed during subsequent DFT optimizations.
We then constructed configuration-coordinate diagrams by incrementally displacing the atoms along the geometric difference vector between the bonded and non-bonded structures.
All relevant spin multiplicities are treated explicitly.
As shown in Fig. S4, this high-throughput protocol yields quantitative insight into the TM-binding stability.

\begin{figure*}[htb!]
    \centering
    \includegraphics[scale=0.9]{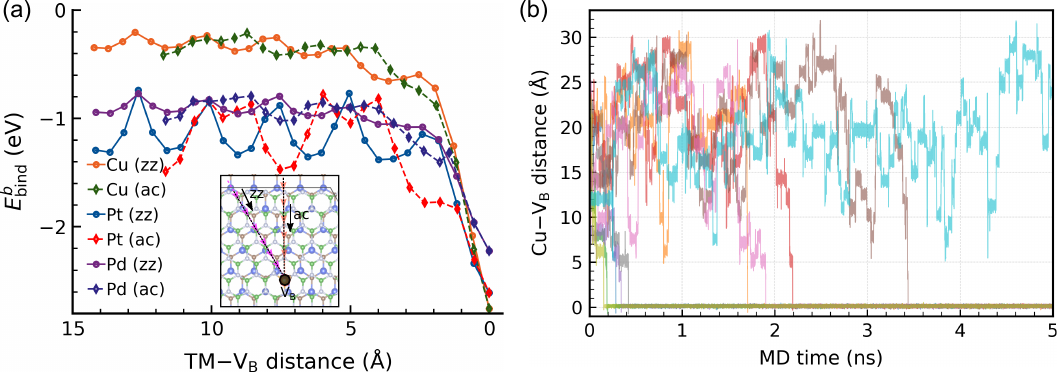}
    \caption{(a) Binding energies of Cu, Pd, and Pt as a function of their distance from the \vb center along the zigzag (zz) and armchair (ac) directions of the hBN layer.
    During the geometry optimizations, the SiC substrate is fixed, while atoms in the hBN layer and the transition metal are allowed to relax along the $z$-direction (perpendicular to the surface).
    (b) Time evolution of the Cu$-$V$_{\mathrm{B}}$ distance extracted from ten independent MLMD simulations at 300 K. Each trace represents one run, and a separation of zero denotes the Cu atom occupying the vacancy site.}
    \label{fig:tmdrag}
\end{figure*}
%
%parag4
There are also the partial charges transferred from a single TM atom to the substrates (denoted as $q_{\mathrm{TM}}^{\rm b}$ and $q_{\mathrm{TM}}^{\rm nb}$ for bonded and non-bonded configurations, respectively) and $E_{\mathrm{bind}}$. Also, two additional energy-related quantities are reported: the energy difference between the bonded and non-bonded systems ($\Delta E^{0} = E^{\rm b} - E^{\rm nb}$) and the transition energy barrier ($\Delta E^{*} = E^{\text{max}} - E^{\rm b}$).
Here, $E^{\rm b}$, $E^{\rm nb}$, and $E^{\text{max}}$ denote the total energy of the optimized structure with interlayer chemical bonds formed, the total energy of the optimized structure without the \ch{Si-N} chemical bonds, and the maximum total energy along the configuration-coordinate diagrams relative to the bonded system, respectively.
These quantities provide insights into the bond formation and dissociation processes.
A more negative $\Delta E^{0}$ indicates a more stable bonded structure, while a smaller $\Delta E^{*}$ suggests an easier transition from the bonded to the non-bonded configuration.

%parag5
Notably, unlike \vb in hBN/SiC, which spontaneously forms chemical bonds, the presence of TMs introduces an energy barrier to interlayer bond formation (see Fig.~\ref{fig:tmbonded-nonbonded} (a)).
Thus, the larger the barrier, the slower the kinetics of structural transition.
We found a direct correlation between the number of $d$-electrons and the value of $\Delta E^{*}$.
The energy barrier ranges from $0.15$ to $3.76$ eV.
Among the TMs studied, the largest barriers are found for Au ($3.76$ eV), Ag ($3.75$ eV), Pd ($2.75$ eV), Cu ($2.68$ eV), and Pt ($2.66$ eV).

%parag6
In contrast to $\Delta E^{*}$, $\Delta E^{0}$ decreases as the number of $d$-electrons increases (see Fig.~\ref{fig:tmbonded-nonbonded} (b)). Negative $\Delta E^{0}$ values show that the bonded configuration is thermodynamically favored for Cu ($-0.11$ eV), Ag ($-1.37$ eV), and Au ($-0.85$ eV).
By comparison, Mo, V, and Cr favor the non-bonded configuration, which lies $3.68$ eV, $3.12$ eV, and $3.03$ eV, respectively.

%parag7
Two different fabrication routes may control the bonding at TM-doped hBN/SiC interfaces,
focused here on Au, Ag, and Cu, but also applicable to other metals:
(i) \emph{Bonded interface}: A \vb-containing hBN monolayer is first deposited onto the SiC substrate, followed by the introduction of TM atoms.
This approach maintains interlayer bonding.
(ii) \emph{Non-bonded interface}: TM atoms are first incorporated into the hBN monolayer, which is subsequently placed onto the SiC substrate. This sequence tends to prevent interlayer bonding.

%parag8
Binding energy is another parameter indicating the strength of metal attachment to the interface and the likelihood of TM rejection.
The data can be categorized into three groups: weakly, moderately, and strongly bound transition metals (Fig.~\ref{fig:tmbonded-nonbonded}(c)).
For instance, Au and Ag belong to the weakly bound group, suggesting limited orbital overlap with the interface.
Comparing the $E_{\mathrm{bind}}^{\rm b}$ values to those of the hBN monolayer (summarized in Table S4) reveals that TM binding to the \vb site is approximately 1.5 to 3.5 times stronger in the monolayer system compared to bonded hBN/SiC.
This difference indicates greater electron sharing between the TMs and the host atoms, resulting in stronger covalent bonding in the hBN monolayer.
It is evident that the formation of interlayer chemical bonds partially compensates for the electron deficiency, reducing the need for intense electron donation from the transition metals.

%parag9
The net Bader charge transferred from the TM atom to the interface ($q_{\mathrm{TM}}$) is a key parameter for gauging electron donation and chemical reactivity at the defect site.
Overall, charge transfer decreases as the number of $d$-electrons increases (Fig.~\ref{fig:tmbonded-nonbonded}(d)).
Accordingly, Pt, Au, and Pd atoms, that donate little charge to the bonded hBN/SiC substrate, retain more electrons that can participate in subsequent electrochemical reactions.
Our results show that TMs donate $1.4-3.6$ times more charge to the substrates in both the non-bonded hBN/SiC and monolayer hBN compared to the bonded hBN/SiC.
For instance, Pt transfers $0.71~e$ to non-bonded hBN/SiC, $0.70~e$ to monolayer hBN, but only $0.20~e$ to bonded hBN/SiC.

%%%%%%%%%%%%%%%%%%%%%%%%%%%%%%%%%%%%%%%%%%%%%%%%%%%%%%%%%%%%%
\subsection{Copper Isolation and Surface Decoration}
\label{cu-isolation}

%
%parag1
The performance and cost of electrocatalysts are key factors driving the development and deployment of electrochemical devices~\cite{ZhouChemCatChem2023,D1TA08252C_rsc,Weon2021_acs}.
Maximizing atomic utilization while minimizing precious metal content requires precise control over atom coordination and spacing to enhance charge transfer, avoid surface passivation, and prevent agglomeration of active species.
However, designing optimal single-atom catalytic platforms remains a major challenge, as isolated metal atoms tend to coalesce into clusters or three-dimensional structures~\cite{WU2023100923,Dong_2021_jpclett,Jeongacsnano_2020,Li_JACSAu_2023}.

%parag2
The hBN/SiC substrate offers a versatile platform for anchoring TM atoms to \vb sites, the chemically active traps of which stabilize adatoms via robust, multidirectional bonds.
Yet practical isolation of single atoms remains unresolved.
To probe the hurdle, we first model Cu, Pt, and Pd adatoms migrating toward a boron vacancy on hBN/SiC (Fig.~\ref{fig:tmdrag}(a)).
Geometry optimizations keep the SiC lattice fixed while allowing hBN and metal atoms to relax only along the surface normal ($z$-axis); this constraint can slightly shift binding energies compared with full relaxation.
Cu, Pt, and Pd are chosen because prior analysis flagged them as the most integration-ready metals.

%parag3
Copper binds strongly at \vb ($-2.73$ eV) yet only weakly to pristine hBN ($E_{\mathrm{bind}}^{\rm b} > -0.4$ eV).
Its diffusion on the pristine surface is virtually barrierless, enabling rapid, thermodynamically favorable migration.
Consistent with recent work on monolayer hBN \cite{Zhang2024_Langmuir}, TMs favor N sites over B sites for binding; for Cu this preference is $\approx 0.22$ eV, and for Pd, adsorption is even stronger ($-0.95$ eV vs $-0.30$ eV for Cu).
Pt shows marked site sensitivity, with the N site about $0.65$ eV more stable than the hexagon center.
Strong vacancy binding, weak affinity for the pristine surface, and low cost make Cu our prime candidate for further MLMD simulations.

%parag4
Figure~\ref{fig:tmdrag}(b) presents a kinetic map of a Cu atom interacting with \vb sites in hBN/SiC at room temperature and dilute concentration ($0.012~\text{nm}^{-2}$).
In nearly all simulations, except one with a cyan trace, Cu atoms, placed at various random initial distances from the vacancy, integrate with the vacancy within $5$ ns.
On the pristine surface, Cu exhibits longer lifetimes at N sites during diffusion, consistent with our earlier binding strength results.

%parag5
By systematically varying the densities of vacancies and Cu atoms, we identified various scenarios (see Movies S4-S6 as examples):
(i) At low density, a single Cu atom rapidly migrates to a \vb vacancy, which acts as a deep potential well that dominates over the flat diffusion landscape of pristine hBN/SiC, enabling single-atom anchoring.
(ii) As Cu density increases, the reduced mean free path between adatoms accelerates aggregation, forming clusters that migrate en bloc to vacancies.
(iii) Cu atoms sequentially accumulate at a single \vb site, forming a cluster directly at the vacancy.
(iv) Cu atoms disperse across multiple \vb sites in varying ratios.
Overall, once metals are trapped in vacancies, their migration ceases.

%%%%%%%%%%%%%%%%%%%%%%%%%%%%%%%%%%%%%%%%%%%%%%%%%%%%%%%%%%%%%%
\section{Summary and Conclusions}
\label{sec:conclusions}
%parag1
Using a combination of density functional theory calculations and machine-learning molecular dynamics, we have mapped the structural, electronic, and (meta-)stable defect landscapes of the vdW hBN/SiC heterostructure and showed how selective point-defect chemistry can transform an otherwise mechanically flexible, electronically inert interface into a versatile platform for single-atom catalysis.

%parag2
Creating a \vb initiates a spontaneous, barrier-free reconstruction that forms up to four interlayer covalent bonds (three \ch{Si-N} and one \ch{C-N}).
The bonding pattern depends on the atomic registry imposed by the Moir\'e pattern.
Decorating \vb with a single transition-metal atom modulates both the stability of these interlayer bonds and the charge reservoir available to the adatom.
The activation barrier for rupturing the \ch{Si-N} bridges rises monotonically with $d$-band filling, peaking for Au.
At the same time, the thermodynamic driving force changes sign across the late $d$-series, favoring the interlayer bonded geometry for Cu, Ag, and Au but the non-bonded geometry for Mo, V, and Cr. Because electron donation decreases as the $d$-band fills, Pt, Au, and Pd emerge as particularly promising active centers that retain their valence electrons for catalytic turnover.

%parag3
Copper, chosen for its natural abundance, modest cost, and favorable binding energetics, diffuses essentially barrier-free across pristine hBN yet becomes irreversibly trapped at a single \vb within nanoseconds.
Machine-learning molecular dynamics shows that, depending on the \vb : Cu density ratio, the system evolves from single-atom isolation to vacancy-directed aggregation to on-site cluster growth.

%parag4
This study suggests two practical routes for tuning surface activity:
(i) deposit \vb-rich hBN first and then metallize it to lock in the bonded configuration, or
(ii) pre-decorate hBN with transition metals before transferring it onto SiC, thereby suppressing \ch{Si-N} bond formation.
By contrast, the latter strategy maximizes adatom charge donation, strengthening binding but lowering chemical reactivity.

%parag5 "outlook1"
Stability concerns associated with freestanding monolayer SiC can be mitigated by transferring hBN; for instance, onto Si-terminated 4H-SiC(0001), as demonstrated in Section IX of the SM. Whereas bulk hexagonal SiC naturally adopts a buckled structure, in which the Si and C sub-lattices occupy slightly offset planes, a freestanding SiC monolayer instead forms a strictly planar honeycomb lattice. Our preliminary results indicate that the 4H-SiC slab closely replicates the electronic and structural properties of monolayer SiC when interfaced with hBN.
We hypothesize that SiC slabs, independent of polytypes or surface orientation, should offer experimentally accessible platforms for stabilizing hBN/SiC interfaces, given that SiC surfaces have demonstrated significant promise as substrates for the epitaxial growth of 2D materials~\cite{mokhov2023growth,zhao2024_nature}.
Yet, further investigation is required to fully realize the potential of this promising coupling.

Several studies have investigated the growth of hBN thin films on Si-terminated SiC substrates~\cite{Lin_PRM_2022,Biswas_AdvMat_2023}, showing that the crystallinity and the formation of uniform, atomically thin layers are highly sensitive to the growth process and surface preparation. In agreement with our results, interfacial defects can significantly reduce the interlayer spacing, lowering film quality. To overcome these issues, we propose a metallization strategy in which a transition-metal precursor is supplied either together with borazine or in separate steps. This approach passivates the reactive Si surface while allowing controlled single-atom deposition, leading in our study to notable improvements in film uniformity and structural quality.

%parag6 "outlook2"
Taken together, these insights position hBN/SiC interface as a defect-programmable, mechanically resilient platform that can immobilize single atoms for catalysis today and, with targeted vacancy engineering, host quantum centers tomorrow.
Its atomically thin geometry further enables deterministic defect decoration and site-selective interlayer bonding, capabilities that could be harnessed in future quantum devices.
Next steps may also include quantifying catalytic turnover under operando electrochemical conditions and, beyond catalysis, extending the strategy to alternative defects (e.g., carbon or oxygen) for qubit or single-photon emitter applications.

\section{Acknowledgments}
\label{sect:ack}
We are grateful to  CSC--IT Center for Science Ltd. and
Aalto Science-IT project for generous grants of computer time.
A.H., Y.W. and T.A-N. have been supported by the Academy of Finland through its QTF Center of Excellence program (project no. 312298). Y.W. and T.A-N. have also been supported in part by the Academy of Finland's Grant No. 353298 under the European Union – NextGenerationEU instrument. T.A-N. further acknowledges support from the Academy of Finland Grant no. 370057.

\section{Data Availability}
The data that support the findings of this study are openly available in A.H.'s GitHub repository (\href{https://github.com/sahashemip/cu-hbn-sic-ml-potential}{GitHub}) and are archived on Zenodo (\href{https://doi.org/10.5281/zenodo.17102508}{Zenodo}).

\bibliographystyle{apsrev4-2}
\bibliography{ars}

\begin{thebibliography}{127}%
\makeatletter
\providecommand \@ifxundefined [1]{%
 \@ifx{#1\undefined}
}%
\providecommand \@ifnum [1]{%
 \ifnum #1\expandafter \@firstoftwo
 \else \expandafter \@secondoftwo
 \fi
}%
\providecommand \@ifx [1]{%
 \ifx #1\expandafter \@firstoftwo
 \else \expandafter \@secondoftwo
 \fi
}%
\providecommand \natexlab [1]{#1}%
\providecommand \enquote  [1]{``#1''}%
\providecommand \bibnamefont  [1]{#1}%
\providecommand \bibfnamefont [1]{#1}%
\providecommand \citenamefont [1]{#1}%
\providecommand \href@noop [0]{\@secondoftwo}%
\providecommand \href [0]{\begingroup \@sanitize@url \@href}%
\providecommand \@href[1]{\@@startlink{#1}\@@href}%
\providecommand \@@href[1]{\endgroup#1\@@endlink}%
\providecommand \@sanitize@url [0]{\catcode `\\12\catcode `\$12\catcode `\&12\catcode `\#12\catcode `\^12\catcode `\_12\catcode `\%12\relax}%
\providecommand \@@startlink[1]{}%
\providecommand \@@endlink[0]{}%
\providecommand \url  [0]{\begingroup\@sanitize@url \@url }%
\providecommand \@url [1]{\endgroup\@href {#1}{\urlprefix }}%
\providecommand \urlprefix  [0]{URL }%
\providecommand \Eprint [0]{\href }%
\providecommand \doibase [0]{https://doi.org/}%
\providecommand \selectlanguage [0]{\@gobble}%
\providecommand \bibinfo  [0]{\@secondoftwo}%
\providecommand \bibfield  [0]{\@secondoftwo}%
\providecommand \translation [1]{[#1]}%
\providecommand \BibitemOpen [0]{}%
\providecommand \bibitemStop [0]{}%
\providecommand \bibitemNoStop [0]{.\EOS\space}%
\providecommand \EOS [0]{\spacefactor3000\relax}%
\providecommand \BibitemShut  [1]{\csname bibitem#1\endcsname}%
\let\auto@bib@innerbib\@empty
%</preamble>
\bibitem [{\citenamefont {Geim}\ and\ \citenamefont {Grigorieva}(2013)}]{geim2013van}%
  \BibitemOpen
  \bibfield  {author} {\bibinfo {author} {\bibfnamefont {A.~K.}\ \bibnamefont {Geim}}\ and\ \bibinfo {author} {\bibfnamefont {I.~V.}\ \bibnamefont {Grigorieva}},\ }\href {https://doi.org/https://doi.org/10.1038/nature12385} {\bibfield  {journal} {\bibinfo  {journal} {Nature}\ }\textbf {\bibinfo {volume} {499}},\ \bibinfo {pages} {419} (\bibinfo {year} {2013})}\BibitemShut {NoStop}%
\bibitem [{\citenamefont {You}\ \emph {et~al.}(2021)\citenamefont {You}, \citenamefont {Gu}, \citenamefont {Su},\ and\ \citenamefont {Feng}}]{jingyang2021}%
  \BibitemOpen
  \bibfield  {author} {\bibinfo {author} {\bibfnamefont {J.-Y.}\ \bibnamefont {You}}, \bibinfo {author} {\bibfnamefont {B.}~\bibnamefont {Gu}}, \bibinfo {author} {\bibfnamefont {G.}~\bibnamefont {Su}},\ and\ \bibinfo {author} {\bibfnamefont {Y.~P.}\ \bibnamefont {Feng}},\ }\href {https://doi.org/10.1103/PhysRevB.103.104503} {\bibfield  {journal} {\bibinfo  {journal} {Phys. Rev. B}\ }\textbf {\bibinfo {volume} {103}},\ \bibinfo {pages} {104503} (\bibinfo {year} {2021})}\BibitemShut {NoStop}%
\bibitem [{\citenamefont {Burch}\ \emph {et~al.}(2018)\citenamefont {Burch}, \citenamefont {Mandrus},\ and\ \citenamefont {Park}}]{burch2018}%
  \BibitemOpen
  \bibfield  {author} {\bibinfo {author} {\bibfnamefont {K.~S.}\ \bibnamefont {Burch}}, \bibinfo {author} {\bibfnamefont {D.}~\bibnamefont {Mandrus}},\ and\ \bibinfo {author} {\bibfnamefont {J.-G.}\ \bibnamefont {Park}},\ }\href {https://doi.org/10.1038/s41586-018-0631-z} {\bibfield  {journal} {\bibinfo  {journal} {Nature}\ }\textbf {\bibinfo {volume} {563}},\ \bibinfo {pages} {47} (\bibinfo {year} {2018})}\BibitemShut {NoStop}%
\bibitem [{\citenamefont {Qi}\ \emph {et~al.}(2023)\citenamefont {Qi}, \citenamefont {Wu}, \citenamefont {Wang}, \citenamefont {Bao}, \citenamefont {Wang}, \citenamefont {Wu}, \citenamefont {Ke}, \citenamefont {Xu},\ and\ \citenamefont {He}}]{Qi_2023}%
  \BibitemOpen
  \bibfield  {author} {\bibinfo {author} {\bibfnamefont {J.}~\bibnamefont {Qi}}, \bibinfo {author} {\bibfnamefont {Z.}~\bibnamefont {Wu}}, \bibinfo {author} {\bibfnamefont {W.}~\bibnamefont {Wang}}, \bibinfo {author} {\bibfnamefont {K.}~\bibnamefont {Bao}}, \bibinfo {author} {\bibfnamefont {L.}~\bibnamefont {Wang}}, \bibinfo {author} {\bibfnamefont {J.}~\bibnamefont {Wu}}, \bibinfo {author} {\bibfnamefont {C.}~\bibnamefont {Ke}}, \bibinfo {author} {\bibfnamefont {Y.}~\bibnamefont {Xu}},\ and\ \bibinfo {author} {\bibfnamefont {Q.}~\bibnamefont {He}},\ }\href {https://doi.org/10.1088/2631-7990/acc8a1} {\bibfield  {journal} {\bibinfo  {journal} {International Journal of Extreme Manufacturing}\ }\textbf {\bibinfo {volume} {5}},\ \bibinfo {pages} {022007} (\bibinfo {year} {2023})}\BibitemShut {NoStop}%
\bibitem [{\citenamefont {Castellanos-Gomez}\ \emph {et~al.}(2022)\citenamefont {Castellanos-Gomez}, \citenamefont {Duan}, \citenamefont {Fei}, \citenamefont {Rodriguez~Gutierrez}, \citenamefont {Huang}, \citenamefont {Huang}, \citenamefont {Quereda}, \citenamefont {Qian}, \citenamefont {Sutter},\ and\ \citenamefont {Sutter}}]{CastellanosGomez2022}%
  \BibitemOpen
  \bibfield  {author} {\bibinfo {author} {\bibfnamefont {A.}~\bibnamefont {Castellanos-Gomez}}, \bibinfo {author} {\bibfnamefont {X.}~\bibnamefont {Duan}}, \bibinfo {author} {\bibfnamefont {Z.}~\bibnamefont {Fei}}, \bibinfo {author} {\bibfnamefont {H.}~\bibnamefont {Rodriguez~Gutierrez}}, \bibinfo {author} {\bibfnamefont {Y.}~\bibnamefont {Huang}}, \bibinfo {author} {\bibfnamefont {X.}~\bibnamefont {Huang}}, \bibinfo {author} {\bibfnamefont {J.}~\bibnamefont {Quereda}}, \bibinfo {author} {\bibfnamefont {Q.}~\bibnamefont {Qian}}, \bibinfo {author} {\bibfnamefont {E.}~\bibnamefont {Sutter}},\ and\ \bibinfo {author} {\bibfnamefont {P.}~\bibnamefont {Sutter}},\ }\href {https://doi.org/10.1038/s43586-022-00139-1} {\bibfield  {journal} {\bibinfo  {journal} {Nature Reviews Methods Primers}\ }\textbf {\bibinfo {volume} {2}},\ \bibinfo {pages} {58} (\bibinfo {year} {2022})}\BibitemShut {NoStop}%
\bibitem [{\citenamefont {Cheng}\ \emph {et~al.}(2018)\citenamefont {Cheng}, \citenamefont {Wang}, \citenamefont {Yin}, \citenamefont {Wang}, \citenamefont {Wen}, \citenamefont {Shifa},\ and\ \citenamefont {He}}]{cheng2018high}%
  \BibitemOpen
  \bibfield  {author} {\bibinfo {author} {\bibfnamefont {R.}~\bibnamefont {Cheng}}, \bibinfo {author} {\bibfnamefont {F.}~\bibnamefont {Wang}}, \bibinfo {author} {\bibfnamefont {L.}~\bibnamefont {Yin}}, \bibinfo {author} {\bibfnamefont {Z.}~\bibnamefont {Wang}}, \bibinfo {author} {\bibfnamefont {Y.}~\bibnamefont {Wen}}, \bibinfo {author} {\bibfnamefont {T.~A.}\ \bibnamefont {Shifa}},\ and\ \bibinfo {author} {\bibfnamefont {J.}~\bibnamefont {He}},\ }\href {https://doi.org/10.1038/s41928-018-0086-0} {\bibfield  {journal} {\bibinfo  {journal} {Nature Electronics}\ }\textbf {\bibinfo {volume} {1}},\ \bibinfo {pages} {356} (\bibinfo {year} {2018})}\BibitemShut {NoStop}%
\bibitem [{\citenamefont {Sun}\ \emph {et~al.}(2022)\citenamefont {Sun}, \citenamefont {Zhu}, \citenamefont {Yi}, \citenamefont {Xiang}, \citenamefont {Ma}, \citenamefont {Liu}, \citenamefont {Zheng}, \citenamefont {Liu}, \citenamefont {You}, \citenamefont {Zhang} \emph {et~al.}}]{sun2022reconfigurable}%
  \BibitemOpen
  \bibfield  {author} {\bibinfo {author} {\bibfnamefont {X.}~\bibnamefont {Sun}}, \bibinfo {author} {\bibfnamefont {C.}~\bibnamefont {Zhu}}, \bibinfo {author} {\bibfnamefont {J.}~\bibnamefont {Yi}}, \bibinfo {author} {\bibfnamefont {L.}~\bibnamefont {Xiang}}, \bibinfo {author} {\bibfnamefont {C.}~\bibnamefont {Ma}}, \bibinfo {author} {\bibfnamefont {H.}~\bibnamefont {Liu}}, \bibinfo {author} {\bibfnamefont {B.}~\bibnamefont {Zheng}}, \bibinfo {author} {\bibfnamefont {Y.}~\bibnamefont {Liu}}, \bibinfo {author} {\bibfnamefont {W.}~\bibnamefont {You}}, \bibinfo {author} {\bibfnamefont {W.}~\bibnamefont {Zhang}}, \emph {et~al.},\ }\href {https://doi.org/10.1038/s41928-022-00858-z} {\bibfield  {journal} {\bibinfo  {journal} {Nature Electronics}\ }\textbf {\bibinfo {volume} {5}},\ \bibinfo {pages} {752} (\bibinfo {year} {2022})}\BibitemShut {NoStop}%
\bibitem [{\citenamefont {Liu}\ \emph {et~al.}(2016)\citenamefont {Liu}, \citenamefont {Weiss}, \citenamefont {Duan}, \citenamefont {Cheng}, \citenamefont {Huang},\ and\ \citenamefont {Duan}}]{liu2016van}%
  \BibitemOpen
  \bibfield  {author} {\bibinfo {author} {\bibfnamefont {Y.}~\bibnamefont {Liu}}, \bibinfo {author} {\bibfnamefont {N.~O.}\ \bibnamefont {Weiss}}, \bibinfo {author} {\bibfnamefont {X.}~\bibnamefont {Duan}}, \bibinfo {author} {\bibfnamefont {H.-C.}\ \bibnamefont {Cheng}}, \bibinfo {author} {\bibfnamefont {Y.}~\bibnamefont {Huang}},\ and\ \bibinfo {author} {\bibfnamefont {X.}~\bibnamefont {Duan}},\ }\href {https://doi.org/10.1038/natrevmats.2016.42} {\bibfield  {journal} {\bibinfo  {journal} {Nature Reviews Materials}\ }\textbf {\bibinfo {volume} {1}},\ \bibinfo {pages} {1} (\bibinfo {year} {2016})}\BibitemShut {NoStop}%
\bibitem [{\citenamefont {Mounet}\ \emph {et~al.}(2018)\citenamefont {Mounet}, \citenamefont {Gibertini}, \citenamefont {Schwaller}, \citenamefont {Campi}, \citenamefont {Merkys}, \citenamefont {Marrazzo}, \citenamefont {Sohier}, \citenamefont {Castelli}, \citenamefont {Cepellotti}, \citenamefont {Pizzi} \emph {et~al.}}]{mounet2018two}%
  \BibitemOpen
  \bibfield  {author} {\bibinfo {author} {\bibfnamefont {N.}~\bibnamefont {Mounet}}, \bibinfo {author} {\bibfnamefont {M.}~\bibnamefont {Gibertini}}, \bibinfo {author} {\bibfnamefont {P.}~\bibnamefont {Schwaller}}, \bibinfo {author} {\bibfnamefont {D.}~\bibnamefont {Campi}}, \bibinfo {author} {\bibfnamefont {A.}~\bibnamefont {Merkys}}, \bibinfo {author} {\bibfnamefont {A.}~\bibnamefont {Marrazzo}}, \bibinfo {author} {\bibfnamefont {T.}~\bibnamefont {Sohier}}, \bibinfo {author} {\bibfnamefont {I.~E.}\ \bibnamefont {Castelli}}, \bibinfo {author} {\bibfnamefont {A.}~\bibnamefont {Cepellotti}}, \bibinfo {author} {\bibfnamefont {G.}~\bibnamefont {Pizzi}}, \emph {et~al.},\ }\href {https://doi.org/https://doi.org/10.1038/s41565-017-0035-5} {\bibfield  {journal} {\bibinfo  {journal} {Nature nanotechnology}\ }\textbf {\bibinfo {volume} {13}},\ \bibinfo {pages} {246} (\bibinfo {year} {2018})}\BibitemShut {NoStop}%
\bibitem [{\citenamefont {Yang}\ \emph {et~al.}(2022)\citenamefont {Yang}, \citenamefont {Valenzuela}, \citenamefont {Chshiev}, \citenamefont {Couet}, \citenamefont {Dieny}, \citenamefont {Dlubak}, \citenamefont {Fert}, \citenamefont {Garello}, \citenamefont {Jamet}, \citenamefont {Jeong} \emph {et~al.}}]{yang2022two}%
  \BibitemOpen
  \bibfield  {author} {\bibinfo {author} {\bibfnamefont {H.}~\bibnamefont {Yang}}, \bibinfo {author} {\bibfnamefont {S.~O.}\ \bibnamefont {Valenzuela}}, \bibinfo {author} {\bibfnamefont {M.}~\bibnamefont {Chshiev}}, \bibinfo {author} {\bibfnamefont {S.}~\bibnamefont {Couet}}, \bibinfo {author} {\bibfnamefont {B.}~\bibnamefont {Dieny}}, \bibinfo {author} {\bibfnamefont {B.}~\bibnamefont {Dlubak}}, \bibinfo {author} {\bibfnamefont {A.}~\bibnamefont {Fert}}, \bibinfo {author} {\bibfnamefont {K.}~\bibnamefont {Garello}}, \bibinfo {author} {\bibfnamefont {M.}~\bibnamefont {Jamet}}, \bibinfo {author} {\bibfnamefont {D.-E.}\ \bibnamefont {Jeong}}, \emph {et~al.},\ }\href {https://doi.org/https://doi.org/10.1038/s41586-022-04768-0} {\bibfield  {journal} {\bibinfo  {journal} {Nature}\ }\textbf {\bibinfo {volume} {606}},\ \bibinfo {pages} {663} (\bibinfo {year} {2022})}\BibitemShut {NoStop}%
\bibitem [{\citenamefont {Liu}\ \emph {et~al.}(2021)\citenamefont {Liu}, \citenamefont {Duan}, \citenamefont {Shin}, \citenamefont {Park}, \citenamefont {Huang},\ and\ \citenamefont {Duan}}]{liu2021promises}%
  \BibitemOpen
  \bibfield  {author} {\bibinfo {author} {\bibfnamefont {Y.}~\bibnamefont {Liu}}, \bibinfo {author} {\bibfnamefont {X.}~\bibnamefont {Duan}}, \bibinfo {author} {\bibfnamefont {H.-J.}\ \bibnamefont {Shin}}, \bibinfo {author} {\bibfnamefont {S.}~\bibnamefont {Park}}, \bibinfo {author} {\bibfnamefont {Y.}~\bibnamefont {Huang}},\ and\ \bibinfo {author} {\bibfnamefont {X.}~\bibnamefont {Duan}},\ }\href {https://doi.org/https://doi.org/10.1038/s41586-021-03339-z} {\bibfield  {journal} {\bibinfo  {journal} {Nature}\ }\textbf {\bibinfo {volume} {591}},\ \bibinfo {pages} {43} (\bibinfo {year} {2021})}\BibitemShut {NoStop}%
\bibitem [{\citenamefont {Hybertsen}(1990)}]{Hybertsen1990_APS}%
  \BibitemOpen
  \bibfield  {author} {\bibinfo {author} {\bibfnamefont {M.~S.}\ \bibnamefont {Hybertsen}},\ }\href {https://doi.org/10.1103/PhysRevLett.64.555} {\bibfield  {journal} {\bibinfo  {journal} {Phys. Rev. Lett.}\ }\textbf {\bibinfo {volume} {64}},\ \bibinfo {pages} {555} (\bibinfo {year} {1990})}\BibitemShut {NoStop}%
\bibitem [{\citenamefont {Lu}\ \emph {et~al.}(2022)\citenamefont {Lu}, \citenamefont {Chen}, \citenamefont {Coupin}, \citenamefont {Sinha},\ and\ \citenamefont {Warner}}]{Lu2022_AM}%
  \BibitemOpen
  \bibfield  {author} {\bibinfo {author} {\bibfnamefont {Y.}~\bibnamefont {Lu}}, \bibinfo {author} {\bibfnamefont {J.}~\bibnamefont {Chen}}, \bibinfo {author} {\bibfnamefont {M.~J.}\ \bibnamefont {Coupin}}, \bibinfo {author} {\bibfnamefont {S.}~\bibnamefont {Sinha}},\ and\ \bibinfo {author} {\bibfnamefont {J.~H.}\ \bibnamefont {Warner}},\ }\href {https://doi.org/https://doi.org/10.1002/adma.202205403} {\bibfield  {journal} {\bibinfo  {journal} {Advanced Materials}\ }\textbf {\bibinfo {volume} {34}},\ \bibinfo {pages} {2205403} (\bibinfo {year} {2022})}\BibitemShut {NoStop}%
\bibitem [{\citenamefont {Li}\ \emph {et~al.}(2024{\natexlab{a}})\citenamefont {Li}, \citenamefont {Brumme},\ and\ \citenamefont {Heine}}]{li2024relaxation}%
  \BibitemOpen
  \bibfield  {author} {\bibinfo {author} {\bibfnamefont {W.}~\bibnamefont {Li}}, \bibinfo {author} {\bibfnamefont {T.}~\bibnamefont {Brumme}},\ and\ \bibinfo {author} {\bibfnamefont {T.}~\bibnamefont {Heine}},\ }\href {https://doi.org/https://doi.org/10.1038/s41699-024-00477-6} {\bibfield  {journal} {\bibinfo  {journal} {npj 2D Materials and Applications}\ }\textbf {\bibinfo {volume} {8}},\ \bibinfo {pages} {43} (\bibinfo {year} {2024}{\natexlab{a}})}\BibitemShut {NoStop}%
\bibitem [{\citenamefont {Wang}\ \emph {et~al.}(2021)\citenamefont {Wang}, \citenamefont {Li}, \citenamefont {Li}, \citenamefont {Chen}, \citenamefont {Pi}, \citenamefont {Zhou},\ and\ \citenamefont {Zhai}}]{Wang2021_AFM}%
  \BibitemOpen
  \bibfield  {author} {\bibinfo {author} {\bibfnamefont {H.}~\bibnamefont {Wang}}, \bibinfo {author} {\bibfnamefont {Z.}~\bibnamefont {Li}}, \bibinfo {author} {\bibfnamefont {D.}~\bibnamefont {Li}}, \bibinfo {author} {\bibfnamefont {P.}~\bibnamefont {Chen}}, \bibinfo {author} {\bibfnamefont {L.}~\bibnamefont {Pi}}, \bibinfo {author} {\bibfnamefont {X.}~\bibnamefont {Zhou}},\ and\ \bibinfo {author} {\bibfnamefont {T.}~\bibnamefont {Zhai}},\ }\href {https://doi.org/https://doi.org/10.1002/adfm.202103106} {\bibfield  {journal} {\bibinfo  {journal} {Advanced Functional Materials}\ }\textbf {\bibinfo {volume} {31}},\ \bibinfo {pages} {2103106} (\bibinfo {year} {2021})}\BibitemShut {NoStop}%
\bibitem [{\citenamefont {Lotsch}(2015)}]{Lotsch2015_armr}%
  \BibitemOpen
  \bibfield  {author} {\bibinfo {author} {\bibfnamefont {B.~V.}\ \bibnamefont {Lotsch}},\ }\href {https://doi.org/https://doi.org/10.1146/annurev-matsci-070214-020934} {\bibfield  {journal} {\bibinfo  {journal} {Annual Review of Materials Research}\ }\textbf {\bibinfo {volume} {45}},\ \bibinfo {pages} {85} (\bibinfo {year} {2015})}\BibitemShut {NoStop}%
\bibitem [{\citenamefont {Futaba}(2023)}]{futaba2023hexagonal}%
  \BibitemOpen
  \bibfield  {author} {\bibinfo {author} {\bibfnamefont {D.~N.}\ \bibnamefont {Futaba}},\ }\href {https://doi.org/10.1038/s41928-023-00917-z} {\bibfield  {journal} {\bibinfo  {journal} {Nature Electronics}\ }\textbf {\bibinfo {volume} {6}},\ \bibinfo {pages} {104} (\bibinfo {year} {2023})}\BibitemShut {NoStop}%
\bibitem [{\citenamefont {Kubota}\ \emph {et~al.}(2007)\citenamefont {Kubota}, \citenamefont {Watanabe}, \citenamefont {Tsuda},\ and\ \citenamefont {Taniguchi}}]{Kubota2007_sci}%
  \BibitemOpen
  \bibfield  {author} {\bibinfo {author} {\bibfnamefont {Y.}~\bibnamefont {Kubota}}, \bibinfo {author} {\bibfnamefont {K.}~\bibnamefont {Watanabe}}, \bibinfo {author} {\bibfnamefont {O.}~\bibnamefont {Tsuda}},\ and\ \bibinfo {author} {\bibfnamefont {T.}~\bibnamefont {Taniguchi}},\ }\href {https://doi.org/10.1126/science.1144216} {\bibfield  {journal} {\bibinfo  {journal} {Science}\ }\textbf {\bibinfo {volume} {317}},\ \bibinfo {pages} {932} (\bibinfo {year} {2007})}\BibitemShut {NoStop}%
\bibitem [{\citenamefont {Wang}\ \emph {et~al.}(2016)\citenamefont {Wang}, \citenamefont {Guo}, \citenamefont {Dong}, \citenamefont {Aiyiti}, \citenamefont {Xu},\ and\ \citenamefont {Li}}]{wang2016superior}%
  \BibitemOpen
  \bibfield  {author} {\bibinfo {author} {\bibfnamefont {C.}~\bibnamefont {Wang}}, \bibinfo {author} {\bibfnamefont {J.}~\bibnamefont {Guo}}, \bibinfo {author} {\bibfnamefont {L.}~\bibnamefont {Dong}}, \bibinfo {author} {\bibfnamefont {A.}~\bibnamefont {Aiyiti}}, \bibinfo {author} {\bibfnamefont {X.}~\bibnamefont {Xu}},\ and\ \bibinfo {author} {\bibfnamefont {B.}~\bibnamefont {Li}},\ }\href {https://doi.org/https://doi.org/10.1038/srep25334} {\bibfield  {journal} {\bibinfo  {journal} {Scientific reports}\ }\textbf {\bibinfo {volume} {6}},\ \bibinfo {pages} {25334} (\bibinfo {year} {2016})}\BibitemShut {NoStop}%
\bibitem [{\citenamefont {Xu}\ \emph {et~al.}(2022)\citenamefont {Xu}, \citenamefont {Ding}, \citenamefont {Xu}, \citenamefont {Huang}, \citenamefont {Wei}, \citenamefont {Chen}, \citenamefont {Lan}, \citenamefont {Pan}, \citenamefont {Cheng},\ and\ \citenamefont {Liu}}]{xu2022magnetically}%
  \BibitemOpen
  \bibfield  {author} {\bibinfo {author} {\bibfnamefont {H.}~\bibnamefont {Xu}}, \bibinfo {author} {\bibfnamefont {B.}~\bibnamefont {Ding}}, \bibinfo {author} {\bibfnamefont {Y.}~\bibnamefont {Xu}}, \bibinfo {author} {\bibfnamefont {Z.}~\bibnamefont {Huang}}, \bibinfo {author} {\bibfnamefont {D.}~\bibnamefont {Wei}}, \bibinfo {author} {\bibfnamefont {S.}~\bibnamefont {Chen}}, \bibinfo {author} {\bibfnamefont {T.}~\bibnamefont {Lan}}, \bibinfo {author} {\bibfnamefont {Y.}~\bibnamefont {Pan}}, \bibinfo {author} {\bibfnamefont {H.-M.}\ \bibnamefont {Cheng}},\ and\ \bibinfo {author} {\bibfnamefont {B.}~\bibnamefont {Liu}},\ }\href {https://doi.org/10.1038/s41565-022-01186-1} {\bibfield  {journal} {\bibinfo  {journal} {Nature nanotechnology}\ }\textbf {\bibinfo {volume} {17}},\ \bibinfo {pages} {1091} (\bibinfo {year} {2022})}\BibitemShut {NoStop}%
\bibitem [{\citenamefont {Ding}\ \emph {et~al.}(2016)\citenamefont {Ding}, \citenamefont {Chen},\ and\ \citenamefont {Wu}}]{ding2016mechanical}%
  \BibitemOpen
  \bibfield  {author} {\bibinfo {author} {\bibfnamefont {N.}~\bibnamefont {Ding}}, \bibinfo {author} {\bibfnamefont {X.}~\bibnamefont {Chen}},\ and\ \bibinfo {author} {\bibfnamefont {C.-M.~L.}\ \bibnamefont {Wu}},\ }\href {https://doi.org/https://doi.org/10.1038/srep31499} {\bibfield  {journal} {\bibinfo  {journal} {Scientific reports}\ }\textbf {\bibinfo {volume} {6}},\ \bibinfo {pages} {31499} (\bibinfo {year} {2016})}\BibitemShut {NoStop}%
\bibitem [{\citenamefont {Moon}\ \emph {et~al.}(2023)\citenamefont {Moon}, \citenamefont {Kim}, \citenamefont {Park}, \citenamefont {Im}, \citenamefont {Kim}, \citenamefont {Hwang},\ and\ \citenamefont {Kim}}]{Moon2023_AM}%
  \BibitemOpen
  \bibfield  {author} {\bibinfo {author} {\bibfnamefont {S.}~\bibnamefont {Moon}}, \bibinfo {author} {\bibfnamefont {J.}~\bibnamefont {Kim}}, \bibinfo {author} {\bibfnamefont {J.}~\bibnamefont {Park}}, \bibinfo {author} {\bibfnamefont {S.}~\bibnamefont {Im}}, \bibinfo {author} {\bibfnamefont {J.}~\bibnamefont {Kim}}, \bibinfo {author} {\bibfnamefont {I.}~\bibnamefont {Hwang}},\ and\ \bibinfo {author} {\bibfnamefont {J.~K.}\ \bibnamefont {Kim}},\ }\href {https://doi.org/https://doi.org/10.1002/adma.202204161} {\bibfield  {journal} {\bibinfo  {journal} {Advanced Materials}\ }\textbf {\bibinfo {volume} {35}},\ \bibinfo {pages} {2204161} (\bibinfo {year} {2023})}\BibitemShut {NoStop}%
\bibitem [{\citenamefont {Li}\ \emph {et~al.}(2022)\citenamefont {Li}, \citenamefont {Huang}, \citenamefont {Chen}, \citenamefont {Yin}, \citenamefont {Zhang}, \citenamefont {Yao}, \citenamefont {Shen}, \citenamefont {Wu},\ and\ \citenamefont {Huang}}]{LI2022101486}%
  \BibitemOpen
  \bibfield  {author} {\bibinfo {author} {\bibfnamefont {M.}~\bibnamefont {Li}}, \bibinfo {author} {\bibfnamefont {G.}~\bibnamefont {Huang}}, \bibinfo {author} {\bibfnamefont {X.}~\bibnamefont {Chen}}, \bibinfo {author} {\bibfnamefont {J.}~\bibnamefont {Yin}}, \bibinfo {author} {\bibfnamefont {P.}~\bibnamefont {Zhang}}, \bibinfo {author} {\bibfnamefont {Y.}~\bibnamefont {Yao}}, \bibinfo {author} {\bibfnamefont {J.}~\bibnamefont {Shen}}, \bibinfo {author} {\bibfnamefont {Y.}~\bibnamefont {Wu}},\ and\ \bibinfo {author} {\bibfnamefont {J.}~\bibnamefont {Huang}},\ }\href {https://doi.org/https://doi.org/10.1016/j.nantod.2022.101486} {\bibfield  {journal} {\bibinfo  {journal} {Nano Today}\ }\textbf {\bibinfo {volume} {44}},\ \bibinfo {pages} {101486} (\bibinfo {year} {2022})}\BibitemShut {NoStop}%
\bibitem [{\citenamefont {Gong}\ \emph {et~al.}(2021)\citenamefont {Gong}, \citenamefont {Xu}, \citenamefont {Li}, \citenamefont {Zhang}, \citenamefont {Aharonovich},\ and\ \citenamefont {Zhang}}]{Gong2021_ACS}%
  \BibitemOpen
  \bibfield  {author} {\bibinfo {author} {\bibfnamefont {Y.}~\bibnamefont {Gong}}, \bibinfo {author} {\bibfnamefont {Z.-Q.}\ \bibnamefont {Xu}}, \bibinfo {author} {\bibfnamefont {D.}~\bibnamefont {Li}}, \bibinfo {author} {\bibfnamefont {J.}~\bibnamefont {Zhang}}, \bibinfo {author} {\bibfnamefont {I.}~\bibnamefont {Aharonovich}},\ and\ \bibinfo {author} {\bibfnamefont {Y.}~\bibnamefont {Zhang}},\ }\href {https://doi.org/10.1021/acsenergylett.0c02427} {\bibfield  {journal} {\bibinfo  {journal} {ACS Energy Letters}\ }\textbf {\bibinfo {volume} {6}},\ \bibinfo {pages} {985} (\bibinfo {year} {2021})}\BibitemShut {NoStop}%
\bibitem [{\citenamefont {Zahoor}\ \emph {et~al.}(2024)\citenamefont {Zahoor}, \citenamefont {Khan}, \citenamefont {Ismail}, \citenamefont {Qiao}, \citenamefont {Haneef}, \citenamefont {Akbar}, \citenamefont {Bououdina}, \citenamefont {Zeng},\ and\ \citenamefont {Ali}}]{ZAHOOR20243}%
  \BibitemOpen
  \bibfield  {author} {\bibinfo {author} {\bibfnamefont {M.}~\bibnamefont {Zahoor}}, \bibinfo {author} {\bibfnamefont {S.}~\bibnamefont {Khan}}, \bibinfo {author} {\bibfnamefont {P.~M.}\ \bibnamefont {Ismail}}, \bibinfo {author} {\bibfnamefont {L.}~\bibnamefont {Qiao}}, \bibinfo {author} {\bibfnamefont {M.}~\bibnamefont {Haneef}}, \bibinfo {author} {\bibfnamefont {J.}~\bibnamefont {Akbar}}, \bibinfo {author} {\bibfnamefont {M.}~\bibnamefont {Bououdina}}, \bibinfo {author} {\bibfnamefont {C.}~\bibnamefont {Zeng}},\ and\ \bibinfo {author} {\bibfnamefont {S.}~\bibnamefont {Ali}},\ }in\ \href {https://doi.org/https://doi.org/10.1016/B978-0-443-18843-5.00024-0} {\emph {\bibinfo {booktitle} {Hexagonal Boron Nitride}}},\ \bibinfo {series and number} {Micro and Nano Technologies},\ \bibinfo {editor} {edited by\ \bibinfo {editor} {\bibfnamefont {K.}~\bibnamefont {Deshmukh}}, \bibinfo {editor} {\bibfnamefont {M.}~\bibnamefont {Pandey}},\ and\ \bibinfo {editor} {\bibfnamefont {C.}~\bibnamefont {{Mustansar Hussain}}}}\
  (\bibinfo  {publisher} {Elsevier},\ \bibinfo {year} {2024})\ pp.\ \bibinfo {pages} {3--28}\BibitemShut {NoStop}%
\bibitem [{\citenamefont {Balta}\ and\ \citenamefont {Simsek}(2024)}]{BALTA2024205}%
  \BibitemOpen
  \bibfield  {author} {\bibinfo {author} {\bibfnamefont {Z.}~\bibnamefont {Balta}}\ and\ \bibinfo {author} {\bibfnamefont {E.~B.}\ \bibnamefont {Simsek}},\ }in\ \href {https://doi.org/https://doi.org/10.1016/B978-0-443-18843-5.00010-0} {\emph {\bibinfo {booktitle} {Hexagonal Boron Nitride}}},\ \bibinfo {series and number} {Micro and Nano Technologies},\ \bibinfo {editor} {edited by\ \bibinfo {editor} {\bibfnamefont {K.}~\bibnamefont {Deshmukh}}, \bibinfo {editor} {\bibfnamefont {M.}~\bibnamefont {Pandey}},\ and\ \bibinfo {editor} {\bibfnamefont {C.}~\bibnamefont {{Mustansar Hussain}}}}\ (\bibinfo  {publisher} {Elsevier},\ \bibinfo {year} {2024})\ pp.\ \bibinfo {pages} {205--233}\BibitemShut {NoStop}%
\bibitem [{\citenamefont {Dean}\ \emph {et~al.}(2010)\citenamefont {Dean}, \citenamefont {Young}, \citenamefont {Meric}, \citenamefont {Lee}, \citenamefont {Wang}, \citenamefont {Sorgenfrei}, \citenamefont {Watanabe}, \citenamefont {Taniguchi}, \citenamefont {Kim}, \citenamefont {Shepard} \emph {et~al.}}]{dean2010_nature}%
  \BibitemOpen
  \bibfield  {author} {\bibinfo {author} {\bibfnamefont {C.~R.}\ \bibnamefont {Dean}}, \bibinfo {author} {\bibfnamefont {A.~F.}\ \bibnamefont {Young}}, \bibinfo {author} {\bibfnamefont {I.}~\bibnamefont {Meric}}, \bibinfo {author} {\bibfnamefont {C.}~\bibnamefont {Lee}}, \bibinfo {author} {\bibfnamefont {L.}~\bibnamefont {Wang}}, \bibinfo {author} {\bibfnamefont {S.}~\bibnamefont {Sorgenfrei}}, \bibinfo {author} {\bibfnamefont {K.}~\bibnamefont {Watanabe}}, \bibinfo {author} {\bibfnamefont {T.}~\bibnamefont {Taniguchi}}, \bibinfo {author} {\bibfnamefont {P.}~\bibnamefont {Kim}}, \bibinfo {author} {\bibfnamefont {K.~L.}\ \bibnamefont {Shepard}}, \emph {et~al.},\ }\href {https://doi.org/10.1038/nnano.2010.172} {\bibfield  {journal} {\bibinfo  {journal} {Nature nanotechnology}\ }\textbf {\bibinfo {volume} {5}},\ \bibinfo {pages} {722} (\bibinfo {year} {2010})}\BibitemShut {NoStop}%
\bibitem [{\citenamefont {Yankowitz}\ \emph {et~al.}(2019)\citenamefont {Yankowitz}, \citenamefont {Ma}, \citenamefont {Jarillo-Herrero},\ and\ \citenamefont {LeRoy}}]{yankowitz2019van}%
  \BibitemOpen
  \bibfield  {author} {\bibinfo {author} {\bibfnamefont {M.}~\bibnamefont {Yankowitz}}, \bibinfo {author} {\bibfnamefont {Q.}~\bibnamefont {Ma}}, \bibinfo {author} {\bibfnamefont {P.}~\bibnamefont {Jarillo-Herrero}},\ and\ \bibinfo {author} {\bibfnamefont {B.~J.}\ \bibnamefont {LeRoy}},\ }\href {https://doi.org/10.1038/s42254-018-0016-0} {\bibfield  {journal} {\bibinfo  {journal} {Nature Reviews Physics}\ }\textbf {\bibinfo {volume} {1}},\ \bibinfo {pages} {112} (\bibinfo {year} {2019})}\BibitemShut {NoStop}%
\bibitem [{\citenamefont {Juma}\ \emph {et~al.}(2021)\citenamefont {Juma}, \citenamefont {Kim}, \citenamefont {Jariwala},\ and\ \citenamefont {Behura}}]{juma2021direct}%
  \BibitemOpen
  \bibfield  {author} {\bibinfo {author} {\bibfnamefont {I.~G.}\ \bibnamefont {Juma}}, \bibinfo {author} {\bibfnamefont {G.}~\bibnamefont {Kim}}, \bibinfo {author} {\bibfnamefont {D.}~\bibnamefont {Jariwala}},\ and\ \bibinfo {author} {\bibfnamefont {S.~K.}\ \bibnamefont {Behura}},\ }\bibfield  {journal} {\bibinfo  {journal} {IScience}\ }\textbf {\bibinfo {volume} {24}},\ \href {https://doi.org/10.1016/j.isci.2021.103374} {10.1016/j.isci.2021.103374} (\bibinfo {year} {2021})\BibitemShut {NoStop}%
\bibitem [{\citenamefont {Ogawa}\ \emph {et~al.}(2022)\citenamefont {Ogawa}, \citenamefont {Fukushima},\ and\ \citenamefont {Shimatani}}]{Ogawa2022}%
  \BibitemOpen
  \bibfield  {author} {\bibinfo {author} {\bibfnamefont {S.}~\bibnamefont {Ogawa}}, \bibinfo {author} {\bibfnamefont {S.}~\bibnamefont {Fukushima}},\ and\ \bibinfo {author} {\bibfnamefont {M.}~\bibnamefont {Shimatani}},\ }\href {https://doi.org/10.1364/JOSAB.472600} {\bibfield  {journal} {\bibinfo  {journal} {J. Opt. Soc. Am. B}\ }\textbf {\bibinfo {volume} {39}},\ \bibinfo {pages} {3149} (\bibinfo {year} {2022})}\BibitemShut {NoStop}%
\bibitem [{\citenamefont {Wang}\ \emph {et~al.}(2019)\citenamefont {Wang}, \citenamefont {Wang}, \citenamefont {Yin}, \citenamefont {Tóvári}, \citenamefont {Yang}, \citenamefont {Lin}, \citenamefont {Holwill}, \citenamefont {Birkbeck}, \citenamefont {Perello}, \citenamefont {Xu}, \citenamefont {Zultak}, \citenamefont {Gorbachev}, \citenamefont {Kretinin}, \citenamefont {Taniguchi}, \citenamefont {Watanabe}, \citenamefont {Morozov}, \citenamefont {Anđelković}, \citenamefont {Milovanović}, \citenamefont {Covaci}, \citenamefont {Peeters}, \citenamefont {Mishchenko}, \citenamefont {Geim}, \citenamefont {Novoselov}, \citenamefont {Fal’ko}, \citenamefont {Knothe},\ and\ \citenamefont {Woods}}]{Zihao2019_sci}%
  \BibitemOpen
  \bibfield  {author} {\bibinfo {author} {\bibfnamefont {Z.}~\bibnamefont {Wang}}, \bibinfo {author} {\bibfnamefont {Y.~B.}\ \bibnamefont {Wang}}, \bibinfo {author} {\bibfnamefont {J.}~\bibnamefont {Yin}}, \bibinfo {author} {\bibfnamefont {E.}~\bibnamefont {Tóvári}}, \bibinfo {author} {\bibfnamefont {Y.}~\bibnamefont {Yang}}, \bibinfo {author} {\bibfnamefont {L.}~\bibnamefont {Lin}}, \bibinfo {author} {\bibfnamefont {M.}~\bibnamefont {Holwill}}, \bibinfo {author} {\bibfnamefont {J.}~\bibnamefont {Birkbeck}}, \bibinfo {author} {\bibfnamefont {D.~J.}\ \bibnamefont {Perello}}, \bibinfo {author} {\bibfnamefont {S.}~\bibnamefont {Xu}}, \bibinfo {author} {\bibfnamefont {J.}~\bibnamefont {Zultak}}, \bibinfo {author} {\bibfnamefont {R.~V.}\ \bibnamefont {Gorbachev}}, \bibinfo {author} {\bibfnamefont {A.~V.}\ \bibnamefont {Kretinin}}, \bibinfo {author} {\bibfnamefont {T.}~\bibnamefont {Taniguchi}}, \bibinfo {author} {\bibfnamefont {K.}~\bibnamefont {Watanabe}}, \bibinfo {author} {\bibfnamefont {S.~V.}\ \bibnamefont
  {Morozov}}, \bibinfo {author} {\bibfnamefont {M.}~\bibnamefont {Anđelković}}, \bibinfo {author} {\bibfnamefont {S.~P.}\ \bibnamefont {Milovanović}}, \bibinfo {author} {\bibfnamefont {L.}~\bibnamefont {Covaci}}, \bibinfo {author} {\bibfnamefont {F.~M.}\ \bibnamefont {Peeters}}, \bibinfo {author} {\bibfnamefont {A.}~\bibnamefont {Mishchenko}}, \bibinfo {author} {\bibfnamefont {A.~K.}\ \bibnamefont {Geim}}, \bibinfo {author} {\bibfnamefont {K.~S.}\ \bibnamefont {Novoselov}}, \bibinfo {author} {\bibfnamefont {V.~I.}\ \bibnamefont {Fal’ko}}, \bibinfo {author} {\bibfnamefont {A.}~\bibnamefont {Knothe}},\ and\ \bibinfo {author} {\bibfnamefont {C.~R.}\ \bibnamefont {Woods}},\ }\href {https://doi.org/10.1126/sciadv.aay8897} {\bibfield  {journal} {\bibinfo  {journal} {Science Advances}\ }\textbf {\bibinfo {volume} {5}},\ \bibinfo {pages} {eaay8897} (\bibinfo {year} {2019})}\BibitemShut {NoStop}%
\bibitem [{\citenamefont {Moore}\ \emph {et~al.}(2021)\citenamefont {Moore}, \citenamefont {Ciccarino}, \citenamefont {Halbertal}, \citenamefont {McGilly}, \citenamefont {Finney}, \citenamefont {Yao}, \citenamefont {Shao}, \citenamefont {Ni}, \citenamefont {Sternbach}, \citenamefont {Telford} \emph {et~al.}}]{moore2021nanoscale}%
  \BibitemOpen
  \bibfield  {author} {\bibinfo {author} {\bibfnamefont {S.}~\bibnamefont {Moore}}, \bibinfo {author} {\bibfnamefont {C.}~\bibnamefont {Ciccarino}}, \bibinfo {author} {\bibfnamefont {D.}~\bibnamefont {Halbertal}}, \bibinfo {author} {\bibfnamefont {L.}~\bibnamefont {McGilly}}, \bibinfo {author} {\bibfnamefont {N.}~\bibnamefont {Finney}}, \bibinfo {author} {\bibfnamefont {K.}~\bibnamefont {Yao}}, \bibinfo {author} {\bibfnamefont {Y.}~\bibnamefont {Shao}}, \bibinfo {author} {\bibfnamefont {G.}~\bibnamefont {Ni}}, \bibinfo {author} {\bibfnamefont {A.}~\bibnamefont {Sternbach}}, \bibinfo {author} {\bibfnamefont {E.}~\bibnamefont {Telford}}, \emph {et~al.},\ }\href {https://doi.org/https://doi.org/10.1038/s41467-021-26072-7} {\bibfield  {journal} {\bibinfo  {journal} {Nature Communications}\ }\textbf {\bibinfo {volume} {12}},\ \bibinfo {pages} {5741} (\bibinfo {year} {2021})}\BibitemShut {NoStop}%
\bibitem [{\citenamefont {Kamat}\ \emph {et~al.}(2024)\citenamefont {Kamat}, \citenamefont {Sharpe}, \citenamefont {Pendharkar}, \citenamefont {Hu}, \citenamefont {Tran}, \citenamefont {Zaborski}, \citenamefont {Hocking}, \citenamefont {Finney}, \citenamefont {Watanabe}, \citenamefont {Taniguchi}, \citenamefont {Kastner}, \citenamefont {Mannix}, \citenamefont {Heinz},\ and\ \citenamefont {Goldhaber-Gordon}}]{Rupini2024_pnas}%
  \BibitemOpen
  \bibfield  {author} {\bibinfo {author} {\bibfnamefont {R.~V.}\ \bibnamefont {Kamat}}, \bibinfo {author} {\bibfnamefont {A.~L.}\ \bibnamefont {Sharpe}}, \bibinfo {author} {\bibfnamefont {M.}~\bibnamefont {Pendharkar}}, \bibinfo {author} {\bibfnamefont {J.}~\bibnamefont {Hu}}, \bibinfo {author} {\bibfnamefont {S.~J.}\ \bibnamefont {Tran}}, \bibinfo {author} {\bibfnamefont {G.}~\bibnamefont {Zaborski}}, \bibinfo {author} {\bibfnamefont {M.}~\bibnamefont {Hocking}}, \bibinfo {author} {\bibfnamefont {J.}~\bibnamefont {Finney}}, \bibinfo {author} {\bibfnamefont {K.}~\bibnamefont {Watanabe}}, \bibinfo {author} {\bibfnamefont {T.}~\bibnamefont {Taniguchi}}, \bibinfo {author} {\bibfnamefont {M.~A.}\ \bibnamefont {Kastner}}, \bibinfo {author} {\bibfnamefont {A.~J.}\ \bibnamefont {Mannix}}, \bibinfo {author} {\bibfnamefont {T.}~\bibnamefont {Heinz}},\ and\ \bibinfo {author} {\bibfnamefont {D.}~\bibnamefont {Goldhaber-Gordon}},\ }\href {https://doi.org/10.1073/pnas.2410993121} {\bibfield  {journal} {\bibinfo  {journal}
  {Proceedings of the National Academy of Sciences}\ }\textbf {\bibinfo {volume} {121}},\ \bibinfo {pages} {e2410993121} (\bibinfo {year} {2024})}\BibitemShut {NoStop}%
\bibitem [{\citenamefont {O’Sullivan}\ \emph {et~al.}(2025)\citenamefont {O’Sullivan}, \citenamefont {Grobert},\ and\ \citenamefont {Swart}}]{OSullivan2025}%
  \BibitemOpen
  \bibfield  {author} {\bibinfo {author} {\bibfnamefont {E.~M.}\ \bibnamefont {O’Sullivan}}, \bibinfo {author} {\bibfnamefont {N.}~\bibnamefont {Grobert}},\ and\ \bibinfo {author} {\bibfnamefont {M.}~\bibnamefont {Swart}},\ }\href {https://doi.org/10.1021/acs.jpcc.4c06121} {\bibfield  {journal} {\bibinfo  {journal} {The Journal of Physical Chemistry C}\ }\textbf {\bibinfo {volume} {129}},\ \bibinfo {pages} {638} (\bibinfo {year} {2025})}\BibitemShut {NoStop}%
\bibitem [{\citenamefont {Huang}\ \emph {et~al.}(2020)\citenamefont {Huang}, \citenamefont {Riccardi}, \citenamefont {Messelot}, \citenamefont {Graef}, \citenamefont {Valmorra}, \citenamefont {Tignon}, \citenamefont {Taniguchi}, \citenamefont {Watanabe}, \citenamefont {Dhillon}, \citenamefont {Placais} \emph {et~al.}}]{huang2020ultra}%
  \BibitemOpen
  \bibfield  {author} {\bibinfo {author} {\bibfnamefont {P.}~\bibnamefont {Huang}}, \bibinfo {author} {\bibfnamefont {E.}~\bibnamefont {Riccardi}}, \bibinfo {author} {\bibfnamefont {S.}~\bibnamefont {Messelot}}, \bibinfo {author} {\bibfnamefont {H.}~\bibnamefont {Graef}}, \bibinfo {author} {\bibfnamefont {F.}~\bibnamefont {Valmorra}}, \bibinfo {author} {\bibfnamefont {J.}~\bibnamefont {Tignon}}, \bibinfo {author} {\bibfnamefont {T.}~\bibnamefont {Taniguchi}}, \bibinfo {author} {\bibfnamefont {K.}~\bibnamefont {Watanabe}}, \bibinfo {author} {\bibfnamefont {S.}~\bibnamefont {Dhillon}}, \bibinfo {author} {\bibfnamefont {B.}~\bibnamefont {Placais}}, \emph {et~al.},\ }\href {https://doi.org/https://doi.org/10.1038/s41467-020-14714-1} {\bibfield  {journal} {\bibinfo  {journal} {Nature communications}\ }\textbf {\bibinfo {volume} {11}},\ \bibinfo {pages} {863} (\bibinfo {year} {2020})}\BibitemShut {NoStop}%
\bibitem [{\citenamefont {Zollner}\ \emph {et~al.}(2021)\citenamefont {Zollner}, \citenamefont {Cummings}, \citenamefont {Roche},\ and\ \citenamefont {Fabian}}]{PhysRevB.103.075129}%
  \BibitemOpen
  \bibfield  {author} {\bibinfo {author} {\bibfnamefont {K.}~\bibnamefont {Zollner}}, \bibinfo {author} {\bibfnamefont {A.~W.}\ \bibnamefont {Cummings}}, \bibinfo {author} {\bibfnamefont {S.}~\bibnamefont {Roche}},\ and\ \bibinfo {author} {\bibfnamefont {J.}~\bibnamefont {Fabian}},\ }\href {https://doi.org/10.1103/PhysRevB.103.075129} {\bibfield  {journal} {\bibinfo  {journal} {Phys. Rev. B}\ }\textbf {\bibinfo {volume} {103}},\ \bibinfo {pages} {075129} (\bibinfo {year} {2021})}\BibitemShut {NoStop}%
\bibitem [{\citenamefont {Li}\ \emph {et~al.}(2023{\natexlab{a}})\citenamefont {Li}, \citenamefont {Ghorbani-Asl}, \citenamefont {Lasek}, \citenamefont {Pathirage}, \citenamefont {Krasheninnikov},\ and\ \citenamefont {Batzill}}]{Jingfeng_acsnano_2023}%
  \BibitemOpen
  \bibfield  {author} {\bibinfo {author} {\bibfnamefont {J.}~\bibnamefont {Li}}, \bibinfo {author} {\bibfnamefont {M.}~\bibnamefont {Ghorbani-Asl}}, \bibinfo {author} {\bibfnamefont {K.}~\bibnamefont {Lasek}}, \bibinfo {author} {\bibfnamefont {V.}~\bibnamefont {Pathirage}}, \bibinfo {author} {\bibfnamefont {A.~V.}\ \bibnamefont {Krasheninnikov}},\ and\ \bibinfo {author} {\bibfnamefont {M.}~\bibnamefont {Batzill}},\ }\href {https://doi.org/10.1021/acsnano.2c12879} {\bibfield  {journal} {\bibinfo  {journal} {ACS Nano}\ }\textbf {\bibinfo {volume} {17}},\ \bibinfo {pages} {5913} (\bibinfo {year} {2023}{\natexlab{a}})},\ \bibinfo {note} {pMID: 36926837}\BibitemShut {NoStop}%
\bibitem [{\citenamefont {Jat}\ \emph {et~al.}(2024)\citenamefont {Jat}, \citenamefont {Tiwari}, \citenamefont {Bajaj}, \citenamefont {Shitut}, \citenamefont {Mandal}, \citenamefont {Watanabe}, \citenamefont {Taniguchi}, \citenamefont {Krishnamurthy}, \citenamefont {Jain},\ and\ \citenamefont {Bid}}]{jat2024_nature}%
  \BibitemOpen
  \bibfield  {author} {\bibinfo {author} {\bibfnamefont {M.~K.}\ \bibnamefont {Jat}}, \bibinfo {author} {\bibfnamefont {P.}~\bibnamefont {Tiwari}}, \bibinfo {author} {\bibfnamefont {R.}~\bibnamefont {Bajaj}}, \bibinfo {author} {\bibfnamefont {I.}~\bibnamefont {Shitut}}, \bibinfo {author} {\bibfnamefont {S.}~\bibnamefont {Mandal}}, \bibinfo {author} {\bibfnamefont {K.}~\bibnamefont {Watanabe}}, \bibinfo {author} {\bibfnamefont {T.}~\bibnamefont {Taniguchi}}, \bibinfo {author} {\bibfnamefont {H.}~\bibnamefont {Krishnamurthy}}, \bibinfo {author} {\bibfnamefont {M.}~\bibnamefont {Jain}},\ and\ \bibinfo {author} {\bibfnamefont {A.}~\bibnamefont {Bid}},\ }\href {https://doi.org/https://doi.org/10.1038/s41467-024-46672-3} {\bibfield  {journal} {\bibinfo  {journal} {Nature Communications}\ }\textbf {\bibinfo {volume} {15}},\ \bibinfo {pages} {2335} (\bibinfo {year} {2024})}\BibitemShut {NoStop}%
\bibitem [{\citenamefont {Ma}\ \emph {et~al.}(2025)\citenamefont {Ma}, \citenamefont {Huang}, \citenamefont {Zhang}, \citenamefont {Hu}, \citenamefont {Zhou}, \citenamefont {Feng}, \citenamefont {Li}, \citenamefont {Chen}, \citenamefont {Lou}, \citenamefont {Zhang} \emph {et~al.}}]{ma2025_nature}%
  \BibitemOpen
  \bibfield  {author} {\bibinfo {author} {\bibfnamefont {Y.}~\bibnamefont {Ma}}, \bibinfo {author} {\bibfnamefont {M.}~\bibnamefont {Huang}}, \bibinfo {author} {\bibfnamefont {X.}~\bibnamefont {Zhang}}, \bibinfo {author} {\bibfnamefont {W.}~\bibnamefont {Hu}}, \bibinfo {author} {\bibfnamefont {Z.}~\bibnamefont {Zhou}}, \bibinfo {author} {\bibfnamefont {K.}~\bibnamefont {Feng}}, \bibinfo {author} {\bibfnamefont {W.}~\bibnamefont {Li}}, \bibinfo {author} {\bibfnamefont {Y.}~\bibnamefont {Chen}}, \bibinfo {author} {\bibfnamefont {C.}~\bibnamefont {Lou}}, \bibinfo {author} {\bibfnamefont {W.}~\bibnamefont {Zhang}}, \emph {et~al.},\ }\href {https://doi.org/https://doi.org/10.1038/s41467-025-57111-2} {\bibfield  {journal} {\bibinfo  {journal} {Nature Communications}\ }\textbf {\bibinfo {volume} {16}},\ \bibinfo {pages} {1860} (\bibinfo {year} {2025})}\BibitemShut {NoStop}%
\bibitem [{\citenamefont {Lu}\ \emph {et~al.}(2025)\citenamefont {Lu}, \citenamefont {Han}, \citenamefont {Yao}, \citenamefont {Hadjri}, \citenamefont {Yang}, \citenamefont {Seo}, \citenamefont {Shi}, \citenamefont {Ye}, \citenamefont {Watanabe}, \citenamefont {Taniguchi} \emph {et~al.}}]{lu2025_nature}%
  \BibitemOpen
  \bibfield  {author} {\bibinfo {author} {\bibfnamefont {Z.}~\bibnamefont {Lu}}, \bibinfo {author} {\bibfnamefont {T.}~\bibnamefont {Han}}, \bibinfo {author} {\bibfnamefont {Y.}~\bibnamefont {Yao}}, \bibinfo {author} {\bibfnamefont {Z.}~\bibnamefont {Hadjri}}, \bibinfo {author} {\bibfnamefont {J.}~\bibnamefont {Yang}}, \bibinfo {author} {\bibfnamefont {J.}~\bibnamefont {Seo}}, \bibinfo {author} {\bibfnamefont {L.}~\bibnamefont {Shi}}, \bibinfo {author} {\bibfnamefont {S.}~\bibnamefont {Ye}}, \bibinfo {author} {\bibfnamefont {K.}~\bibnamefont {Watanabe}}, \bibinfo {author} {\bibfnamefont {T.}~\bibnamefont {Taniguchi}}, \emph {et~al.},\ }\href {https://doi.org/https://doi.org/10.1038/s41586-024-08470-1} {\bibfield  {journal} {\bibinfo  {journal} {Nature}\ ,\ \bibinfo {pages} {1}} (\bibinfo {year} {2025})}\BibitemShut {NoStop}%
\bibitem [{\citenamefont {Hafner}\ \emph {et~al.}(2006)\citenamefont {Hafner}, \citenamefont {Wolverton},\ and\ \citenamefont {Ceder}}]{hafner2006toward}%
  \BibitemOpen
  \bibfield  {author} {\bibinfo {author} {\bibfnamefont {J.}~\bibnamefont {Hafner}}, \bibinfo {author} {\bibfnamefont {C.}~\bibnamefont {Wolverton}},\ and\ \bibinfo {author} {\bibfnamefont {G.}~\bibnamefont {Ceder}},\ }\href {https://doi.org/10.1557/mrs2006.174} {\bibfield  {journal} {\bibinfo  {journal} {MRS bulletin}\ }\textbf {\bibinfo {volume} {31}},\ \bibinfo {pages} {659} (\bibinfo {year} {2006})}\BibitemShut {NoStop}%
\bibitem [{\citenamefont {Dingreville}\ \emph {et~al.}(2016)\citenamefont {Dingreville}, \citenamefont {Karnesky}, \citenamefont {Puel},\ and\ \citenamefont {Schmitt}}]{dingreville2016review}%
  \BibitemOpen
  \bibfield  {author} {\bibinfo {author} {\bibfnamefont {R.}~\bibnamefont {Dingreville}}, \bibinfo {author} {\bibfnamefont {R.~A.}\ \bibnamefont {Karnesky}}, \bibinfo {author} {\bibfnamefont {G.}~\bibnamefont {Puel}},\ and\ \bibinfo {author} {\bibfnamefont {J.-H.}\ \bibnamefont {Schmitt}},\ }\href {https://doi.org/https://doi.org/10.1007/s10853-015-9551-6} {\bibfield  {journal} {\bibinfo  {journal} {Journal of materials science}\ }\textbf {\bibinfo {volume} {51}},\ \bibinfo {pages} {1178} (\bibinfo {year} {2016})}\BibitemShut {NoStop}%
\bibitem [{\citenamefont {Li}\ \emph {et~al.}(2017)\citenamefont {Li}, \citenamefont {Xu}, \citenamefont {Luo}, \citenamefont {Li}, \citenamefont {Huang}, \citenamefont {Wang},\ and\ \citenamefont {Yu}}]{Li2017_RSC}%
  \BibitemOpen
  \bibfield  {author} {\bibinfo {author} {\bibfnamefont {Q.}~\bibnamefont {Li}}, \bibinfo {author} {\bibfnamefont {L.}~\bibnamefont {Xu}}, \bibinfo {author} {\bibfnamefont {K.-W.}\ \bibnamefont {Luo}}, \bibinfo {author} {\bibfnamefont {X.-F.}\ \bibnamefont {Li}}, \bibinfo {author} {\bibfnamefont {W.-Q.}\ \bibnamefont {Huang}}, \bibinfo {author} {\bibfnamefont {L.-L.}\ \bibnamefont {Wang}},\ and\ \bibinfo {author} {\bibfnamefont {Y.-B.}\ \bibnamefont {Yu}},\ }\href {https://doi.org/10.1039/C7TC00562H} {\bibfield  {journal} {\bibinfo  {journal} {J. Mater. Chem. C}\ }\textbf {\bibinfo {volume} {5}},\ \bibinfo {pages} {4426} (\bibinfo {year} {2017})}\BibitemShut {NoStop}%
\bibitem [{\citenamefont {Liu}\ \emph {et~al.}(2025)\citenamefont {Liu}, \citenamefont {Hsu}, \citenamefont {Chen}, \citenamefont {Hsu}, \citenamefont {Lan}, \citenamefont {Chiu},\ and\ \citenamefont {Lin}}]{LIU2025100687}%
  \BibitemOpen
  \bibfield  {author} {\bibinfo {author} {\bibfnamefont {C.-M.}\ \bibnamefont {Liu}}, \bibinfo {author} {\bibfnamefont {S.-Y.}\ \bibnamefont {Hsu}}, \bibinfo {author} {\bibfnamefont {H.-S.}\ \bibnamefont {Chen}}, \bibinfo {author} {\bibfnamefont {C.-C.}\ \bibnamefont {Hsu}}, \bibinfo {author} {\bibfnamefont {Y.-W.}\ \bibnamefont {Lan}}, \bibinfo {author} {\bibfnamefont {H.-C.}\ \bibnamefont {Chiu}},\ and\ \bibinfo {author} {\bibfnamefont {W.-C.}\ \bibnamefont {Lin}},\ }\href {https://doi.org/https://doi.org/10.1016/j.apsadv.2024.100687} {\bibfield  {journal} {\bibinfo  {journal} {Applied Surface Science Advances}\ }\textbf {\bibinfo {volume} {25}},\ \bibinfo {pages} {100687} (\bibinfo {year} {2025})}\BibitemShut {NoStop}%
\bibitem [{\citenamefont {Duan}\ \emph {et~al.}(2025)\citenamefont {Duan}, \citenamefont {Wang}, \citenamefont {Tian},\ and\ \citenamefont {Zhang}}]{DUAN2025118129}%
  \BibitemOpen
  \bibfield  {author} {\bibinfo {author} {\bibfnamefont {J.-X.}\ \bibnamefont {Duan}}, \bibinfo {author} {\bibfnamefont {C.-B.}\ \bibnamefont {Wang}}, \bibinfo {author} {\bibfnamefont {Y.}~\bibnamefont {Tian}},\ and\ \bibinfo {author} {\bibfnamefont {L.-L.}\ \bibnamefont {Zhang}},\ }\href {https://doi.org/https://doi.org/10.1016/j.mseb.2025.118129} {\bibfield  {journal} {\bibinfo  {journal} {Materials Science and Engineering: B}\ }\textbf {\bibinfo {volume} {317}},\ \bibinfo {pages} {118129} (\bibinfo {year} {2025})}\BibitemShut {NoStop}%
\bibitem [{\citenamefont {Zhao}\ \emph {et~al.}(2022)\citenamefont {Zhao}, \citenamefont {Cui}, \citenamefont {Ge}, \citenamefont {Lu}, \citenamefont {Guan}, \citenamefont {Zhang}, \citenamefont {Zhen}, \citenamefont {Sun}, \citenamefont {Wang},\ and\ \citenamefont {Lu}}]{Zhao2022_APL}%
  \BibitemOpen
  \bibfield  {author} {\bibinfo {author} {\bibfnamefont {X.}~\bibnamefont {Zhao}}, \bibinfo {author} {\bibfnamefont {Z.}~\bibnamefont {Cui}}, \bibinfo {author} {\bibfnamefont {A.}~\bibnamefont {Ge}}, \bibinfo {author} {\bibfnamefont {X.}~\bibnamefont {Lu}}, \bibinfo {author} {\bibfnamefont {X.}~\bibnamefont {Guan}}, \bibinfo {author} {\bibfnamefont {J.}~\bibnamefont {Zhang}}, \bibinfo {author} {\bibfnamefont {H.}~\bibnamefont {Zhen}}, \bibinfo {author} {\bibfnamefont {L.}~\bibnamefont {Sun}}, \bibinfo {author} {\bibfnamefont {S.}~\bibnamefont {Wang}},\ and\ \bibinfo {author} {\bibfnamefont {W.}~\bibnamefont {Lu}},\ }\href {https://doi.org/10.1063/5.0118834} {\bibfield  {journal} {\bibinfo  {journal} {Applied Physics Letters}\ }\textbf {\bibinfo {volume} {121}},\ \bibinfo {pages} {231106} (\bibinfo {year} {2022})}\BibitemShut {NoStop}%
\bibitem [{\citenamefont {Li}\ \emph {et~al.}(2024{\natexlab{b}})\citenamefont {Li}, \citenamefont {Jiang}, \citenamefont {Liu}, \citenamefont {Wu}, \citenamefont {Lyu}, \citenamefont {Li}, \citenamefont {Lin}, \citenamefont {Tang}, \citenamefont {Lyu}, \citenamefont {Yang}, \citenamefont {Wu}, \citenamefont {Lu}, \citenamefont {Tan}, \citenamefont {Peng}, \citenamefont {Gao}, \citenamefont {Hu},\ and\ \citenamefont {Gong}}]{doi:10.1021/acs.jpcc.3c07843}%
  \BibitemOpen
  \bibfield  {author} {\bibinfo {author} {\bibfnamefont {Y.}~\bibnamefont {Li}}, \bibinfo {author} {\bibfnamefont {P.}~\bibnamefont {Jiang}}, \bibinfo {author} {\bibfnamefont {X.}~\bibnamefont {Liu}}, \bibinfo {author} {\bibfnamefont {H.}~\bibnamefont {Wu}}, \bibinfo {author} {\bibfnamefont {X.}~\bibnamefont {Lyu}}, \bibinfo {author} {\bibfnamefont {X.}~\bibnamefont {Li}}, \bibinfo {author} {\bibfnamefont {H.}~\bibnamefont {Lin}}, \bibinfo {author} {\bibfnamefont {J.}~\bibnamefont {Tang}}, \bibinfo {author} {\bibfnamefont {Q.}~\bibnamefont {Lyu}}, \bibinfo {author} {\bibfnamefont {H.}~\bibnamefont {Yang}}, \bibinfo {author} {\bibfnamefont {C.}~\bibnamefont {Wu}}, \bibinfo {author} {\bibfnamefont {G.}~\bibnamefont {Lu}}, \bibinfo {author} {\bibfnamefont {P.-H.}\ \bibnamefont {Tan}}, \bibinfo {author} {\bibfnamefont {L.-Y.}\ \bibnamefont {Peng}}, \bibinfo {author} {\bibfnamefont {Y.}~\bibnamefont {Gao}}, \bibinfo {author} {\bibfnamefont {X.}~\bibnamefont {Hu}},\ and\ \bibinfo {author} {\bibfnamefont
  {Q.}~\bibnamefont {Gong}},\ }\href {https://doi.org/10.1021/acs.jpcc.3c07843} {\bibfield  {journal} {\bibinfo  {journal} {The Journal of Physical Chemistry C}\ }\textbf {\bibinfo {volume} {128}},\ \bibinfo {pages} {4286} (\bibinfo {year} {2024}{\natexlab{b}})}\BibitemShut {NoStop}%
\bibitem [{\citenamefont {Le~Ster}\ \emph {et~al.}(2019)\citenamefont {Le~Ster}, \citenamefont {Maerkl}, \citenamefont {Kowalczyk},\ and\ \citenamefont {Brown}}]{PhysRevB.99.075422}%
  \BibitemOpen
  \bibfield  {author} {\bibinfo {author} {\bibfnamefont {M.}~\bibnamefont {Le~Ster}}, \bibinfo {author} {\bibfnamefont {T.}~\bibnamefont {Maerkl}}, \bibinfo {author} {\bibfnamefont {P.~J.}\ \bibnamefont {Kowalczyk}},\ and\ \bibinfo {author} {\bibfnamefont {S.~A.}\ \bibnamefont {Brown}},\ }\href {https://doi.org/10.1103/PhysRevB.99.075422} {\bibfield  {journal} {\bibinfo  {journal} {Phys. Rev. B}\ }\textbf {\bibinfo {volume} {99}},\ \bibinfo {pages} {075422} (\bibinfo {year} {2019})}\BibitemShut {NoStop}%
\bibitem [{\citenamefont {Rakib}\ \emph {et~al.}(2022)\citenamefont {Rakib}, \citenamefont {Pochet}, \citenamefont {Ertekin},\ and\ \citenamefont {Johnson}}]{Rakib2022_JAP}%
  \BibitemOpen
  \bibfield  {author} {\bibinfo {author} {\bibfnamefont {T.}~\bibnamefont {Rakib}}, \bibinfo {author} {\bibfnamefont {P.}~\bibnamefont {Pochet}}, \bibinfo {author} {\bibfnamefont {E.}~\bibnamefont {Ertekin}},\ and\ \bibinfo {author} {\bibfnamefont {H.~T.}\ \bibnamefont {Johnson}},\ }\href {https://doi.org/10.1063/5.0105405} {\bibfield  {journal} {\bibinfo  {journal} {Journal of Applied Physics}\ }\textbf {\bibinfo {volume} {132}},\ \bibinfo {pages} {120901} (\bibinfo {year} {2022})}\BibitemShut {NoStop}%
\bibitem [{\citenamefont {Wang}\ \emph {et~al.}(2024)\citenamefont {Wang}, \citenamefont {Zhao}, \citenamefont {Lv}, \citenamefont {Zhao}, \citenamefont {Wei},\ and\ \citenamefont {Liu}}]{D4CP01673D}%
  \BibitemOpen
  \bibfield  {author} {\bibinfo {author} {\bibfnamefont {X.}~\bibnamefont {Wang}}, \bibinfo {author} {\bibfnamefont {G.}~\bibnamefont {Zhao}}, \bibinfo {author} {\bibfnamefont {X.}~\bibnamefont {Lv}}, \bibinfo {author} {\bibfnamefont {M.}~\bibnamefont {Zhao}}, \bibinfo {author} {\bibfnamefont {W.}~\bibnamefont {Wei}},\ and\ \bibinfo {author} {\bibfnamefont {G.}~\bibnamefont {Liu}},\ }\href {https://doi.org/10.1039/D4CP01673D} {\bibfield  {journal} {\bibinfo  {journal} {Phys. Chem. Chem. Phys.}\ }\textbf {\bibinfo {volume} {26}},\ \bibinfo {pages} {18402} (\bibinfo {year} {2024})}\BibitemShut {NoStop}%
\bibitem [{\citenamefont {Alem}\ \emph {et~al.}(2009)\citenamefont {Alem}, \citenamefont {Erni}, \citenamefont {Kisielowski}, \citenamefont {Rossell}, \citenamefont {Gannett},\ and\ \citenamefont {Zettl}}]{PhysRevB.80.155425}%
  \BibitemOpen
  \bibfield  {author} {\bibinfo {author} {\bibfnamefont {N.}~\bibnamefont {Alem}}, \bibinfo {author} {\bibfnamefont {R.}~\bibnamefont {Erni}}, \bibinfo {author} {\bibfnamefont {C.}~\bibnamefont {Kisielowski}}, \bibinfo {author} {\bibfnamefont {M.~D.}\ \bibnamefont {Rossell}}, \bibinfo {author} {\bibfnamefont {W.}~\bibnamefont {Gannett}},\ and\ \bibinfo {author} {\bibfnamefont {A.}~\bibnamefont {Zettl}},\ }\href {https://doi.org/10.1103/PhysRevB.80.155425} {\bibfield  {journal} {\bibinfo  {journal} {Phys. Rev. B}\ }\textbf {\bibinfo {volume} {80}},\ \bibinfo {pages} {155425} (\bibinfo {year} {2009})}\BibitemShut {NoStop}%
\bibitem [{\citenamefont {Gibb}\ \emph {et~al.}(2013)\citenamefont {Gibb}, \citenamefont {Alem}, \citenamefont {Chen}, \citenamefont {Erickson}, \citenamefont {Ciston}, \citenamefont {Gautam}, \citenamefont {Linck},\ and\ \citenamefont {Zettl}}]{doi:10.1021/ja400637n}%
  \BibitemOpen
  \bibfield  {author} {\bibinfo {author} {\bibfnamefont {A.~L.}\ \bibnamefont {Gibb}}, \bibinfo {author} {\bibfnamefont {N.}~\bibnamefont {Alem}}, \bibinfo {author} {\bibfnamefont {J.-H.}\ \bibnamefont {Chen}}, \bibinfo {author} {\bibfnamefont {K.~J.}\ \bibnamefont {Erickson}}, \bibinfo {author} {\bibfnamefont {J.}~\bibnamefont {Ciston}}, \bibinfo {author} {\bibfnamefont {A.}~\bibnamefont {Gautam}}, \bibinfo {author} {\bibfnamefont {M.}~\bibnamefont {Linck}},\ and\ \bibinfo {author} {\bibfnamefont {A.}~\bibnamefont {Zettl}},\ }\href {https://doi.org/10.1021/ja400637n} {\bibfield  {journal} {\bibinfo  {journal} {Journal of the American Chemical Society}\ }\textbf {\bibinfo {volume} {135}},\ \bibinfo {pages} {6758} (\bibinfo {year} {2013})},\ \bibinfo {note} {pMID: 23550733}\BibitemShut {NoStop}%
\bibitem [{\citenamefont {Alem}\ \emph {et~al.}(2012)\citenamefont {Alem}, \citenamefont {Ramasse}, \citenamefont {Seabourne}, \citenamefont {Yazyev}, \citenamefont {Erickson}, \citenamefont {Sarahan}, \citenamefont {Kisielowski}, \citenamefont {Scott}, \citenamefont {Louie},\ and\ \citenamefont {Zettl}}]{PhysRevLett.109.205502}%
  \BibitemOpen
  \bibfield  {author} {\bibinfo {author} {\bibfnamefont {N.}~\bibnamefont {Alem}}, \bibinfo {author} {\bibfnamefont {Q.~M.}\ \bibnamefont {Ramasse}}, \bibinfo {author} {\bibfnamefont {C.~R.}\ \bibnamefont {Seabourne}}, \bibinfo {author} {\bibfnamefont {O.~V.}\ \bibnamefont {Yazyev}}, \bibinfo {author} {\bibfnamefont {K.}~\bibnamefont {Erickson}}, \bibinfo {author} {\bibfnamefont {M.~C.}\ \bibnamefont {Sarahan}}, \bibinfo {author} {\bibfnamefont {C.}~\bibnamefont {Kisielowski}}, \bibinfo {author} {\bibfnamefont {A.~J.}\ \bibnamefont {Scott}}, \bibinfo {author} {\bibfnamefont {S.~G.}\ \bibnamefont {Louie}},\ and\ \bibinfo {author} {\bibfnamefont {A.}~\bibnamefont {Zettl}},\ }\href {https://doi.org/10.1103/PhysRevLett.109.205502} {\bibfield  {journal} {\bibinfo  {journal} {Phys. Rev. Lett.}\ }\textbf {\bibinfo {volume} {109}},\ \bibinfo {pages} {205502} (\bibinfo {year} {2012})}\BibitemShut {NoStop}%
\bibitem [{\citenamefont {Singh}\ \emph {et~al.}(2018)\citenamefont {Singh}, \citenamefont {Manjanath},\ and\ \citenamefont {Singh}}]{Singh2018_jpcc}%
  \BibitemOpen
  \bibfield  {author} {\bibinfo {author} {\bibfnamefont {A.}~\bibnamefont {Singh}}, \bibinfo {author} {\bibfnamefont {A.}~\bibnamefont {Manjanath}},\ and\ \bibinfo {author} {\bibfnamefont {A.~K.}\ \bibnamefont {Singh}},\ }\href {https://doi.org/10.1021/acs.jpcc.8b08082} {\bibfield  {journal} {\bibinfo  {journal} {The Journal of Physical Chemistry C}\ }\textbf {\bibinfo {volume} {122}},\ \bibinfo {pages} {24475} (\bibinfo {year} {2018})}\BibitemShut {NoStop}%
\bibitem [{\citenamefont {Rhodes}\ \emph {et~al.}(2019)\citenamefont {Rhodes}, \citenamefont {Chae}, \citenamefont {Ribeiro-Palau},\ and\ \citenamefont {Hone}}]{rhodes2019disorder}%
  \BibitemOpen
  \bibfield  {author} {\bibinfo {author} {\bibfnamefont {D.}~\bibnamefont {Rhodes}}, \bibinfo {author} {\bibfnamefont {S.~H.}\ \bibnamefont {Chae}}, \bibinfo {author} {\bibfnamefont {R.}~\bibnamefont {Ribeiro-Palau}},\ and\ \bibinfo {author} {\bibfnamefont {J.}~\bibnamefont {Hone}},\ }\href {https://doi.org/https://doi.org/10.1038/s41563-019-0366-8} {\bibfield  {journal} {\bibinfo  {journal} {Nature materials}\ }\textbf {\bibinfo {volume} {18}},\ \bibinfo {pages} {541} (\bibinfo {year} {2019})}\BibitemShut {NoStop}%
\bibitem [{\citenamefont {Chen}\ \emph {et~al.}(2023{\natexlab{a}})\citenamefont {Chen}, \citenamefont {Jiang}, \citenamefont {Yang},\ and\ \citenamefont {Dai}}]{Chen2023_acr}%
  \BibitemOpen
  \bibfield  {author} {\bibinfo {author} {\bibfnamefont {H.}~\bibnamefont {Chen}}, \bibinfo {author} {\bibfnamefont {D.-e.}\ \bibnamefont {Jiang}}, \bibinfo {author} {\bibfnamefont {Z.}~\bibnamefont {Yang}},\ and\ \bibinfo {author} {\bibfnamefont {S.}~\bibnamefont {Dai}},\ }\href {https://doi.org/10.1021/acs.accounts.2c00564} {\bibfield  {journal} {\bibinfo  {journal} {Accounts of Chemical Research}\ }\textbf {\bibinfo {volume} {56}},\ \bibinfo {pages} {52} (\bibinfo {year} {2023}{\natexlab{a}})},\ \bibinfo {note} {pMID: 36378327}\BibitemShut {NoStop}%
\bibitem [{\citenamefont {He}\ \emph {et~al.}(2014)\citenamefont {He}, \citenamefont {Jiao}, \citenamefont {Zhang}, \citenamefont {Xiao}, \citenamefont {Chen},\ and\ \citenamefont {Sun}}]{doi:10.1021/jp410716q}%
  \BibitemOpen
  \bibfield  {author} {\bibinfo {author} {\bibfnamefont {J.}~\bibnamefont {He}}, \bibinfo {author} {\bibfnamefont {N.}~\bibnamefont {Jiao}}, \bibinfo {author} {\bibfnamefont {C.}~\bibnamefont {Zhang}}, \bibinfo {author} {\bibfnamefont {H.}~\bibnamefont {Xiao}}, \bibinfo {author} {\bibfnamefont {X.}~\bibnamefont {Chen}},\ and\ \bibinfo {author} {\bibfnamefont {L.}~\bibnamefont {Sun}},\ }\href {https://doi.org/10.1021/jp410716q} {\bibfield  {journal} {\bibinfo  {journal} {The Journal of Physical Chemistry C}\ }\textbf {\bibinfo {volume} {118}},\ \bibinfo {pages} {8899} (\bibinfo {year} {2014})}\BibitemShut {NoStop}%
\bibitem [{\citenamefont {Zhang}\ \emph {et~al.}(2020)\citenamefont {Zhang}, \citenamefont {Sun}, \citenamefont {Ruan}, \citenamefont {Zhang}, \citenamefont {Li}, \citenamefont {Zhang}, \citenamefont {Cheng}, \citenamefont {Wang},\ and\ \citenamefont {Wang}}]{Zhang2020_JAP}%
  \BibitemOpen
  \bibfield  {author} {\bibinfo {author} {\bibfnamefont {J.}~\bibnamefont {Zhang}}, \bibinfo {author} {\bibfnamefont {R.}~\bibnamefont {Sun}}, \bibinfo {author} {\bibfnamefont {D.}~\bibnamefont {Ruan}}, \bibinfo {author} {\bibfnamefont {M.}~\bibnamefont {Zhang}}, \bibinfo {author} {\bibfnamefont {Y.}~\bibnamefont {Li}}, \bibinfo {author} {\bibfnamefont {K.}~\bibnamefont {Zhang}}, \bibinfo {author} {\bibfnamefont {F.}~\bibnamefont {Cheng}}, \bibinfo {author} {\bibfnamefont {Z.}~\bibnamefont {Wang}},\ and\ \bibinfo {author} {\bibfnamefont {Z.-M.}\ \bibnamefont {Wang}},\ }\href {https://doi.org/10.1063/5.0021093} {\bibfield  {journal} {\bibinfo  {journal} {Journal of Applied Physics}\ }\textbf {\bibinfo {volume} {128}},\ \bibinfo {pages} {100902} (\bibinfo {year} {2020})}\BibitemShut {NoStop}%
\bibitem [{\citenamefont {Cholsuk}\ \emph {et~al.}(2024)\citenamefont {Cholsuk}, \citenamefont {Zand}, \citenamefont {{\c{C}}akan},\ and\ \citenamefont {Vogl}}]{Cholsuk2024_JPCC}%
  \BibitemOpen
  \bibfield  {author} {\bibinfo {author} {\bibfnamefont {C.}~\bibnamefont {Cholsuk}}, \bibinfo {author} {\bibfnamefont {A.}~\bibnamefont {Zand}}, \bibinfo {author} {\bibfnamefont {A.}~\bibnamefont {{\c{C}}akan}},\ and\ \bibinfo {author} {\bibfnamefont {T.}~\bibnamefont {Vogl}},\ }\href {https://doi.org/10.1021/acs.jpcc.4c03404} {\bibfield  {journal} {\bibinfo  {journal} {The Journal of Physical Chemistry C}\ }\textbf {\bibinfo {volume} {128}},\ \bibinfo {pages} {12716} (\bibinfo {year} {2024})}\BibitemShut {NoStop}%
\bibitem [{\citenamefont {Zeng}\ \emph {et~al.}(2024)\citenamefont {Zeng}, \citenamefont {Zhang}, \citenamefont {Meng}, \citenamefont {Chen}, \citenamefont {Jiang}, \citenamefont {Shi}, \citenamefont {Huang}, \citenamefont {Yin}, \citenamefont {Wu},\ and\ \citenamefont {Zhang}}]{Zeng2024_acsami}%
  \BibitemOpen
  \bibfield  {author} {\bibinfo {author} {\bibfnamefont {L.}~\bibnamefont {Zeng}}, \bibinfo {author} {\bibfnamefont {S.}~\bibnamefont {Zhang}}, \bibinfo {author} {\bibfnamefont {J.}~\bibnamefont {Meng}}, \bibinfo {author} {\bibfnamefont {J.}~\bibnamefont {Chen}}, \bibinfo {author} {\bibfnamefont {J.}~\bibnamefont {Jiang}}, \bibinfo {author} {\bibfnamefont {Y.}~\bibnamefont {Shi}}, \bibinfo {author} {\bibfnamefont {J.}~\bibnamefont {Huang}}, \bibinfo {author} {\bibfnamefont {Z.}~\bibnamefont {Yin}}, \bibinfo {author} {\bibfnamefont {J.}~\bibnamefont {Wu}},\ and\ \bibinfo {author} {\bibfnamefont {X.}~\bibnamefont {Zhang}},\ }\href {https://doi.org/10.1021/acsami.4c02601} {\bibfield  {journal} {\bibinfo  {journal} {ACS Applied Materials \& Interfaces}\ }\textbf {\bibinfo {volume} {16}},\ \bibinfo {pages} {24899} (\bibinfo {year} {2024})},\ \bibinfo {note} {pMID: 38687622}\BibitemShut {NoStop}%
\bibitem [{\citenamefont {Smart}\ \emph {et~al.}(2021)\citenamefont {Smart}, \citenamefont {Li}, \citenamefont {Xu},\ and\ \citenamefont {Ping}}]{smart2021intersystem}%
  \BibitemOpen
  \bibfield  {author} {\bibinfo {author} {\bibfnamefont {T.~J.}\ \bibnamefont {Smart}}, \bibinfo {author} {\bibfnamefont {K.}~\bibnamefont {Li}}, \bibinfo {author} {\bibfnamefont {J.}~\bibnamefont {Xu}},\ and\ \bibinfo {author} {\bibfnamefont {Y.}~\bibnamefont {Ping}},\ }\href {https://doi.org/https://doi.org/10.1038/s41524-021-00525-5} {\bibfield  {journal} {\bibinfo  {journal} {npj Computational Materials}\ }\textbf {\bibinfo {volume} {7}},\ \bibinfo {pages} {59} (\bibinfo {year} {2021})}\BibitemShut {NoStop}%
\bibitem [{\citenamefont {Längle}\ \emph {et~al.}(2024)\citenamefont {Längle}, \citenamefont {Mayer}, \citenamefont {Madsen}, \citenamefont {Propst}, \citenamefont {Bo}, \citenamefont {Kofler}, \citenamefont {Hana}, \citenamefont {Mangler}, \citenamefont {Susi},\ and\ \citenamefont {Kotakoski}}]{langle2024}%
  \BibitemOpen
  \bibfield  {author} {\bibinfo {author} {\bibfnamefont {M.}~\bibnamefont {Längle}}, \bibinfo {author} {\bibfnamefont {B.~M.}\ \bibnamefont {Mayer}}, \bibinfo {author} {\bibfnamefont {J.}~\bibnamefont {Madsen}}, \bibinfo {author} {\bibfnamefont {D.}~\bibnamefont {Propst}}, \bibinfo {author} {\bibfnamefont {A.}~\bibnamefont {Bo}}, \bibinfo {author} {\bibfnamefont {C.}~\bibnamefont {Kofler}}, \bibinfo {author} {\bibfnamefont {V.}~\bibnamefont {Hana}}, \bibinfo {author} {\bibfnamefont {C.}~\bibnamefont {Mangler}}, \bibinfo {author} {\bibfnamefont {T.}~\bibnamefont {Susi}},\ and\ \bibinfo {author} {\bibfnamefont {J.}~\bibnamefont {Kotakoski}},\ }\href {https://arxiv.org/abs/2404.07166} {\bibinfo {title} {Defect-engineering hexagonal boron nitride using low-energy ar+ irradiation}} (\bibinfo {year} {2024}),\ \Eprint {https://arxiv.org/abs/2404.07166} {arXiv:2404.07166 [cond-mat.mtrl-sci]} \BibitemShut {NoStop}%
\bibitem [{\citenamefont {Gong}\ \emph {et~al.}(2023)\citenamefont {Gong}, \citenamefont {He}, \citenamefont {Gao}, \citenamefont {Ju}, \citenamefont {Liu}, \citenamefont {Ye}, \citenamefont {Henriksen}, \citenamefont {Li},\ and\ \citenamefont {Zu}}]{gong2023coherent}%
  \BibitemOpen
  \bibfield  {author} {\bibinfo {author} {\bibfnamefont {R.}~\bibnamefont {Gong}}, \bibinfo {author} {\bibfnamefont {G.}~\bibnamefont {He}}, \bibinfo {author} {\bibfnamefont {X.}~\bibnamefont {Gao}}, \bibinfo {author} {\bibfnamefont {P.}~\bibnamefont {Ju}}, \bibinfo {author} {\bibfnamefont {Z.}~\bibnamefont {Liu}}, \bibinfo {author} {\bibfnamefont {B.}~\bibnamefont {Ye}}, \bibinfo {author} {\bibfnamefont {E.~A.}\ \bibnamefont {Henriksen}}, \bibinfo {author} {\bibfnamefont {T.}~\bibnamefont {Li}},\ and\ \bibinfo {author} {\bibfnamefont {C.}~\bibnamefont {Zu}},\ }\href {https://doi.org/https://doi.org/10.1038/s41467-023-39115-y} {\bibfield  {journal} {\bibinfo  {journal} {Nature Communications}\ }\textbf {\bibinfo {volume} {14}},\ \bibinfo {pages} {3299} (\bibinfo {year} {2023})}\BibitemShut {NoStop}%
\bibitem [{\citenamefont {Jin}\ \emph {et~al.}(2009)\citenamefont {Jin}, \citenamefont {Lin}, \citenamefont {Suenaga},\ and\ \citenamefont {Iijima}}]{Jin2009_PRL}%
  \BibitemOpen
  \bibfield  {author} {\bibinfo {author} {\bibfnamefont {C.}~\bibnamefont {Jin}}, \bibinfo {author} {\bibfnamefont {F.}~\bibnamefont {Lin}}, \bibinfo {author} {\bibfnamefont {K.}~\bibnamefont {Suenaga}},\ and\ \bibinfo {author} {\bibfnamefont {S.}~\bibnamefont {Iijima}},\ }\href {https://doi.org/10.1103/PhysRevLett.102.195505} {\bibfield  {journal} {\bibinfo  {journal} {Phys. Rev. Lett.}\ }\textbf {\bibinfo {volume} {102}},\ \bibinfo {pages} {195505} (\bibinfo {year} {2009})}\BibitemShut {NoStop}%
\bibitem [{\citenamefont {Zhu}\ \emph {et~al.}(2017)\citenamefont {Zhu}, \citenamefont {Wu}, \citenamefont {Foo}, \citenamefont {Gao}, \citenamefont {Zhou}, \citenamefont {Liu}, \citenamefont {Veith}, \citenamefont {Wu}, \citenamefont {Browning}, \citenamefont {Lee} \emph {et~al.}}]{zhu2017taming}%
  \BibitemOpen
  \bibfield  {author} {\bibinfo {author} {\bibfnamefont {W.}~\bibnamefont {Zhu}}, \bibinfo {author} {\bibfnamefont {Z.}~\bibnamefont {Wu}}, \bibinfo {author} {\bibfnamefont {G.~S.}\ \bibnamefont {Foo}}, \bibinfo {author} {\bibfnamefont {X.}~\bibnamefont {Gao}}, \bibinfo {author} {\bibfnamefont {M.}~\bibnamefont {Zhou}}, \bibinfo {author} {\bibfnamefont {B.}~\bibnamefont {Liu}}, \bibinfo {author} {\bibfnamefont {G.~M.}\ \bibnamefont {Veith}}, \bibinfo {author} {\bibfnamefont {P.}~\bibnamefont {Wu}}, \bibinfo {author} {\bibfnamefont {K.~L.}\ \bibnamefont {Browning}}, \bibinfo {author} {\bibfnamefont {H.~N.}\ \bibnamefont {Lee}}, \emph {et~al.},\ }\href {https://doi.org/https://doi.org/10.1038/ncomms15291} {\bibfield  {journal} {\bibinfo  {journal} {Nature communications}\ }\textbf {\bibinfo {volume} {8}},\ \bibinfo {pages} {15291} (\bibinfo {year} {2017})}\BibitemShut {NoStop}%
\bibitem [{\citenamefont {Chen}\ \emph {et~al.}(2023{\natexlab{b}})\citenamefont {Chen}, \citenamefont {Kang}, \citenamefont {Pu}, \citenamefont {Tian}, \citenamefont {Wan}, \citenamefont {Xu}, \citenamefont {Yu}, \citenamefont {Jie},\ and\ \citenamefont {Zhao}}]{D2NR07234C}%
  \BibitemOpen
  \bibfield  {author} {\bibinfo {author} {\bibfnamefont {H.}~\bibnamefont {Chen}}, \bibinfo {author} {\bibfnamefont {Y.}~\bibnamefont {Kang}}, \bibinfo {author} {\bibfnamefont {D.}~\bibnamefont {Pu}}, \bibinfo {author} {\bibfnamefont {M.}~\bibnamefont {Tian}}, \bibinfo {author} {\bibfnamefont {N.}~\bibnamefont {Wan}}, \bibinfo {author} {\bibfnamefont {Y.}~\bibnamefont {Xu}}, \bibinfo {author} {\bibfnamefont {B.}~\bibnamefont {Yu}}, \bibinfo {author} {\bibfnamefont {W.}~\bibnamefont {Jie}},\ and\ \bibinfo {author} {\bibfnamefont {Y.}~\bibnamefont {Zhao}},\ }\href {https://doi.org/10.1039/D2NR07234C} {\bibfield  {journal} {\bibinfo  {journal} {Nanoscale}\ }\textbf {\bibinfo {volume} {15}},\ \bibinfo {pages} {4309} (\bibinfo {year} {2023}{\natexlab{b}})}\BibitemShut {NoStop}%
\bibitem [{\citenamefont {Babar}\ \emph {et~al.}(2024)\citenamefont {Babar}, \citenamefont {Barcza}, \citenamefont {Pershin}, \citenamefont {Park}, \citenamefont {Bulancea~Lindvall}, \citenamefont {Thiering}, \citenamefont {Legeza}, \citenamefont {Warner}, \citenamefont {Abrikosov}, \citenamefont {Gali} \emph {et~al.}}]{babar2024low}%
  \BibitemOpen
  \bibfield  {author} {\bibinfo {author} {\bibfnamefont {R.}~\bibnamefont {Babar}}, \bibinfo {author} {\bibfnamefont {G.}~\bibnamefont {Barcza}}, \bibinfo {author} {\bibfnamefont {A.}~\bibnamefont {Pershin}}, \bibinfo {author} {\bibfnamefont {H.}~\bibnamefont {Park}}, \bibinfo {author} {\bibfnamefont {O.}~\bibnamefont {Bulancea~Lindvall}}, \bibinfo {author} {\bibfnamefont {G.}~\bibnamefont {Thiering}}, \bibinfo {author} {\bibfnamefont {{\"O}.}~\bibnamefont {Legeza}}, \bibinfo {author} {\bibfnamefont {J.~H.}\ \bibnamefont {Warner}}, \bibinfo {author} {\bibfnamefont {I.~A.}\ \bibnamefont {Abrikosov}}, \bibinfo {author} {\bibfnamefont {A.}~\bibnamefont {Gali}}, \emph {et~al.},\ }\href {https://doi.org/https://doi.org/10.1038/s41524-024-01361-z} {\bibfield  {journal} {\bibinfo  {journal} {npj Computational Materials}\ }\textbf {\bibinfo {volume} {10}},\ \bibinfo {pages} {184} (\bibinfo {year} {2024})}\BibitemShut {NoStop}%
\bibitem [{\citenamefont {Huang}\ \emph {et~al.}(2022)\citenamefont {Huang}, \citenamefont {Grzeszczyk}, \citenamefont {Vaklinova}, \citenamefont {Watanabe}, \citenamefont {Taniguchi}, \citenamefont {Novoselov},\ and\ \citenamefont {Koperski}}]{PhysRevB.106.014107}%
  \BibitemOpen
  \bibfield  {author} {\bibinfo {author} {\bibfnamefont {P.}~\bibnamefont {Huang}}, \bibinfo {author} {\bibfnamefont {M.}~\bibnamefont {Grzeszczyk}}, \bibinfo {author} {\bibfnamefont {K.}~\bibnamefont {Vaklinova}}, \bibinfo {author} {\bibfnamefont {K.}~\bibnamefont {Watanabe}}, \bibinfo {author} {\bibfnamefont {T.}~\bibnamefont {Taniguchi}}, \bibinfo {author} {\bibfnamefont {K.~S.}\ \bibnamefont {Novoselov}},\ and\ \bibinfo {author} {\bibfnamefont {M.}~\bibnamefont {Koperski}},\ }\href {https://doi.org/10.1103/PhysRevB.106.014107} {\bibfield  {journal} {\bibinfo  {journal} {Phys. Rev. B}\ }\textbf {\bibinfo {volume} {106}},\ \bibinfo {pages} {014107} (\bibinfo {year} {2022})}\BibitemShut {NoStop}%
\bibitem [{\citenamefont {Chen}\ \emph {et~al.}(2018)\citenamefont {Chen}, \citenamefont {Li}, \citenamefont {Xu}, \citenamefont {Pan}, \citenamefont {Fu},\ and\ \citenamefont {Bao}}]{C7TA08515J}%
  \BibitemOpen
  \bibfield  {author} {\bibinfo {author} {\bibfnamefont {S.}~\bibnamefont {Chen}}, \bibinfo {author} {\bibfnamefont {P.}~\bibnamefont {Li}}, \bibinfo {author} {\bibfnamefont {S.}~\bibnamefont {Xu}}, \bibinfo {author} {\bibfnamefont {X.}~\bibnamefont {Pan}}, \bibinfo {author} {\bibfnamefont {Q.}~\bibnamefont {Fu}},\ and\ \bibinfo {author} {\bibfnamefont {X.}~\bibnamefont {Bao}},\ }\href {https://doi.org/10.1039/C7TA08515J} {\bibfield  {journal} {\bibinfo  {journal} {J. Mater. Chem. A}\ }\textbf {\bibinfo {volume} {6}},\ \bibinfo {pages} {1832} (\bibinfo {year} {2018})}\BibitemShut {NoStop}%
\bibitem [{\citenamefont {Berseneva}\ \emph {et~al.}(2013)\citenamefont {Berseneva}, \citenamefont {Gulans}, \citenamefont {Krasheninnikov},\ and\ \citenamefont {Nieminen}}]{PhysRevB.87.035404}%
  \BibitemOpen
  \bibfield  {author} {\bibinfo {author} {\bibfnamefont {N.}~\bibnamefont {Berseneva}}, \bibinfo {author} {\bibfnamefont {A.}~\bibnamefont {Gulans}}, \bibinfo {author} {\bibfnamefont {A.~V.}\ \bibnamefont {Krasheninnikov}},\ and\ \bibinfo {author} {\bibfnamefont {R.~M.}\ \bibnamefont {Nieminen}},\ }\href {https://doi.org/10.1103/PhysRevB.87.035404} {\bibfield  {journal} {\bibinfo  {journal} {Phys. Rev. B}\ }\textbf {\bibinfo {volume} {87}},\ \bibinfo {pages} {035404} (\bibinfo {year} {2013})}\BibitemShut {NoStop}%
\bibitem [{\citenamefont {Liu}\ \emph {et~al.}(2019)\citenamefont {Liu}, \citenamefont {Nattestad}, \citenamefont {Naficy}, \citenamefont {Han}, \citenamefont {Casillas}, \citenamefont {Angeloski}, \citenamefont {Sun},\ and\ \citenamefont {Huang}}]{Liu2019_AOM}%
  \BibitemOpen
  \bibfield  {author} {\bibinfo {author} {\bibfnamefont {F.}~\bibnamefont {Liu}}, \bibinfo {author} {\bibfnamefont {A.}~\bibnamefont {Nattestad}}, \bibinfo {author} {\bibfnamefont {S.}~\bibnamefont {Naficy}}, \bibinfo {author} {\bibfnamefont {R.}~\bibnamefont {Han}}, \bibinfo {author} {\bibfnamefont {G.}~\bibnamefont {Casillas}}, \bibinfo {author} {\bibfnamefont {A.}~\bibnamefont {Angeloski}}, \bibinfo {author} {\bibfnamefont {X.}~\bibnamefont {Sun}},\ and\ \bibinfo {author} {\bibfnamefont {Z.}~\bibnamefont {Huang}},\ }\href {https://doi.org/https://doi.org/10.1002/adom.201901380} {\bibfield  {journal} {\bibinfo  {journal} {Advanced Optical Materials}\ }\textbf {\bibinfo {volume} {7}},\ \bibinfo {pages} {1901380} (\bibinfo {year} {2019})}\BibitemShut {NoStop}%
\bibitem [{\citenamefont {Asif}\ \emph {et~al.}(2021)\citenamefont {Asif}, \citenamefont {Hussain}, \citenamefont {Kashif}, \citenamefont {Tayyab},\ and\ \citenamefont {Rafique}}]{asif2021computational}%
  \BibitemOpen
  \bibfield  {author} {\bibinfo {author} {\bibfnamefont {Q.~u.~A.}\ \bibnamefont {Asif}}, \bibinfo {author} {\bibfnamefont {A.}~\bibnamefont {Hussain}}, \bibinfo {author} {\bibfnamefont {M.}~\bibnamefont {Kashif}}, \bibinfo {author} {\bibfnamefont {M.}~\bibnamefont {Tayyab}},\ and\ \bibinfo {author} {\bibfnamefont {H.~M.}\ \bibnamefont {Rafique}},\ }\href {https://doi.org/https://doi.org/10.1007/s00894-021-04938-3} {\bibfield  {journal} {\bibinfo  {journal} {Journal of Molecular Modeling}\ }\textbf {\bibinfo {volume} {27}},\ \bibinfo {pages} {1} (\bibinfo {year} {2021})}\BibitemShut {NoStop}%
\bibitem [{\citenamefont {Ghorbani-Asl}\ \emph {et~al.}(2022)\citenamefont {Ghorbani-Asl}, \citenamefont {Kretschmer},\ and\ \citenamefont {Krasheninnikov}}]{GHORBANIASL2022259}%
  \BibitemOpen
  \bibfield  {author} {\bibinfo {author} {\bibfnamefont {M.}~\bibnamefont {Ghorbani-Asl}}, \bibinfo {author} {\bibfnamefont {S.}~\bibnamefont {Kretschmer}},\ and\ \bibinfo {author} {\bibfnamefont {A.~V.}\ \bibnamefont {Krasheninnikov}},\ }in\ \href {https://doi.org/https://doi.org/10.1016/B978-0-12-820292-0.00015-X} {\emph {\bibinfo {booktitle} {Defects in Two-Dimensional Materials}}},\ \bibinfo {series and number} {Materials Today},\ \bibinfo {editor} {edited by\ \bibinfo {editor} {\bibfnamefont {R.}~\bibnamefont {Addou}}\ and\ \bibinfo {editor} {\bibfnamefont {L.}~\bibnamefont {Colombo}}}\ (\bibinfo  {publisher} {Elsevier},\ \bibinfo {year} {2022})\ pp.\ \bibinfo {pages} {259--301}\BibitemShut {NoStop}%
\bibitem [{\citenamefont {Bui}\ \emph {et~al.}(2023)\citenamefont {Bui}, \citenamefont {Leuthner}, \citenamefont {Madsen}, \citenamefont {Monazam}, \citenamefont {Chirita}, \citenamefont {Postl}, \citenamefont {Mangler}, \citenamefont {Kotakoski},\ and\ \citenamefont {Susi}}]{Bui_small_2023}%
  \BibitemOpen
  \bibfield  {author} {\bibinfo {author} {\bibfnamefont {T.~A.}\ \bibnamefont {Bui}}, \bibinfo {author} {\bibfnamefont {G.~T.}\ \bibnamefont {Leuthner}}, \bibinfo {author} {\bibfnamefont {J.}~\bibnamefont {Madsen}}, \bibinfo {author} {\bibfnamefont {M.~R.~A.}\ \bibnamefont {Monazam}}, \bibinfo {author} {\bibfnamefont {A.~I.}\ \bibnamefont {Chirita}}, \bibinfo {author} {\bibfnamefont {A.}~\bibnamefont {Postl}}, \bibinfo {author} {\bibfnamefont {C.}~\bibnamefont {Mangler}}, \bibinfo {author} {\bibfnamefont {J.}~\bibnamefont {Kotakoski}},\ and\ \bibinfo {author} {\bibfnamefont {T.}~\bibnamefont {Susi}},\ }\href {https://doi.org/https://doi.org/10.1002/smll.202301926} {\bibfield  {journal} {\bibinfo  {journal} {Small}\ }\textbf {\bibinfo {volume} {19}},\ \bibinfo {pages} {2301926} (\bibinfo {year} {2023})}\BibitemShut {NoStop}%
\bibitem [{\citenamefont {Kohlrausch}\ \emph {et~al.}(2025)\citenamefont {Kohlrausch}, \citenamefont {Ghaderzadeh}, \citenamefont {Aliev}, \citenamefont {Popov}, \citenamefont {Saad}, \citenamefont {Alharbi}, \citenamefont {Ramasse}, \citenamefont {Rance}, \citenamefont {Danaie}, \citenamefont {Thangamuthu}, \citenamefont {Young}, \citenamefont {Plummer}, \citenamefont {Morgan}, \citenamefont {Theis}, \citenamefont {Besley}, \citenamefont {Khlobystov},\ and\ \citenamefont {Alves~Fernandes}}]{Kohlrausch_AdSci_2025}%
  \BibitemOpen
  \bibfield  {author} {\bibinfo {author} {\bibfnamefont {E.~C.}\ \bibnamefont {Kohlrausch}}, \bibinfo {author} {\bibfnamefont {S.}~\bibnamefont {Ghaderzadeh}}, \bibinfo {author} {\bibfnamefont {G.~N.}\ \bibnamefont {Aliev}}, \bibinfo {author} {\bibfnamefont {I.}~\bibnamefont {Popov}}, \bibinfo {author} {\bibfnamefont {F.}~\bibnamefont {Saad}}, \bibinfo {author} {\bibfnamefont {E.}~\bibnamefont {Alharbi}}, \bibinfo {author} {\bibfnamefont {Q.~M.}\ \bibnamefont {Ramasse}}, \bibinfo {author} {\bibfnamefont {G.~A.}\ \bibnamefont {Rance}}, \bibinfo {author} {\bibfnamefont {M.}~\bibnamefont {Danaie}}, \bibinfo {author} {\bibfnamefont {M.}~\bibnamefont {Thangamuthu}}, \bibinfo {author} {\bibfnamefont {M.}~\bibnamefont {Young}}, \bibinfo {author} {\bibfnamefont {R.}~\bibnamefont {Plummer}}, \bibinfo {author} {\bibfnamefont {D.~J.}\ \bibnamefont {Morgan}}, \bibinfo {author} {\bibfnamefont {W.}~\bibnamefont {Theis}}, \bibinfo {author} {\bibfnamefont {E.}~\bibnamefont {Besley}}, \bibinfo {author} {\bibfnamefont {A.~N.}\
  \bibnamefont {Khlobystov}},\ and\ \bibinfo {author} {\bibfnamefont {J.}~\bibnamefont {Alves~Fernandes}},\ }\href {https://doi.org/https://doi.org/10.1002/advs.202508034} {\bibfield  {journal} {\bibinfo  {journal} {Advanced Science}\ }\textbf {\bibinfo {volume} {n/a}},\ \bibinfo {pages} {e08034} (\bibinfo {year} {2025})}\BibitemShut {NoStop}%
\bibitem [{\citenamefont {Yu}\ \emph {et~al.}(2017)\citenamefont {Yu}, \citenamefont {Li}, \citenamefont {Lai}, \citenamefont {Hao},\ and\ \citenamefont {Zhang}}]{Yu_2017_iop}%
  \BibitemOpen
  \bibfield  {author} {\bibinfo {author} {\bibfnamefont {S.}~\bibnamefont {Yu}}, \bibinfo {author} {\bibfnamefont {L.}~\bibnamefont {Li}}, \bibinfo {author} {\bibfnamefont {Z.}~\bibnamefont {Lai}}, \bibinfo {author} {\bibfnamefont {J.}~\bibnamefont {Hao}},\ and\ \bibinfo {author} {\bibfnamefont {K.}~\bibnamefont {Zhang}},\ }\href {https://doi.org/10.1088/2053-1591/aa93bf} {\bibfield  {journal} {\bibinfo  {journal} {Materials Research Express}\ }\textbf {\bibinfo {volume} {4}},\ \bibinfo {pages} {116302} (\bibinfo {year} {2017})}\BibitemShut {NoStop}%
\bibitem [{\citenamefont {Lian}\ \emph {et~al.}(2024)\citenamefont {Lian}, \citenamefont {Tang}, \citenamefont {Liao}, \citenamefont {Guo}, \citenamefont {Zhang},\ and\ \citenamefont {Gao}}]{lian2024theoretical}%
  \BibitemOpen
  \bibfield  {author} {\bibinfo {author} {\bibfnamefont {X.}~\bibnamefont {Lian}}, \bibinfo {author} {\bibfnamefont {X.}~\bibnamefont {Tang}}, \bibinfo {author} {\bibfnamefont {H.}~\bibnamefont {Liao}}, \bibinfo {author} {\bibfnamefont {W.}~\bibnamefont {Guo}}, \bibinfo {author} {\bibfnamefont {Y.}~\bibnamefont {Zhang}},\ and\ \bibinfo {author} {\bibfnamefont {G.}~\bibnamefont {Gao}},\ }\href@noop {} {\bibfield  {journal} {\bibinfo  {journal} {Reaction Kinetics, Mechanisms and Catalysis}\ }\textbf {\bibinfo {volume} {137}},\ \bibinfo {pages} {3241} (\bibinfo {year} {2024})}\BibitemShut {NoStop}%
\bibitem [{\citenamefont {Zhong}\ \emph {et~al.}(2022)\citenamefont {Zhong}, \citenamefont {Wu}, \citenamefont {Yu}, \citenamefont {Shen}, \citenamefont {Yan},\ and\ \citenamefont {Xu}}]{zhong2022first}%
  \BibitemOpen
  \bibfield  {author} {\bibinfo {author} {\bibfnamefont {S.-Y.}\ \bibnamefont {Zhong}}, \bibinfo {author} {\bibfnamefont {S.-Y.}\ \bibnamefont {Wu}}, \bibinfo {author} {\bibfnamefont {X.-Y.}\ \bibnamefont {Yu}}, \bibinfo {author} {\bibfnamefont {G.-Q.}\ \bibnamefont {Shen}}, \bibinfo {author} {\bibfnamefont {L.}~\bibnamefont {Yan}},\ and\ \bibinfo {author} {\bibfnamefont {K.-L.}\ \bibnamefont {Xu}},\ }\href {https://doi.org/https://doi.org/10.1007/s10563-021-09350-8} {\bibfield  {journal} {\bibinfo  {journal} {Catalysis Surveys from Asia}\ }\textbf {\bibinfo {volume} {26}},\ \bibinfo {pages} {69} (\bibinfo {year} {2022})}\BibitemShut {NoStop}%
\bibitem [{\citenamefont {Wu}\ \emph {et~al.}(2013)\citenamefont {Wu}, \citenamefont {Yin},\ and\ \citenamefont {Zhang}}]{C3RA23132A}%
  \BibitemOpen
  \bibfield  {author} {\bibinfo {author} {\bibfnamefont {J.}~\bibnamefont {Wu}}, \bibinfo {author} {\bibfnamefont {L.}~\bibnamefont {Yin}},\ and\ \bibinfo {author} {\bibfnamefont {L.}~\bibnamefont {Zhang}},\ }\href {https://doi.org/10.1039/C3RA23132A} {\bibfield  {journal} {\bibinfo  {journal} {RSC Adv.}\ }\textbf {\bibinfo {volume} {3}},\ \bibinfo {pages} {7408} (\bibinfo {year} {2013})}\BibitemShut {NoStop}%
\bibitem [{\citenamefont {Khalil}\ \emph {et~al.}(2023)\citenamefont {Khalil}, \citenamefont {Ernandes}, \citenamefont {Avila}, \citenamefont {Rousseau}, \citenamefont {Dudin}, \citenamefont {Zhigadlo}, \citenamefont {Cassabois}, \citenamefont {Gil}, \citenamefont {Oehler}, \citenamefont {Chaste},\ and\ \citenamefont {Ouerghi}}]{D2NA00843B}%
  \BibitemOpen
  \bibfield  {author} {\bibinfo {author} {\bibfnamefont {L.}~\bibnamefont {Khalil}}, \bibinfo {author} {\bibfnamefont {C.}~\bibnamefont {Ernandes}}, \bibinfo {author} {\bibfnamefont {J.}~\bibnamefont {Avila}}, \bibinfo {author} {\bibfnamefont {A.}~\bibnamefont {Rousseau}}, \bibinfo {author} {\bibfnamefont {P.}~\bibnamefont {Dudin}}, \bibinfo {author} {\bibfnamefont {N.~D.}\ \bibnamefont {Zhigadlo}}, \bibinfo {author} {\bibfnamefont {G.}~\bibnamefont {Cassabois}}, \bibinfo {author} {\bibfnamefont {B.}~\bibnamefont {Gil}}, \bibinfo {author} {\bibfnamefont {F.}~\bibnamefont {Oehler}}, \bibinfo {author} {\bibfnamefont {J.}~\bibnamefont {Chaste}},\ and\ \bibinfo {author} {\bibfnamefont {A.}~\bibnamefont {Ouerghi}},\ }\href {https://doi.org/10.1039/D2NA00843B} {\bibfield  {journal} {\bibinfo  {journal} {Nanoscale Adv.}\ }\textbf {\bibinfo {volume} {5}},\ \bibinfo {pages} {3225} (\bibinfo {year} {2023})}\BibitemShut {NoStop}%
\bibitem [{\citenamefont {Zhang}\ \emph {et~al.}(2024)\citenamefont {Zhang}, \citenamefont {Qin}, \citenamefont {Zhu}, \citenamefont {Wang},\ and\ \citenamefont {Cao}}]{Zhang2024_Langmuir}%
  \BibitemOpen
  \bibfield  {author} {\bibinfo {author} {\bibfnamefont {Y.}~\bibnamefont {Zhang}}, \bibinfo {author} {\bibfnamefont {C.}~\bibnamefont {Qin}}, \bibinfo {author} {\bibfnamefont {L.}~\bibnamefont {Zhu}}, \bibinfo {author} {\bibfnamefont {Y.}~\bibnamefont {Wang}},\ and\ \bibinfo {author} {\bibfnamefont {J.}~\bibnamefont {Cao}},\ }\href {https://doi.org/10.1021/acs.langmuir.3c03282} {\bibfield  {journal} {\bibinfo  {journal} {Langmuir}\ }\textbf {\bibinfo {volume} {40}},\ \bibinfo {pages} {1058} (\bibinfo {year} {2024})},\ \bibinfo {note} {pMID: 38146207}\BibitemShut {NoStop}%
\bibitem [{\citenamefont {Kozubek}\ \emph {et~al.}(2018)\citenamefont {Kozubek}, \citenamefont {Ernst}, \citenamefont {Herbig}, \citenamefont {Michely},\ and\ \citenamefont {Schleberger}}]{Kozubek2018_acs}%
  \BibitemOpen
  \bibfield  {author} {\bibinfo {author} {\bibfnamefont {R.}~\bibnamefont {Kozubek}}, \bibinfo {author} {\bibfnamefont {P.}~\bibnamefont {Ernst}}, \bibinfo {author} {\bibfnamefont {C.}~\bibnamefont {Herbig}}, \bibinfo {author} {\bibfnamefont {T.}~\bibnamefont {Michely}},\ and\ \bibinfo {author} {\bibfnamefont {M.}~\bibnamefont {Schleberger}},\ }\href {https://doi.org/10.1021/acsanm.8b00903} {\bibfield  {journal} {\bibinfo  {journal} {ACS Applied Nano Materials}\ }\textbf {\bibinfo {volume} {1}},\ \bibinfo {pages} {3765} (\bibinfo {year} {2018})}\BibitemShut {NoStop}%
\bibitem [{\citenamefont {Lu}\ \emph {et~al.}(2023)\citenamefont {Lu}, \citenamefont {Li}, \citenamefont {Xu}, \citenamefont {Zhou}, \citenamefont {Xiao}, \citenamefont {Jiang}, \citenamefont {Li}, \citenamefont {Hu}, \citenamefont {Gong},\ and\ \citenamefont {Cao}}]{lu2023_nature}%
  \BibitemOpen
  \bibfield  {author} {\bibinfo {author} {\bibfnamefont {Y.}~\bibnamefont {Lu}}, \bibinfo {author} {\bibfnamefont {B.}~\bibnamefont {Li}}, \bibinfo {author} {\bibfnamefont {N.}~\bibnamefont {Xu}}, \bibinfo {author} {\bibfnamefont {Z.}~\bibnamefont {Zhou}}, \bibinfo {author} {\bibfnamefont {Y.}~\bibnamefont {Xiao}}, \bibinfo {author} {\bibfnamefont {Y.}~\bibnamefont {Jiang}}, \bibinfo {author} {\bibfnamefont {T.}~\bibnamefont {Li}}, \bibinfo {author} {\bibfnamefont {S.}~\bibnamefont {Hu}}, \bibinfo {author} {\bibfnamefont {Y.}~\bibnamefont {Gong}},\ and\ \bibinfo {author} {\bibfnamefont {Y.}~\bibnamefont {Cao}},\ }\href {https://doi.org/https://doi.org/10.1038/s41467-023-42696-3} {\bibfield  {journal} {\bibinfo  {journal} {Nature Communications}\ }\textbf {\bibinfo {volume} {14}},\ \bibinfo {pages} {6965} (\bibinfo {year} {2023})}\BibitemShut {NoStop}%
\bibitem [{\citenamefont {Gottscholl}\ \emph {et~al.}(2021)\citenamefont {Gottscholl}, \citenamefont {Diez}, \citenamefont {Soltamov}, \citenamefont {Kasper}, \citenamefont {Krau{\ss}e}, \citenamefont {Sperlich}, \citenamefont {Kianinia}, \citenamefont {Bradac}, \citenamefont {Aharonovich},\ and\ \citenamefont {Dyakonov}}]{gottscholl2021spin}%
  \BibitemOpen
  \bibfield  {author} {\bibinfo {author} {\bibfnamefont {A.}~\bibnamefont {Gottscholl}}, \bibinfo {author} {\bibfnamefont {M.}~\bibnamefont {Diez}}, \bibinfo {author} {\bibfnamefont {V.}~\bibnamefont {Soltamov}}, \bibinfo {author} {\bibfnamefont {C.}~\bibnamefont {Kasper}}, \bibinfo {author} {\bibfnamefont {D.}~\bibnamefont {Krau{\ss}e}}, \bibinfo {author} {\bibfnamefont {A.}~\bibnamefont {Sperlich}}, \bibinfo {author} {\bibfnamefont {M.}~\bibnamefont {Kianinia}}, \bibinfo {author} {\bibfnamefont {C.}~\bibnamefont {Bradac}}, \bibinfo {author} {\bibfnamefont {I.}~\bibnamefont {Aharonovich}},\ and\ \bibinfo {author} {\bibfnamefont {V.}~\bibnamefont {Dyakonov}},\ }\href {https://doi.org/https://doi.org/10.1038/s41467-021-24725-1} {\bibfield  {journal} {\bibinfo  {journal} {Nature communications}\ }\textbf {\bibinfo {volume} {12}},\ \bibinfo {pages} {4480} (\bibinfo {year} {2021})}\BibitemShut {NoStop}%
\bibitem [{\citenamefont {Fang}\ \emph {et~al.}(2024)\citenamefont {Fang}, \citenamefont {Wang}, \citenamefont {Marie},\ and\ \citenamefont {Sun}}]{fang2024quantum}%
  \BibitemOpen
  \bibfield  {author} {\bibinfo {author} {\bibfnamefont {H.-H.}\ \bibnamefont {Fang}}, \bibinfo {author} {\bibfnamefont {X.-J.}\ \bibnamefont {Wang}}, \bibinfo {author} {\bibfnamefont {X.}~\bibnamefont {Marie}},\ and\ \bibinfo {author} {\bibfnamefont {H.-B.}\ \bibnamefont {Sun}},\ }\href {https://doi.org/https://doi.org/10.1038/s41377-024-01630-y} {\bibfield  {journal} {\bibinfo  {journal} {Light: Science \& Applications}\ }\textbf {\bibinfo {volume} {13}},\ \bibinfo {pages} {303} (\bibinfo {year} {2024})}\BibitemShut {NoStop}%
\bibitem [{\citenamefont {Fischer}\ \emph {et~al.}(2021)\citenamefont {Fischer}, \citenamefont {Caridad}, \citenamefont {Sajid}, \citenamefont {Ghaderzadeh}, \citenamefont {Ghorbani-Asl}, \citenamefont {Gammelgaard}, \citenamefont {B{\o}ggild}, \citenamefont {Thygesen}, \citenamefont {Krasheninnikov}, \citenamefont {Xiao} \emph {et~al.}}]{fischer2021controlled}%
  \BibitemOpen
  \bibfield  {author} {\bibinfo {author} {\bibfnamefont {M.}~\bibnamefont {Fischer}}, \bibinfo {author} {\bibfnamefont {J.~M.}\ \bibnamefont {Caridad}}, \bibinfo {author} {\bibfnamefont {A.}~\bibnamefont {Sajid}}, \bibinfo {author} {\bibfnamefont {S.}~\bibnamefont {Ghaderzadeh}}, \bibinfo {author} {\bibfnamefont {M.}~\bibnamefont {Ghorbani-Asl}}, \bibinfo {author} {\bibfnamefont {L.}~\bibnamefont {Gammelgaard}}, \bibinfo {author} {\bibfnamefont {P.}~\bibnamefont {B{\o}ggild}}, \bibinfo {author} {\bibfnamefont {K.~S.}\ \bibnamefont {Thygesen}}, \bibinfo {author} {\bibfnamefont {A.~V.}\ \bibnamefont {Krasheninnikov}}, \bibinfo {author} {\bibfnamefont {S.}~\bibnamefont {Xiao}}, \emph {et~al.},\ }\href {https://doi.org/10.1126/sciadv.abe7138} {\bibfield  {journal} {\bibinfo  {journal} {Science Advances}\ }\textbf {\bibinfo {volume} {7}},\ \bibinfo {pages} {eabe7138} (\bibinfo {year} {2021})}\BibitemShut {NoStop}%
\bibitem [{\citenamefont {White}\ \emph {et~al.}(2022)\citenamefont {White}, \citenamefont {Yang}, \citenamefont {Dontschuk}, \citenamefont {Li}, \citenamefont {Xu}, \citenamefont {Kianinia}, \citenamefont {Stacey}, \citenamefont {Toth},\ and\ \citenamefont {Aharonovich}}]{white2022electrical}%
  \BibitemOpen
  \bibfield  {author} {\bibinfo {author} {\bibfnamefont {S.~J.}\ \bibnamefont {White}}, \bibinfo {author} {\bibfnamefont {T.}~\bibnamefont {Yang}}, \bibinfo {author} {\bibfnamefont {N.}~\bibnamefont {Dontschuk}}, \bibinfo {author} {\bibfnamefont {C.}~\bibnamefont {Li}}, \bibinfo {author} {\bibfnamefont {Z.-Q.}\ \bibnamefont {Xu}}, \bibinfo {author} {\bibfnamefont {M.}~\bibnamefont {Kianinia}}, \bibinfo {author} {\bibfnamefont {A.}~\bibnamefont {Stacey}}, \bibinfo {author} {\bibfnamefont {M.}~\bibnamefont {Toth}},\ and\ \bibinfo {author} {\bibfnamefont {I.}~\bibnamefont {Aharonovich}},\ }\href {https://doi.org/https://doi.org/10.1038/s41377-024-01491-5} {\bibfield  {journal} {\bibinfo  {journal} {Light: Science \& Applications}\ }\textbf {\bibinfo {volume} {11}},\ \bibinfo {pages} {186} (\bibinfo {year} {2022})}\BibitemShut {NoStop}%
\bibitem [{\citenamefont {Polley}\ \emph {et~al.}(2023)\citenamefont {Polley}, \citenamefont {Fedderwitz}, \citenamefont {Balasubramanian}, \citenamefont {Zakharov}, \citenamefont {Yakimova}, \citenamefont {B\"acke}, \citenamefont {Ekman}, \citenamefont {Dash}, \citenamefont {Kubatkin},\ and\ \citenamefont {Lara-Avila}}]{Polley2023_PRL}%
  \BibitemOpen
  \bibfield  {author} {\bibinfo {author} {\bibfnamefont {C.~M.}\ \bibnamefont {Polley}}, \bibinfo {author} {\bibfnamefont {H.}~\bibnamefont {Fedderwitz}}, \bibinfo {author} {\bibfnamefont {T.}~\bibnamefont {Balasubramanian}}, \bibinfo {author} {\bibfnamefont {A.~A.}\ \bibnamefont {Zakharov}}, \bibinfo {author} {\bibfnamefont {R.}~\bibnamefont {Yakimova}}, \bibinfo {author} {\bibfnamefont {O.}~\bibnamefont {B\"acke}}, \bibinfo {author} {\bibfnamefont {J.}~\bibnamefont {Ekman}}, \bibinfo {author} {\bibfnamefont {S.~P.}\ \bibnamefont {Dash}}, \bibinfo {author} {\bibfnamefont {S.}~\bibnamefont {Kubatkin}},\ and\ \bibinfo {author} {\bibfnamefont {S.}~\bibnamefont {Lara-Avila}},\ }\href {https://doi.org/10.1103/PhysRevLett.130.076203} {\bibfield  {journal} {\bibinfo  {journal} {Phys. Rev. Lett.}\ }\textbf {\bibinfo {volume} {130}},\ \bibinfo {pages} {076203} (\bibinfo {year} {2023})}\BibitemShut {NoStop}%
\bibitem [{\citenamefont {Da}\ \emph {et~al.}(2024)\citenamefont {Da}, \citenamefont {Luo}, \citenamefont {Lei}, \citenamefont {Ji},\ and\ \citenamefont {Zhou}}]{Da_2024}%
  \BibitemOpen
  \bibfield  {author} {\bibinfo {author} {\bibfnamefont {Y.}~\bibnamefont {Da}}, \bibinfo {author} {\bibfnamefont {R.}~\bibnamefont {Luo}}, \bibinfo {author} {\bibfnamefont {B.}~\bibnamefont {Lei}}, \bibinfo {author} {\bibfnamefont {W.}~\bibnamefont {Ji}},\ and\ \bibinfo {author} {\bibfnamefont {W.}~\bibnamefont {Zhou}},\ }\href {https://doi.org/10.1088/1674-1056/ad6132} {\bibfield  {journal} {\bibinfo  {journal} {Chinese Physics B}\ }\textbf {\bibinfo {volume} {33}},\ \bibinfo {pages} {086802} (\bibinfo {year} {2024})}\BibitemShut {NoStop}%
\bibitem [{\citenamefont {Mokhov}\ \emph {et~al.}(2023)\citenamefont {Mokhov}, \citenamefont {Davydov}, \citenamefont {Smirnov},\ and\ \citenamefont {Nagaluk}}]{mokhov2023growth}%
  \BibitemOpen
  \bibfield  {author} {\bibinfo {author} {\bibfnamefont {E.}~\bibnamefont {Mokhov}}, \bibinfo {author} {\bibfnamefont {V.~Y.}\ \bibnamefont {Davydov}}, \bibinfo {author} {\bibfnamefont {A.}~\bibnamefont {Smirnov}},\ and\ \bibinfo {author} {\bibfnamefont {S.}~\bibnamefont {Nagaluk}},\ }\href {https://doi.org/https://doi.org/10.1134/S1063782623080092} {\bibfield  {journal} {\bibinfo  {journal} {Semiconductors}\ }\textbf {\bibinfo {volume} {57}},\ \bibinfo {pages} {483} (\bibinfo {year} {2023})}\BibitemShut {NoStop}%
\bibitem [{\citenamefont {Zhao}\ \emph {et~al.}(2024)\citenamefont {Zhao}, \citenamefont {Ji}, \citenamefont {Li}, \citenamefont {Li}, \citenamefont {Zhang}, \citenamefont {Tian}, \citenamefont {Yu}, \citenamefont {Bian}, \citenamefont {Hao}, \citenamefont {Xiao} \emph {et~al.}}]{zhao2024_nature}%
  \BibitemOpen
  \bibfield  {author} {\bibinfo {author} {\bibfnamefont {J.}~\bibnamefont {Zhao}}, \bibinfo {author} {\bibfnamefont {P.}~\bibnamefont {Ji}}, \bibinfo {author} {\bibfnamefont {Y.}~\bibnamefont {Li}}, \bibinfo {author} {\bibfnamefont {R.}~\bibnamefont {Li}}, \bibinfo {author} {\bibfnamefont {K.}~\bibnamefont {Zhang}}, \bibinfo {author} {\bibfnamefont {H.}~\bibnamefont {Tian}}, \bibinfo {author} {\bibfnamefont {K.}~\bibnamefont {Yu}}, \bibinfo {author} {\bibfnamefont {B.}~\bibnamefont {Bian}}, \bibinfo {author} {\bibfnamefont {L.}~\bibnamefont {Hao}}, \bibinfo {author} {\bibfnamefont {X.}~\bibnamefont {Xiao}}, \emph {et~al.},\ }\href {https://doi.org/https://doi.org/10.1038/s41586-023-06811-0} {\bibfield  {journal} {\bibinfo  {journal} {Nature}\ }\textbf {\bibinfo {volume} {625}},\ \bibinfo {pages} {60} (\bibinfo {year} {2024})}\BibitemShut {NoStop}%
\bibitem [{\citenamefont {Kresse}\ and\ \citenamefont {Hafner}(1993)}]{kres1}%
  \BibitemOpen
  \bibfield  {author} {\bibinfo {author} {\bibfnamefont {G.}~\bibnamefont {Kresse}}\ and\ \bibinfo {author} {\bibfnamefont {J.}~\bibnamefont {Hafner}},\ }\href {https://doi.org/10.1103/PhysRevB.47.558} {\bibfield  {journal} {\bibinfo  {journal} {Phys. Rev. B}\ }\textbf {\bibinfo {volume} {47}},\ \bibinfo {pages} {558} (\bibinfo {year} {1993})}\BibitemShut {NoStop}%
\bibitem [{\citenamefont {Bl\"ochl}(1994)}]{PAW1994}%
  \BibitemOpen
  \bibfield  {author} {\bibinfo {author} {\bibfnamefont {P.~E.}\ \bibnamefont {Bl\"ochl}},\ }\href {https://doi.org/10.1103/PhysRevB.50.17953} {\bibfield  {journal} {\bibinfo  {journal} {Phys. Rev. B}\ }\textbf {\bibinfo {volume} {50}},\ \bibinfo {pages} {17953} (\bibinfo {year} {1994})}\BibitemShut {NoStop}%
\bibitem [{\citenamefont {Perdew}\ \emph {et~al.}(1996)\citenamefont {Perdew}, \citenamefont {Burke},\ and\ \citenamefont {Ernzerhof}}]{Perdew1996_PRL}%
  \BibitemOpen
  \bibfield  {author} {\bibinfo {author} {\bibfnamefont {J.~P.}\ \bibnamefont {Perdew}}, \bibinfo {author} {\bibfnamefont {K.}~\bibnamefont {Burke}},\ and\ \bibinfo {author} {\bibfnamefont {M.}~\bibnamefont {Ernzerhof}},\ }\href {https://doi.org/10.1103/PhysRevLett.77.3865} {\bibfield  {journal} {\bibinfo  {journal} {Phys. Rev. Lett.}\ }\textbf {\bibinfo {volume} {77}},\ \bibinfo {pages} {3865} (\bibinfo {year} {1996})}\BibitemShut {NoStop}%
\bibitem [{\citenamefont {Heyd}\ \emph {et~al.}(2003)\citenamefont {Heyd}, \citenamefont {Scuseria},\ and\ \citenamefont {Ernzerhof}}]{Heyd2003_JCP}%
  \BibitemOpen
  \bibfield  {author} {\bibinfo {author} {\bibfnamefont {J.}~\bibnamefont {Heyd}}, \bibinfo {author} {\bibfnamefont {G.~E.}\ \bibnamefont {Scuseria}},\ and\ \bibinfo {author} {\bibfnamefont {M.}~\bibnamefont {Ernzerhof}},\ }\href {https://doi.org/10.1063/1.1564060} {\bibfield  {journal} {\bibinfo  {journal} {The Journal of Chemical Physics}\ }\textbf {\bibinfo {volume} {118}},\ \bibinfo {pages} {8207} (\bibinfo {year} {2003})}\BibitemShut {NoStop}%
\bibitem [{\citenamefont {Heyd}\ and\ \citenamefont {Scuseria}(2004)}]{Heyd2004_JCP}%
  \BibitemOpen
  \bibfield  {author} {\bibinfo {author} {\bibfnamefont {J.}~\bibnamefont {Heyd}}\ and\ \bibinfo {author} {\bibfnamefont {G.~E.}\ \bibnamefont {Scuseria}},\ }\href {https://doi.org/10.1063/1.1760074} {\bibfield  {journal} {\bibinfo  {journal} {The Journal of Chemical Physics}\ }\textbf {\bibinfo {volume} {121}},\ \bibinfo {pages} {1187} (\bibinfo {year} {2004})}\BibitemShut {NoStop}%
\bibitem [{sup()}]{supplemental_material}%
  \BibitemOpen
  \href@noop {} {}\bibinfo {note} {See Supplemental Material at [URL will be inserted by publisher]. The Supplemental Material also contains Refs.~\cite{kres1,Pease1952AnXS,paszkowicz2002lattice,KORSAKS2015976,Polley2023_PRL,Chahal2022_JPCC,C7TC04300G,Chimene2015_AM}.}\BibitemShut {Stop}%
\bibitem [{\citenamefont {Grimme}\ \emph {et~al.}(2011)\citenamefont {Grimme}, \citenamefont {Ehrlich},\ and\ \citenamefont {Goerigk}}]{Grimme2011_JCC}%
  \BibitemOpen
  \bibfield  {author} {\bibinfo {author} {\bibfnamefont {S.}~\bibnamefont {Grimme}}, \bibinfo {author} {\bibfnamefont {S.}~\bibnamefont {Ehrlich}},\ and\ \bibinfo {author} {\bibfnamefont {L.}~\bibnamefont {Goerigk}},\ }\href {https://doi.org/https://doi.org/10.1002/jcc.21759} {\bibfield  {journal} {\bibinfo  {journal} {Journal of Computational Chemistry}\ }\textbf {\bibinfo {volume} {32}},\ \bibinfo {pages} {1456} (\bibinfo {year} {2011})}\BibitemShut {NoStop}%
\bibitem [{\citenamefont {Pease}(1952)}]{Pease1952AnXS}%
  \BibitemOpen
  \bibfield  {author} {\bibinfo {author} {\bibfnamefont {R.~S.}\ \bibnamefont {Pease}},\ }\href {https://api.semanticscholar.org/CorpusID:98194613} {\bibfield  {journal} {\bibinfo  {journal} {Acta Crystallographica}\ }\textbf {\bibinfo {volume} {5}},\ \bibinfo {pages} {356} (\bibinfo {year} {1952})}\BibitemShut {NoStop}%
\bibitem [{\citenamefont {Paszkowicz}\ \emph {et~al.}(2002)\citenamefont {Paszkowicz}, \citenamefont {Pelka}, \citenamefont {Knapp}, \citenamefont {Szyszko},\ and\ \citenamefont {Podsiadlo}}]{paszkowicz2002lattice}%
  \BibitemOpen
  \bibfield  {author} {\bibinfo {author} {\bibfnamefont {W.}~\bibnamefont {Paszkowicz}}, \bibinfo {author} {\bibfnamefont {J.}~\bibnamefont {Pelka}}, \bibinfo {author} {\bibfnamefont {M.}~\bibnamefont {Knapp}}, \bibinfo {author} {\bibfnamefont {T.}~\bibnamefont {Szyszko}},\ and\ \bibinfo {author} {\bibfnamefont {S.}~\bibnamefont {Podsiadlo}},\ }\href {https://doi.org/10.1007/s003390100999} {\bibfield  {journal} {\bibinfo  {journal} {Applied Physics A}\ }\textbf {\bibinfo {volume} {75}},\ \bibinfo {pages} {431} (\bibinfo {year} {2002})}\BibitemShut {NoStop}%
\bibitem [{\citenamefont {Korsaks}(2015)}]{KORSAKS2015976}%
  \BibitemOpen
  \bibfield  {author} {\bibinfo {author} {\bibfnamefont {V.}~\bibnamefont {Korsaks}},\ }\href {https://doi.org/https://doi.org/10.1016/j.materresbull.2015.06.032} {\bibfield  {journal} {\bibinfo  {journal} {Materials Research Bulletin}\ }\textbf {\bibinfo {volume} {70}},\ \bibinfo {pages} {976} (\bibinfo {year} {2015})}\BibitemShut {NoStop}%
\bibitem [{\citenamefont {Chahal}\ \emph {et~al.}(2022)\citenamefont {Chahal}, \citenamefont {Sahu}, \citenamefont {Kar}, \citenamefont {Ray}, \citenamefont {Biju},\ and\ \citenamefont {Kumar}}]{Chahal2022_JPCC}%
  \BibitemOpen
  \bibfield  {author} {\bibinfo {author} {\bibfnamefont {S.}~\bibnamefont {Chahal}}, \bibinfo {author} {\bibfnamefont {T.~K.}\ \bibnamefont {Sahu}}, \bibinfo {author} {\bibfnamefont {S.}~\bibnamefont {Kar}}, \bibinfo {author} {\bibfnamefont {S.~J.}\ \bibnamefont {Ray}}, \bibinfo {author} {\bibfnamefont {V.}~\bibnamefont {Biju}},\ and\ \bibinfo {author} {\bibfnamefont {P.}~\bibnamefont {Kumar}},\ }\href {https://doi.org/10.1021/acs.jpcc.2c06693} {\bibfield  {journal} {\bibinfo  {journal} {The Journal of Physical Chemistry C}\ }\textbf {\bibinfo {volume} {126}},\ \bibinfo {pages} {21084} (\bibinfo {year} {2022})}\BibitemShut {NoStop}%
\bibitem [{\citenamefont {Zhang}\ \emph {et~al.}(2017)\citenamefont {Zhang}, \citenamefont {Feng}, \citenamefont {Wang}, \citenamefont {Yang},\ and\ \citenamefont {Wang}}]{C7TC04300G}%
  \BibitemOpen
  \bibfield  {author} {\bibinfo {author} {\bibfnamefont {K.}~\bibnamefont {Zhang}}, \bibinfo {author} {\bibfnamefont {Y.}~\bibnamefont {Feng}}, \bibinfo {author} {\bibfnamefont {F.}~\bibnamefont {Wang}}, \bibinfo {author} {\bibfnamefont {Z.}~\bibnamefont {Yang}},\ and\ \bibinfo {author} {\bibfnamefont {J.}~\bibnamefont {Wang}},\ }\href {https://doi.org/10.1039/C7TC04300G} {\bibfield  {journal} {\bibinfo  {journal} {J. Mater. Chem. C}\ }\textbf {\bibinfo {volume} {5}},\ \bibinfo {pages} {11992} (\bibinfo {year} {2017})}\BibitemShut {NoStop}%
\bibitem [{\citenamefont {Chimene}\ \emph {et~al.}(2015)\citenamefont {Chimene}, \citenamefont {Alge},\ and\ \citenamefont {Gaharwar}}]{Chimene2015_AM}%
  \BibitemOpen
  \bibfield  {author} {\bibinfo {author} {\bibfnamefont {D.}~\bibnamefont {Chimene}}, \bibinfo {author} {\bibfnamefont {D.~L.}\ \bibnamefont {Alge}},\ and\ \bibinfo {author} {\bibfnamefont {A.~K.}\ \bibnamefont {Gaharwar}},\ }\href {https://doi.org/https://doi.org/10.1002/adma.201502422} {\bibfield  {journal} {\bibinfo  {journal} {Advanced Materials}\ }\textbf {\bibinfo {volume} {27}},\ \bibinfo {pages} {7261} (\bibinfo {year} {2015})}\BibitemShut {NoStop}%
\bibitem [{\citenamefont {Togo}\ and\ \citenamefont {Tanaka}(2015)}]{TOGO2015}%
  \BibitemOpen
  \bibfield  {author} {\bibinfo {author} {\bibfnamefont {A.}~\bibnamefont {Togo}}\ and\ \bibinfo {author} {\bibfnamefont {I.}~\bibnamefont {Tanaka}},\ }\href {https://doi.org/https://doi.org/10.1016/j.scriptamat.2015.07.021} {\bibfield  {journal} {\bibinfo  {journal} {Scripta Materialia}\ }\textbf {\bibinfo {volume} {108}},\ \bibinfo {pages} {1 } (\bibinfo {year} {2015})}\BibitemShut {NoStop}%
\bibitem [{\citenamefont {Fan}\ \emph {et~al.}(2017)\citenamefont {Fan}, \citenamefont {Chen}, \citenamefont {Vierimaa},\ and\ \citenamefont {Harju}}]{FAN201710}%
  \BibitemOpen
  \bibfield  {author} {\bibinfo {author} {\bibfnamefont {Z.}~\bibnamefont {Fan}}, \bibinfo {author} {\bibfnamefont {W.}~\bibnamefont {Chen}}, \bibinfo {author} {\bibfnamefont {V.}~\bibnamefont {Vierimaa}},\ and\ \bibinfo {author} {\bibfnamefont {A.}~\bibnamefont {Harju}},\ }\href {https://doi.org/https://doi.org/10.1016/j.cpc.2017.05.003} {\bibfield  {journal} {\bibinfo  {journal} {Computer Physics Communications}\ }\textbf {\bibinfo {volume} {218}},\ \bibinfo {pages} {10} (\bibinfo {year} {2017})}\BibitemShut {NoStop}%
\bibitem [{\citenamefont {Ying}\ and\ \citenamefont {Fan}(2023)}]{Ying_2024}%
  \BibitemOpen
  \bibfield  {author} {\bibinfo {author} {\bibfnamefont {P.}~\bibnamefont {Ying}}\ and\ \bibinfo {author} {\bibfnamefont {Z.}~\bibnamefont {Fan}},\ }\href {https://doi.org/10.1088/1361-648X/ad1278} {\bibfield  {journal} {\bibinfo  {journal} {Journal of Physics: Condensed Matter}\ }\textbf {\bibinfo {volume} {36}},\ \bibinfo {pages} {125901} (\bibinfo {year} {2023})}\BibitemShut {NoStop}%
\bibitem [{\citenamefont {Mouhat}\ and\ \citenamefont {Coudert}(2014)}]{Mouhat2014_PRB}%
  \BibitemOpen
  \bibfield  {author} {\bibinfo {author} {\bibfnamefont {F.}~\bibnamefont {Mouhat}}\ and\ \bibinfo {author} {\bibfnamefont {F.~m. c.-X.}\ \bibnamefont {Coudert}},\ }\href {https://doi.org/10.1103/PhysRevB.90.224104} {\bibfield  {journal} {\bibinfo  {journal} {Phys. Rev. B}\ }\textbf {\bibinfo {volume} {90}},\ \bibinfo {pages} {224104} (\bibinfo {year} {2014})}\BibitemShut {NoStop}%
\bibitem [{\citenamefont {Tang}\ \emph {et~al.}(2009)\citenamefont {Tang}, \citenamefont {Sanville},\ and\ \citenamefont {Henkelman}}]{Tang_2009}%
  \BibitemOpen
  \bibfield  {author} {\bibinfo {author} {\bibfnamefont {W.}~\bibnamefont {Tang}}, \bibinfo {author} {\bibfnamefont {E.}~\bibnamefont {Sanville}},\ and\ \bibinfo {author} {\bibfnamefont {G.}~\bibnamefont {Henkelman}},\ }\href {https://doi.org/10.1088/0953-8984/21/8/084204} {\bibfield  {journal} {\bibinfo  {journal} {Journal of Physics: Condensed Matter}\ }\textbf {\bibinfo {volume} {21}},\ \bibinfo {pages} {084204} (\bibinfo {year} {2009})}\BibitemShut {NoStop}%
\bibitem [{\citenamefont {Lin}\ \emph {et~al.}(2021)\citenamefont {Lin}, \citenamefont {Motoyama}, \citenamefont {Kretschmer}, \citenamefont {Ghaderzadeh}, \citenamefont {Ghorbani-Asl}, \citenamefont {Araki}, \citenamefont {Krasheninnikov}, \citenamefont {Ago},\ and\ \citenamefont {Suenaga}}]{Lin2021_AdMat}%
  \BibitemOpen
  \bibfield  {author} {\bibinfo {author} {\bibfnamefont {Y.-C.}\ \bibnamefont {Lin}}, \bibinfo {author} {\bibfnamefont {A.}~\bibnamefont {Motoyama}}, \bibinfo {author} {\bibfnamefont {S.}~\bibnamefont {Kretschmer}}, \bibinfo {author} {\bibfnamefont {S.}~\bibnamefont {Ghaderzadeh}}, \bibinfo {author} {\bibfnamefont {M.}~\bibnamefont {Ghorbani-Asl}}, \bibinfo {author} {\bibfnamefont {Y.}~\bibnamefont {Araki}}, \bibinfo {author} {\bibfnamefont {A.~V.}\ \bibnamefont {Krasheninnikov}}, \bibinfo {author} {\bibfnamefont {H.}~\bibnamefont {Ago}},\ and\ \bibinfo {author} {\bibfnamefont {K.}~\bibnamefont {Suenaga}},\ }\href {https://doi.org/https://doi.org/10.1002/adma.202105898} {\bibfield  {journal} {\bibinfo  {journal} {Advanced Materials}\ }\textbf {\bibinfo {volume} {33}},\ \bibinfo {pages} {2105898} (\bibinfo {year} {2021})}\BibitemShut {NoStop}%
\bibitem [{\citenamefont {Back}\ and\ \citenamefont {Siahrostami}(2019)}]{Back2019_nanoadv}%
  \BibitemOpen
  \bibfield  {author} {\bibinfo {author} {\bibfnamefont {S.}~\bibnamefont {Back}}\ and\ \bibinfo {author} {\bibfnamefont {S.}~\bibnamefont {Siahrostami}},\ }\href {https://doi.org/10.1039/C8NA00059J} {\bibfield  {journal} {\bibinfo  {journal} {Nanoscale Adv.}\ }\textbf {\bibinfo {volume} {1}},\ \bibinfo {pages} {132} (\bibinfo {year} {2019})}\BibitemShut {NoStop}%
\bibitem [{\citenamefont {de~Oliveira}\ and\ \citenamefont {Miwa}(2015)}]{deOliveria2015jcp}%
  \BibitemOpen
  \bibfield  {author} {\bibinfo {author} {\bibfnamefont {I.~S.~S.}\ \bibnamefont {de~Oliveira}}\ and\ \bibinfo {author} {\bibfnamefont {R.~H.}\ \bibnamefont {Miwa}},\ }\href {https://doi.org/10.1063/1.4906435} {\bibfield  {journal} {\bibinfo  {journal} {The Journal of Chemical Physics}\ }\textbf {\bibinfo {volume} {142}},\ \bibinfo {pages} {044301} (\bibinfo {year} {2015})}\BibitemShut {NoStop}%
\bibitem [{\citenamefont {Vlassiouk}\ \emph {et~al.}(2025)\citenamefont {Vlassiouk}, \citenamefont {Wu}, \citenamefont {Puretzky}, \citenamefont {Liang}, \citenamefont {Lasseter}, \citenamefont {Dryzhakov}, \citenamefont {Gallagher}, \citenamefont {Ghosh}, \citenamefont {Lavrik}, \citenamefont {Dyck}, \citenamefont {Lupini}, \citenamefont {Checa}, \citenamefont {Collins}, \citenamefont {Zhao}, \citenamefont {Likhi}, \citenamefont {Xiao}, \citenamefont {Ivanov}, \citenamefont {Glasgow}, \citenamefont {Tselev}, \citenamefont {Lawrie}, \citenamefont {Smirnov},\ and\ \citenamefont {Randolph}}]{Vlassiouk2025_arXiv}%
  \BibitemOpen
  \bibfield  {author} {\bibinfo {author} {\bibfnamefont {I.~V.}\ \bibnamefont {Vlassiouk}}, \bibinfo {author} {\bibfnamefont {Y.-C.}\ \bibnamefont {Wu}}, \bibinfo {author} {\bibfnamefont {A.}~\bibnamefont {Puretzky}}, \bibinfo {author} {\bibfnamefont {L.}~\bibnamefont {Liang}}, \bibinfo {author} {\bibfnamefont {J.}~\bibnamefont {Lasseter}}, \bibinfo {author} {\bibfnamefont {B.}~\bibnamefont {Dryzhakov}}, \bibinfo {author} {\bibfnamefont {I.}~\bibnamefont {Gallagher}}, \bibinfo {author} {\bibfnamefont {S.}~\bibnamefont {Ghosh}}, \bibinfo {author} {\bibfnamefont {N.}~\bibnamefont {Lavrik}}, \bibinfo {author} {\bibfnamefont {O.}~\bibnamefont {Dyck}}, \bibinfo {author} {\bibfnamefont {A.~R.}\ \bibnamefont {Lupini}}, \bibinfo {author} {\bibfnamefont {M.}~\bibnamefont {Checa}}, \bibinfo {author} {\bibfnamefont {L.}~\bibnamefont {Collins}}, \bibinfo {author} {\bibfnamefont {H.}~\bibnamefont {Zhao}}, \bibinfo {author} {\bibfnamefont {F.}~\bibnamefont {Likhi}}, \bibinfo {author} {\bibfnamefont {K.}~\bibnamefont
  {Xiao}}, \bibinfo {author} {\bibfnamefont {I.}~\bibnamefont {Ivanov}}, \bibinfo {author} {\bibfnamefont {D.}~\bibnamefont {Glasgow}}, \bibinfo {author} {\bibfnamefont {A.}~\bibnamefont {Tselev}}, \bibinfo {author} {\bibfnamefont {B.}~\bibnamefont {Lawrie}}, \bibinfo {author} {\bibfnamefont {S.}~\bibnamefont {Smirnov}},\ and\ \bibinfo {author} {\bibfnamefont {S.}~\bibnamefont {Randolph}},\ }\href {https://arxiv.org/abs/2503.18894} {\bibinfo {title} {Defect engineering in large-scale cvd-grown hexagonal boron nitride: Formation, spectroscopy, and spin relaxation dynamics}} (\bibinfo {year} {2025}),\ \Eprint {https://arxiv.org/abs/2503.18894} {arXiv:2503.18894 [cond-mat.mtrl-sci]} \BibitemShut {NoStop}%
\bibitem [{\citenamefont {Thiering}\ and\ \citenamefont {Gali}(2016)}]{Thiering_PRB2016}%
  \BibitemOpen
  \bibfield  {author} {\bibinfo {author} {\bibfnamefont {G.~m.~H.}\ \bibnamefont {Thiering}}\ and\ \bibinfo {author} {\bibfnamefont {A.}~\bibnamefont {Gali}},\ }\href {https://doi.org/10.1103/PhysRevB.94.125202} {\bibfield  {journal} {\bibinfo  {journal} {Phys. Rev. B}\ }\textbf {\bibinfo {volume} {94}},\ \bibinfo {pages} {125202} (\bibinfo {year} {2016})}\BibitemShut {NoStop}%
\bibitem [{\citenamefont {Rasool}\ \emph {et~al.}(2015)\citenamefont {Rasool}, \citenamefont {Ophus},\ and\ \citenamefont {Zettl}}]{Rasool2015_AM}%
  \BibitemOpen
  \bibfield  {author} {\bibinfo {author} {\bibfnamefont {H.~I.}\ \bibnamefont {Rasool}}, \bibinfo {author} {\bibfnamefont {C.}~\bibnamefont {Ophus}},\ and\ \bibinfo {author} {\bibfnamefont {A.}~\bibnamefont {Zettl}},\ }\href {https://doi.org/https://doi.org/10.1002/adma.201500231} {\bibfield  {journal} {\bibinfo  {journal} {Advanced Materials}\ }\textbf {\bibinfo {volume} {27}},\ \bibinfo {pages} {5771} (\bibinfo {year} {2015})}\BibitemShut {NoStop}%
\bibitem [{\citenamefont {Linder\"alv}\ \emph {et~al.}(2021)\citenamefont {Linder\"alv}, \citenamefont {Wieczorek},\ and\ \citenamefont {Erhart}}]{Linderalv2021_PRB}%
  \BibitemOpen
  \bibfield  {author} {\bibinfo {author} {\bibfnamefont {C.}~\bibnamefont {Linder\"alv}}, \bibinfo {author} {\bibfnamefont {W.}~\bibnamefont {Wieczorek}},\ and\ \bibinfo {author} {\bibfnamefont {P.}~\bibnamefont {Erhart}},\ }\href {https://doi.org/10.1103/PhysRevB.103.115421} {\bibfield  {journal} {\bibinfo  {journal} {Phys. Rev. B}\ }\textbf {\bibinfo {volume} {103}},\ \bibinfo {pages} {115421} (\bibinfo {year} {2021})}\BibitemShut {NoStop}%
\bibitem [{\citenamefont {Weston}\ \emph {et~al.}(2018)\citenamefont {Weston}, \citenamefont {Wickramaratne}, \citenamefont {Mackoit}, \citenamefont {Alkauskas},\ and\ \citenamefont {Van~de Walle}}]{Weston2018_aps}%
  \BibitemOpen
  \bibfield  {author} {\bibinfo {author} {\bibfnamefont {L.}~\bibnamefont {Weston}}, \bibinfo {author} {\bibfnamefont {D.}~\bibnamefont {Wickramaratne}}, \bibinfo {author} {\bibfnamefont {M.}~\bibnamefont {Mackoit}}, \bibinfo {author} {\bibfnamefont {A.}~\bibnamefont {Alkauskas}},\ and\ \bibinfo {author} {\bibfnamefont {C.~G.}\ \bibnamefont {Van~de Walle}},\ }\href {https://doi.org/10.1103/PhysRevB.97.214104} {\bibfield  {journal} {\bibinfo  {journal} {Phys. Rev. B}\ }\textbf {\bibinfo {volume} {97}},\ \bibinfo {pages} {214104} (\bibinfo {year} {2018})}\BibitemShut {NoStop}%
\bibitem [{\citenamefont {Alem}\ \emph {et~al.}(2011)\citenamefont {Alem}, \citenamefont {Yazyev}, \citenamefont {Kisielowski}, \citenamefont {Denes}, \citenamefont {Dahmen}, \citenamefont {Hartel}, \citenamefont {Haider}, \citenamefont {Bischoff}, \citenamefont {Jiang}, \citenamefont {Louie},\ and\ \citenamefont {Zettl}}]{PhysRevLett.106.126102}%
  \BibitemOpen
  \bibfield  {author} {\bibinfo {author} {\bibfnamefont {N.}~\bibnamefont {Alem}}, \bibinfo {author} {\bibfnamefont {O.~V.}\ \bibnamefont {Yazyev}}, \bibinfo {author} {\bibfnamefont {C.}~\bibnamefont {Kisielowski}}, \bibinfo {author} {\bibfnamefont {P.}~\bibnamefont {Denes}}, \bibinfo {author} {\bibfnamefont {U.}~\bibnamefont {Dahmen}}, \bibinfo {author} {\bibfnamefont {P.}~\bibnamefont {Hartel}}, \bibinfo {author} {\bibfnamefont {M.}~\bibnamefont {Haider}}, \bibinfo {author} {\bibfnamefont {M.}~\bibnamefont {Bischoff}}, \bibinfo {author} {\bibfnamefont {B.}~\bibnamefont {Jiang}}, \bibinfo {author} {\bibfnamefont {S.~G.}\ \bibnamefont {Louie}},\ and\ \bibinfo {author} {\bibfnamefont {A.}~\bibnamefont {Zettl}},\ }\href {https://doi.org/10.1103/PhysRevLett.106.126102} {\bibfield  {journal} {\bibinfo  {journal} {Phys. Rev. Lett.}\ }\textbf {\bibinfo {volume} {106}},\ \bibinfo {pages} {126102} (\bibinfo {year} {2011})}\BibitemShut {NoStop}%
\bibitem [{\citenamefont {Zhou}\ \emph {et~al.}(2023)\citenamefont {Zhou}, \citenamefont {Jiang}, \citenamefont {Ji}, \citenamefont {Lang}, \citenamefont {Fang},\ and\ \citenamefont {Wu}}]{ZhouChemCatChem2023}%
  \BibitemOpen
  \bibfield  {author} {\bibinfo {author} {\bibfnamefont {Y.}~\bibnamefont {Zhou}}, \bibinfo {author} {\bibfnamefont {Y.}~\bibnamefont {Jiang}}, \bibinfo {author} {\bibfnamefont {Y.}~\bibnamefont {Ji}}, \bibinfo {author} {\bibfnamefont {R.}~\bibnamefont {Lang}}, \bibinfo {author} {\bibfnamefont {Y.}~\bibnamefont {Fang}},\ and\ \bibinfo {author} {\bibfnamefont {C.-D.}\ \bibnamefont {Wu}},\ }\href {https://doi.org/https://doi.org/10.1002/cctc.202201176} {\bibfield  {journal} {\bibinfo  {journal} {ChemCatChem}\ }\textbf {\bibinfo {volume} {15}},\ \bibinfo {pages} {e202201176} (\bibinfo {year} {2023})}\BibitemShut {NoStop}%
\bibitem [{\citenamefont {Liu}\ \emph {et~al.}(2022)\citenamefont {Liu}, \citenamefont {Wan}, \citenamefont {Kong}, \citenamefont {Han},\ and\ \citenamefont {Xiong}}]{D1TA08252C_rsc}%
  \BibitemOpen
  \bibfield  {author} {\bibinfo {author} {\bibfnamefont {D.}~\bibnamefont {Liu}}, \bibinfo {author} {\bibfnamefont {X.}~\bibnamefont {Wan}}, \bibinfo {author} {\bibfnamefont {T.}~\bibnamefont {Kong}}, \bibinfo {author} {\bibfnamefont {W.}~\bibnamefont {Han}},\ and\ \bibinfo {author} {\bibfnamefont {Y.}~\bibnamefont {Xiong}},\ }\href {https://doi.org/10.1039/D1TA08252C} {\bibfield  {journal} {\bibinfo  {journal} {J. Mater. Chem. A}\ }\textbf {\bibinfo {volume} {10}},\ \bibinfo {pages} {5878} (\bibinfo {year} {2022})}\BibitemShut {NoStop}%
\bibitem [{\citenamefont {Weon}\ \emph {et~al.}(2021)\citenamefont {Weon}, \citenamefont {Huang}, \citenamefont {Rigby}, \citenamefont {Chu}, \citenamefont {Wu},\ and\ \citenamefont {Kim}}]{Weon2021_acs}%
  \BibitemOpen
  \bibfield  {author} {\bibinfo {author} {\bibfnamefont {S.}~\bibnamefont {Weon}}, \bibinfo {author} {\bibfnamefont {D.}~\bibnamefont {Huang}}, \bibinfo {author} {\bibfnamefont {K.}~\bibnamefont {Rigby}}, \bibinfo {author} {\bibfnamefont {C.}~\bibnamefont {Chu}}, \bibinfo {author} {\bibfnamefont {X.}~\bibnamefont {Wu}},\ and\ \bibinfo {author} {\bibfnamefont {J.-H.}\ \bibnamefont {Kim}},\ }\href {https://doi.org/10.1021/acsestengg.0c00136} {\bibfield  {journal} {\bibinfo  {journal} {ACS ES\&T Engineering}\ }\textbf {\bibinfo {volume} {1}},\ \bibinfo {pages} {157} (\bibinfo {year} {2021})}\BibitemShut {NoStop}%
\bibitem [{\citenamefont {Wu}\ \emph {et~al.}(2023)\citenamefont {Wu}, \citenamefont {Shi}, \citenamefont {Li},\ and\ \citenamefont {Guo}}]{WU2023100923}%
  \BibitemOpen
  \bibfield  {author} {\bibinfo {author} {\bibfnamefont {J.}~\bibnamefont {Wu}}, \bibinfo {author} {\bibfnamefont {H.}~\bibnamefont {Shi}}, \bibinfo {author} {\bibfnamefont {K.}~\bibnamefont {Li}},\ and\ \bibinfo {author} {\bibfnamefont {X.}~\bibnamefont {Guo}},\ }\href {https://doi.org/https://doi.org/10.1016/j.coche.2023.100923} {\bibfield  {journal} {\bibinfo  {journal} {Current Opinion in Chemical Engineering}\ }\textbf {\bibinfo {volume} {40}},\ \bibinfo {pages} {100923} (\bibinfo {year} {2023})}\BibitemShut {NoStop}%
\bibitem [{\citenamefont {Dong}\ \emph {et~al.}(2021)\citenamefont {Dong}, \citenamefont {Gao},\ and\ \citenamefont {Fu}}]{Dong_2021_jpclett}%
  \BibitemOpen
  \bibfield  {author} {\bibinfo {author} {\bibfnamefont {J.}~\bibnamefont {Dong}}, \bibinfo {author} {\bibfnamefont {L.}~\bibnamefont {Gao}},\ and\ \bibinfo {author} {\bibfnamefont {Q.}~\bibnamefont {Fu}},\ }\href {https://doi.org/10.1021/acs.jpclett.1c02626} {\bibfield  {journal} {\bibinfo  {journal} {The Journal of Physical Chemistry Letters}\ }\textbf {\bibinfo {volume} {12}},\ \bibinfo {pages} {9608} (\bibinfo {year} {2021})},\ \bibinfo {note} {pMID: 34585925}\BibitemShut {NoStop}%
\bibitem [{\citenamefont {Jeong}\ \emph {et~al.}(2020)\citenamefont {Jeong}, \citenamefont {Shin},\ and\ \citenamefont {Lee}}]{Jeongacsnano_2020}%
  \BibitemOpen
  \bibfield  {author} {\bibinfo {author} {\bibfnamefont {H.}~\bibnamefont {Jeong}}, \bibinfo {author} {\bibfnamefont {S.}~\bibnamefont {Shin}},\ and\ \bibinfo {author} {\bibfnamefont {H.}~\bibnamefont {Lee}},\ }\href {https://doi.org/10.1021/acsnano.0c06610} {\bibfield  {journal} {\bibinfo  {journal} {ACS Nano}\ }\textbf {\bibinfo {volume} {14}},\ \bibinfo {pages} {14355} (\bibinfo {year} {2020})},\ \bibinfo {note} {pMID: 33140947}\BibitemShut {NoStop}%
\bibitem [{\citenamefont {Li}\ \emph {et~al.}(2023{\natexlab{b}})\citenamefont {Li}, \citenamefont {Chen}, \citenamefont {Xu}, \citenamefont {Zhang}, \citenamefont {Wei}, \citenamefont {Zhao}, \citenamefont {Wu},\ and\ \citenamefont {Chen}}]{Li_JACSAu_2023}%
  \BibitemOpen
  \bibfield  {author} {\bibinfo {author} {\bibfnamefont {J.}~\bibnamefont {Li}}, \bibinfo {author} {\bibfnamefont {C.}~\bibnamefont {Chen}}, \bibinfo {author} {\bibfnamefont {L.}~\bibnamefont {Xu}}, \bibinfo {author} {\bibfnamefont {Y.}~\bibnamefont {Zhang}}, \bibinfo {author} {\bibfnamefont {W.}~\bibnamefont {Wei}}, \bibinfo {author} {\bibfnamefont {E.}~\bibnamefont {Zhao}}, \bibinfo {author} {\bibfnamefont {Y.}~\bibnamefont {Wu}},\ and\ \bibinfo {author} {\bibfnamefont {C.}~\bibnamefont {Chen}},\ }\href {https://doi.org/10.1021/jacsau.3c00001} {\bibfield  {journal} {\bibinfo  {journal} {JACS Au}\ }\textbf {\bibinfo {volume} {3}},\ \bibinfo {pages} {736} (\bibinfo {year} {2023}{\natexlab{b}})}\BibitemShut {NoStop}%
\bibitem [{\citenamefont {Lin}\ \emph {et~al.}(2022)\citenamefont {Lin}, \citenamefont {Franke}, \citenamefont {Parhizkar}, \citenamefont {Raths}, \citenamefont {Wen-zhe Yu}, \citenamefont {Lee}, \citenamefont {Soubatch}, \citenamefont {Blum}, \citenamefont {Tautz}, \citenamefont {Kumpf},\ and\ \citenamefont {Bocquet}}]{Lin_PRM_2022}%
  \BibitemOpen
  \bibfield  {author} {\bibinfo {author} {\bibfnamefont {Y.-R.}\ \bibnamefont {Lin}}, \bibinfo {author} {\bibfnamefont {M.}~\bibnamefont {Franke}}, \bibinfo {author} {\bibfnamefont {S.}~\bibnamefont {Parhizkar}}, \bibinfo {author} {\bibfnamefont {M.}~\bibnamefont {Raths}}, \bibinfo {author} {\bibfnamefont {V.}~\bibnamefont {Wen-zhe Yu}}, \bibinfo {author} {\bibfnamefont {T.-L.}\ \bibnamefont {Lee}}, \bibinfo {author} {\bibfnamefont {S.}~\bibnamefont {Soubatch}}, \bibinfo {author} {\bibfnamefont {V.}~\bibnamefont {Blum}}, \bibinfo {author} {\bibfnamefont {F.~S.}\ \bibnamefont {Tautz}}, \bibinfo {author} {\bibfnamefont {C.}~\bibnamefont {Kumpf}},\ and\ \bibinfo {author} {\bibfnamefont {F.~m. c.~C.}\ \bibnamefont {Bocquet}},\ }\href {https://doi.org/10.1103/PhysRevMaterials.6.064002} {\bibfield  {journal} {\bibinfo  {journal} {Phys. Rev. Mater.}\ }\textbf {\bibinfo {volume} {6}},\ \bibinfo {pages} {064002} (\bibinfo {year} {2022})}\BibitemShut {NoStop}%
\bibitem [{\citenamefont {Biswas}\ \emph {et~al.}(2023)\citenamefont {Biswas}, \citenamefont {Xu}, \citenamefont {Alvarez}, \citenamefont {Zhang}, \citenamefont {Christiansen-Salameh}, \citenamefont {Puthirath}, \citenamefont {Burns}, \citenamefont {Hachtel}, \citenamefont {Li}, \citenamefont {Iyengar}, \citenamefont {Gray}, \citenamefont {Li}, \citenamefont {Zhang}, \citenamefont {Kannan}, \citenamefont {Elkins}, \citenamefont {Pieshkov}, \citenamefont {Vajtai}, \citenamefont {Birdwell}, \citenamefont {Neupane}, \citenamefont {Garratt}, \citenamefont {Ivanov}, \citenamefont {Pate}, \citenamefont {Zhao}, \citenamefont {Zhu}, \citenamefont {Tian}, \citenamefont {Rubio},\ and\ \citenamefont {Ajayan}}]{Biswas_AdvMat_2023}%
  \BibitemOpen
  \bibfield  {author} {\bibinfo {author} {\bibfnamefont {A.}~\bibnamefont {Biswas}}, \bibinfo {author} {\bibfnamefont {R.}~\bibnamefont {Xu}}, \bibinfo {author} {\bibfnamefont {G.~A.}\ \bibnamefont {Alvarez}}, \bibinfo {author} {\bibfnamefont {J.}~\bibnamefont {Zhang}}, \bibinfo {author} {\bibfnamefont {J.}~\bibnamefont {Christiansen-Salameh}}, \bibinfo {author} {\bibfnamefont {A.~B.}\ \bibnamefont {Puthirath}}, \bibinfo {author} {\bibfnamefont {K.}~\bibnamefont {Burns}}, \bibinfo {author} {\bibfnamefont {J.~A.}\ \bibnamefont {Hachtel}}, \bibinfo {author} {\bibfnamefont {T.}~\bibnamefont {Li}}, \bibinfo {author} {\bibfnamefont {S.~A.}\ \bibnamefont {Iyengar}}, \bibinfo {author} {\bibfnamefont {T.}~\bibnamefont {Gray}}, \bibinfo {author} {\bibfnamefont {C.}~\bibnamefont {Li}}, \bibinfo {author} {\bibfnamefont {X.}~\bibnamefont {Zhang}}, \bibinfo {author} {\bibfnamefont {H.}~\bibnamefont {Kannan}}, \bibinfo {author} {\bibfnamefont {J.}~\bibnamefont {Elkins}}, \bibinfo {author} {\bibfnamefont {T.~S.}\
  \bibnamefont {Pieshkov}}, \bibinfo {author} {\bibfnamefont {R.}~\bibnamefont {Vajtai}}, \bibinfo {author} {\bibfnamefont {A.~G.}\ \bibnamefont {Birdwell}}, \bibinfo {author} {\bibfnamefont {M.~R.}\ \bibnamefont {Neupane}}, \bibinfo {author} {\bibfnamefont {E.~J.}\ \bibnamefont {Garratt}}, \bibinfo {author} {\bibfnamefont {T.~G.}\ \bibnamefont {Ivanov}}, \bibinfo {author} {\bibfnamefont {B.~B.}\ \bibnamefont {Pate}}, \bibinfo {author} {\bibfnamefont {Y.}~\bibnamefont {Zhao}}, \bibinfo {author} {\bibfnamefont {H.}~\bibnamefont {Zhu}}, \bibinfo {author} {\bibfnamefont {Z.}~\bibnamefont {Tian}}, \bibinfo {author} {\bibfnamefont {A.}~\bibnamefont {Rubio}},\ and\ \bibinfo {author} {\bibfnamefont {P.~M.}\ \bibnamefont {Ajayan}},\ }\href {https://doi.org/https://doi.org/10.1002/adma.202304624} {\bibfield  {journal} {\bibinfo  {journal} {Advanced Materials}\ }\textbf {\bibinfo {volume} {35}},\ \bibinfo {pages} {2304624} (\bibinfo {year} {2023})}\BibitemShut {NoStop}%
\end{thebibliography}%


\begin{thebibliography}{6}%
\makeatletter
\providecommand \@ifxundefined [1]{%
 \@ifx{#1\undefined}
}%
\providecommand \@ifnum [1]{%
 \ifnum #1\expandafter \@firstoftwo
 \else \expandafter \@secondoftwo
 \fi
}%
\providecommand \@ifx [1]{%
 \ifx #1\expandafter \@firstoftwo
 \else \expandafter \@secondoftwo
 \fi
}%
\providecommand \natexlab [1]{#1}%
\providecommand \enquote  [1]{``#1''}%
\providecommand \bibnamefont  [1]{#1}%
\providecommand \bibfnamefont [1]{#1}%
\providecommand \citenamefont [1]{#1}%
\providecommand \href@noop [0]{\@secondoftwo}%
\providecommand \href [0]{\begingroup \@sanitize@url \@href}%
\providecommand \@href[1]{\@@startlink{#1}\@@href}%
\providecommand \@@href[1]{\endgroup#1\@@endlink}%
\providecommand \@sanitize@url [0]{\catcode `\\12\catcode `\$12\catcode `\&12\catcode `\#12\catcode `\^12\catcode `\_12\catcode `\%12\relax}%
\providecommand \@@startlink[1]{}%
\providecommand \@@endlink[0]{}%
\providecommand \url  [0]{\begingroup\@sanitize@url \@url }%
\providecommand \@url [1]{\endgroup\@href {#1}{\urlprefix }}%
\providecommand \urlprefix  [0]{URL }%
\providecommand \Eprint [0]{\href }%
\providecommand \doibase [0]{https://doi.org/}%
\providecommand \selectlanguage [0]{\@gobble}%
\providecommand \bibinfo  [0]{\@secondoftwo}%
\providecommand \bibfield  [0]{\@secondoftwo}%
\providecommand \translation [1]{[#1]}%
\providecommand \BibitemOpen [0]{}%
\providecommand \bibitemStop [0]{}%
\providecommand \bibitemNoStop [0]{.\EOS\space}%
\providecommand \EOS [0]{\spacefactor3000\relax}%
\providecommand \BibitemShut  [1]{\csname bibitem#1\endcsname}%
\let\auto@bib@innerbib\@empty
%</preamble>
\bibitem [{\citenamefont {Kresse}\ and\ \citenamefont {Hafner}(1993)}]{kres1}%
  \BibitemOpen
  \bibfield  {author} {\bibinfo {author} {\bibfnamefont {G.}~\bibnamefont {Kresse}}\ and\ \bibinfo {author} {\bibfnamefont {J.}~\bibnamefont {Hafner}},\ }\bibfield  {title} {\bibinfo {title} {{Ab Initio Molecular Dynamics For Liquid Metals}},\ }\href {https://doi.org/10.1103/PhysRevB.47.558} {\bibfield  {journal} {\bibinfo  {journal} {Phys. Rev. B}\ }\textbf {\bibinfo {volume} {47}},\ \bibinfo {pages} {558} (\bibinfo {year} {1993})}\BibitemShut {NoStop}%
\bibitem [{\citenamefont {Korsaks}(2015)}]{KORSAKS2015976}%
  \BibitemOpen
  \bibfield  {author} {\bibinfo {author} {\bibfnamefont {V.}~\bibnamefont {Korsaks}},\ }\bibfield  {title} {\bibinfo {title} {Hexagonal boron nitride luminescence dependent on vacuum level and surrounding gases},\ }\href {https://doi.org/https://doi.org/10.1016/j.materresbull.2015.06.032} {\bibfield  {journal} {\bibinfo  {journal} {Materials Research Bulletin}\ }\textbf {\bibinfo {volume} {70}},\ \bibinfo {pages} {976} (\bibinfo {year} {2015})}\BibitemShut {NoStop}%
\bibitem [{\citenamefont {Pease}(1952)}]{Pease1952AnXS}%
  \BibitemOpen
  \bibfield  {author} {\bibinfo {author} {\bibfnamefont {R.~S.}\ \bibnamefont {Pease}},\ }\bibfield  {title} {\bibinfo {title} {An x‐ray study of boron nitride},\ }\href {https://api.semanticscholar.org/CorpusID:98194613} {\bibfield  {journal} {\bibinfo  {journal} {Acta Crystallographica}\ }\textbf {\bibinfo {volume} {5}},\ \bibinfo {pages} {356} (\bibinfo {year} {1952})}\BibitemShut {NoStop}%
\bibitem [{\citenamefont {Paszkowicz}\ \emph {et~al.}(2002)\citenamefont {Paszkowicz}, \citenamefont {Pelka}, \citenamefont {Knapp}, \citenamefont {Szyszko},\ and\ \citenamefont {Podsiadlo}}]{paszkowicz2002lattice}%
  \BibitemOpen
  \bibfield  {author} {\bibinfo {author} {\bibfnamefont {W.}~\bibnamefont {Paszkowicz}}, \bibinfo {author} {\bibfnamefont {J.}~\bibnamefont {Pelka}}, \bibinfo {author} {\bibfnamefont {M.}~\bibnamefont {Knapp}}, \bibinfo {author} {\bibfnamefont {T.}~\bibnamefont {Szyszko}},\ and\ \bibinfo {author} {\bibfnamefont {S.}~\bibnamefont {Podsiadlo}},\ }\bibfield  {title} {\bibinfo {title} {Lattice parameters and anisotropic thermal expansion of hexagonal boron nitride in the 10--297.5 k temperature range},\ }\href {https://doi.org/10.1007/s003390100999} {\bibfield  {journal} {\bibinfo  {journal} {Applied Physics A}\ }\textbf {\bibinfo {volume} {75}},\ \bibinfo {pages} {431} (\bibinfo {year} {2002})}\BibitemShut {NoStop}%
\bibitem [{\citenamefont {Polley}\ \emph {et~al.}(2023)\citenamefont {Polley}, \citenamefont {Fedderwitz}, \citenamefont {Balasubramanian}, \citenamefont {Zakharov}, \citenamefont {Yakimova}, \citenamefont {B\"acke}, \citenamefont {Ekman}, \citenamefont {Dash}, \citenamefont {Kubatkin},\ and\ \citenamefont {Lara-Avila}}]{Polley2023_PRL}%
  \BibitemOpen
  \bibfield  {author} {\bibinfo {author} {\bibfnamefont {C.~M.}\ \bibnamefont {Polley}}, \bibinfo {author} {\bibfnamefont {H.}~\bibnamefont {Fedderwitz}}, \bibinfo {author} {\bibfnamefont {T.}~\bibnamefont {Balasubramanian}}, \bibinfo {author} {\bibfnamefont {A.~A.}\ \bibnamefont {Zakharov}}, \bibinfo {author} {\bibfnamefont {R.}~\bibnamefont {Yakimova}}, \bibinfo {author} {\bibfnamefont {O.}~\bibnamefont {B\"acke}}, \bibinfo {author} {\bibfnamefont {J.}~\bibnamefont {Ekman}}, \bibinfo {author} {\bibfnamefont {S.~P.}\ \bibnamefont {Dash}}, \bibinfo {author} {\bibfnamefont {S.}~\bibnamefont {Kubatkin}},\ and\ \bibinfo {author} {\bibfnamefont {S.}~\bibnamefont {Lara-Avila}},\ }\bibfield  {title} {\bibinfo {title} {Bottom-up growth of monolayer honeycomb sic},\ }\href {https://doi.org/10.1103/PhysRevLett.130.076203} {\bibfield  {journal} {\bibinfo  {journal} {Phys. Rev. Lett.}\ }\textbf {\bibinfo {volume} {130}},\ \bibinfo {pages} {076203} (\bibinfo {year} {2023})}\BibitemShut {NoStop}%
\bibitem [{\citenamefont {Freysoldt}\ and\ \citenamefont {Neugebauer}(2018)}]{Freysoldt2018}%
  \BibitemOpen
  \bibfield  {author} {\bibinfo {author} {\bibfnamefont {C.}~\bibnamefont {Freysoldt}}\ and\ \bibinfo {author} {\bibfnamefont {J.}~\bibnamefont {Neugebauer}},\ }\bibfield  {title} {\bibinfo {title} {First-principles calculations for charged defects at surfaces, interfaces, and two-dimensional materials in the presence of electric fields},\ }\href {https://doi.org/10.1103/PhysRevB.97.205425} {\bibfield  {journal} {\bibinfo  {journal} {Phys. Rev. B}\ }\textbf {\bibinfo {volume} {97}},\ \bibinfo {pages} {205425} (\bibinfo {year} {2018})}\BibitemShut {NoStop}%
\end{thebibliography}%

\end{document}

%\title{Insights into Interlayer Bonding in hBN/SiC Heterostructures:{\texorpdfstring{\\}{}} Toward Transition-Metal Single-Atom Isolation (Supplemental Material)}

\title{Stabilisation of hBN/SiC Heterostructures with Vacancies and Transition-Metal Atoms (Supplemental Material)}

\author{Arsalan Hashemi}
\email{arsalan.hashemi@aalto.fi}
\affiliation{MSP Group, Quantum Technology Finland Center of Excellence, Department of Applied Physics, Aalto University, FI-00076 Espoo, Finland}
\author{Nima Ghafari Cherati}
\affiliation{HUN-REN Wigner Research Centre for Physics, PO Box 49, H-1525 Budapest, Hungary}
\affiliation{Department of Atomic Physics, Institute of Physics, Budapest University of Technology and Economics, M\H{u}egyetem rkp. 3, H-1111 Budapest, Hungary}
\author{Sadegh Ghaderzadeh}
\affiliation{School of Chemistry, University of Nottingham, Nottingham NG7 2RD, UK}
\author{Yanzhou Wang}
\affiliation{MSP Group, Quantum Technology Finland Center of Excellence, Department of Applied Physics, Aalto University, FI-00076 Espoo, Finland}
\author{Mahdi Ghorbani-Asl}
\affiliation{Institute of Ion Beam Physics and Materials Research, Helmholtz-Zentrum Dresden-Rossendorf, 01328 Dresden, Germany}
\author{Tapio Ala-Nissila}
\email{tapio.ala-nissila@aalto.fi}
\affiliation{MSP Group, Quantum Technology Finland Center of Excellence, Department of Applied Physics, Aalto University, FI-00076 Espoo, Finland}
\affiliation{Interdisciplinary Centre for Mathematical Modelling and Department of Mathematical Sciences, Loughborough University, Loughborough, Leicestershire LE11 3TU, United Kingdom}

%\date{\today}

\begin{abstract}
See the Supplemental Material for: validation of van der Waals (vdW) methods; machine-learning datasets; fundamental properties of hBN, SiC, and the hBN/SiC interface; molecular-dynamics trajectory analyses; transition-metal adsorption energy profiles; and the influence of bulk SiC(0001) on defective hBN/SiC interfaces. Supplemental movies provide additional visualization where appropriate.
\end{abstract}

\maketitle

\tableofcontents

\section{Different Dispersion Interaction Methods}
\label{sec:stacking}

We benchmark several van der Waals (vdW) correction schemes implemented in VASP~\cite{kres1} (i.e., IVDW) to select and justify the one ultimately used for the hBN/SiC heterostructure.
All quantities were calculated for the AA$'$ stacking order of hBN and SiC structures, where N(C) and Si(B) are positioned over each other.
The bilayer binding energy ($E_{\rm b}$) is defined as:
\begin{equation}
    E_{\rm b} = \frac{E_{\rm{bilayer}} - 2 E_{\rm{monolayer}}}{A},
\end{equation}
where $E_{\rm{bilayer}}$ and $E_{\rm{monolayer}}$ are the bilayer unit cell energy and the monolayer unit cell energy, respectively.
Here, $A$ denotes the basal area of the computational cell.

Our results show that the interlayer distance ($d_{zz}$) and lattice constant ($a$) predicted using IVDW values of 3, 4, and 12 are in close agreement with the available experimental values~\cite{KORSAKS2015976,Pease1952AnXS,paszkowicz2002lattice,Polley2023_PRL}.

Bilayer binding energy calculations show that the vdW interactions modeled with the Grimme DFT-D3 scheme (Becke–Johnson damping, IVDW = 12) are almost twice as strong in SiC as in hBN, pointing to much stronger interlayer binding in SiC. This is the largest vdW-dispersion difference observed among all the methods examined.
This difference can be attributed to the larger atomic dipole moments in the SiC layers.
Additionally, the binding energy $E_{\rm b}$ is proportional to the interlayer distance $d_{zz}$.
\begin{table}[ht!]
\centering
\caption{Calculated lattice constant ($a$), interlayer spacing ($d_{zz}$), and bilayer binding energy ($E_{\rm b}$) for h-BN and SiC, obtained using the various vdW dispersions (IVDWs) implemented in VASP.}
\label{tab:infohbnsic}
\begin{tabular}{lccccccc}
    \hline
      & \multicolumn{3}{c}{hBN} & &\multicolumn{3}{c}{SiC}  \\\cline{2-4}\cline{6-8}\noalign{\vskip 2pt}
 IVDW & $a$ (\AA) & $d_{zz}$ (\AA) & $E_{\rm b}$ (meV/\AA$^2$) & & $a$ (\AA) &  $d_{zz}$ (\AA) & $E_{\rm b}$ (meV/\AA$^2$)  \\ \hline
 $3$    & $2.51$ & $3.29$ &$-24.7$&& $3.09$ & $3.22$ &$-31.5$ \\ 
 $4$    & $2.50$ & $3.30$ &$-21.8$&& $3.08$ & $3.23$ &$-21.3$ \\ 
 $10$   & $2.51$ & $3.13$ &$-25.2$&& $3.09$ & $3.21$ &$-26.8$ \\ 
 $11$   & $2.51$ & $3.41$ &$-16.1$&& $3.10$ & $3.48$ &$-19.1$ \\
 $12$   & $2.51$ & $3.26$ &$-16.9$&& $3.09$ & $3.23$ &$-28.7$ \\
 $20$   & $2.51$ & $3.39$ &$-27.8$&& $3.10$ & $3.49$ &$-31.6$ \\
 $21$   & $2.51$ & $3.47$ &$-19.1$&& $3.09$ & $3.57$ &$-20.1$ \\ \hline
\end{tabular}
\end{table}
%

%%%%%%%%%%%%%%%%%%%%%%%%%%%%%%%%%%%%%%%%%%%%%%%%%%%%%%%%%%
\section{Defect Formation Energy Calculations}
In order to assess the charge-state $q$ stability of \vb and \vn, we compute the defect formation energy $E_\text{f}$ as:
\begin{equation}
\label{eq:formation}
    E^q_\text{f} = E^q_\text{defect} - E_{\rm{pristine}} - \sum_{i} n_i \mu_i + q(E_\text{Fermi}) +E^q_\text{corr},
\end{equation}
where $ E^q_\text{defect} $ is the total energy of the hBN/SiC heterostructure containing the defect in charge state $q$, and $ E_\text{pristine} $ is the total energy of the pristine system.
The term $n_i = -1$ represents that a single atom was removed relative to the pristine structure, and $ \mu_i $ is the corresponding atomic chemical potential.
The chemical potentials are defined based on the growth environment. In N-rich conditions, $ \mu_{\rm N} $ is derived from N$_2$ gas, and the B chemical potential is given by $ \mu_{\rm{B}} = \mu_{\rm{BN}} - \mu_{\rm{N}} $, where $ \mu_{\rm{BN}} $ is the chemical potential of bulk hBN.
In N-poor conditions, $ \mu_{\rm{B}} $ is taken from bulk B, and $\mu_{\rm{N}}$ is computed as $\mu_{\rm{BN}} - \mu_{\rm{B}}$.
Additionally, $E_\text{Fermi}$ represents the Fermi energy level, which corresponds to the chemical potential of the electron reservoir.
It is aligned with the valence band maximum (VBM) energy of the pristine hBN/SiC platform.
Finally, $E^q_\text{corr}$ is a correction term applied to the total energies of charged supercells, accounting for spurious electrostatic interactions introduced by periodic boundary conditions~\cite{Freysoldt2018}.

%%%%%%%%%%%%%%%%%%%%%%%%%%%%%%%%%%%%%%%%%%%%%%%%%%%%%%%%%%
\section{Dataset Construction and NEP Potential Training}
DFT calculations employ two different parameters compared to those described in the "Computational Details" section of the main manuscript.
For ab initio molecular dynamics (AIMD) simulations, $E^{\rm CUT}$ was set to $400$ eV, and $dE$ was set to \(10^{-4}\) eV.  
Single-shot energy minimization runs were performed with higher precision using $E^{\rm CUT} = 500$ eV and $dE = 10^{-7}$ eV to compile the dataset.
The canonical ($NVT$) ensemble using the Nos\'{e}-Hoover thermostat was used.
The time step was $0.5$ fs.
Below, we describe the distinct groups of data points; further details are available in A.H.'s GitHub repository (\href{https://github.com/sahashemip/cu-hbn-sic-ml-potential}{GitHub}).

\subsection*{1. Relaxation of Monovacancy (\texorpdfstring{$V_\mathrm{B}$}{V\_B}) Defects}

Lattice mismatch within the hBN/SiC heterostructure supercell gives rise to 15 symmetry-distinct boron vacancy (\(V_\mathrm{B}\)) sites. For each unique site, we created the corresponding defective structure and performed full geometry optimizations.
Optimization trajectories were stored in VASP \texttt{XDATCAR} files.
Some geometries were selected, while each was randomly rattled, strained, and deformed.
A total of 1113 data points underwent single‑point DFT calculations and were incorporated into the dataset.

\subsection*{2. AIMD for the Most Stable Monovacancies}

For the most stable \(V_\mathrm{B}\) configuration (four interlayer chemical bonds), we carried out AIMD simulations at target temperatures \(T=\) 300, 500, 700, and \SI{900}{K}.
Each trajectory comprised 6000 time steps.
Subsequently, frames were sampled every 40 steps using, producing 600 additional (unperturbed) structures.

\subsection*{3. Pristine Supercell Configurations}

To capture pristine interfacial variability, 120 additional defect–free structures were generated via
(i) atomic rattling, (ii) systematic in-plane sliding of the hBN layer, and (iii) modulation of the interlayer separation.
Interlayer sliding was implemented by translating the hBN sheet from fractional displacement \((0,0)\) to \((a_x/2, a_y/2)\) in the rectangular supercell, where \(a_x=\SI{3.09}{\angstrom}\) and \(a_y=\SI{5.35}{\angstrom}\).

\subsection*{4. Iterative Data Enhancement via Initial NEP-Guided MLMD}

An initial NEP model was trained on the 1833 configurations from Sections~1–3.
Using this preliminary model, we conducted machine-learning molecular dynamics (MLMD) in the canonical ($NVT$) ensemble with a Berendsen thermostat over a temperature range \SIrange{200}{1000}{K}.
Configurations exhibiting (i) structural instability, (ii) local distortions, or (iii) non–physical behavior were extracted and appended to the training pool after DFT validation.
This procedure contributed about 900 additional (curated) structures comprising pristine and monovacancy-containing systems.
Rectangular supercells, used for MLMD, contained the same number of atoms as the original triangular cell.

\subsection*{5. Di- and Tri-Vacancy Defect Set}

To sample elevated vacancy concentrations, we generated 400 unique multi–vacancy configurations by random removal of two or three boron atoms (di- and tri-vacancies).
All structures were relaxed with the previous step's trained potential and subsequently subjected to single‑point DFT calculations.

\subsection*{6. AIMD Simulations of Distinct Monovacancy-Cu Coupling Configurations}
AIMD simulations were run in the $NVT$ ensemble at $300$ K for four model systems in which a single Cu atom was initially positioned at progressively greater radial offsets from the center of the boron monovacancy, \vb.
The four models also differ in the number of \ch{N-Si} interlayer bonds, enabling a direct comparison of bonding motifs.
Each trajectory was propagated for 6000 steps with a 0.5 fs timestep; every 300th frame was saved, providing 80 statistically independent snapshots per system.
Each snapshot was visualized, subjected to single-point DFT energy evaluation, and then incorporated into the overall dataset.
\begin{figure*}[hbt!]
    \centering
    \includegraphics[scale=0.875]{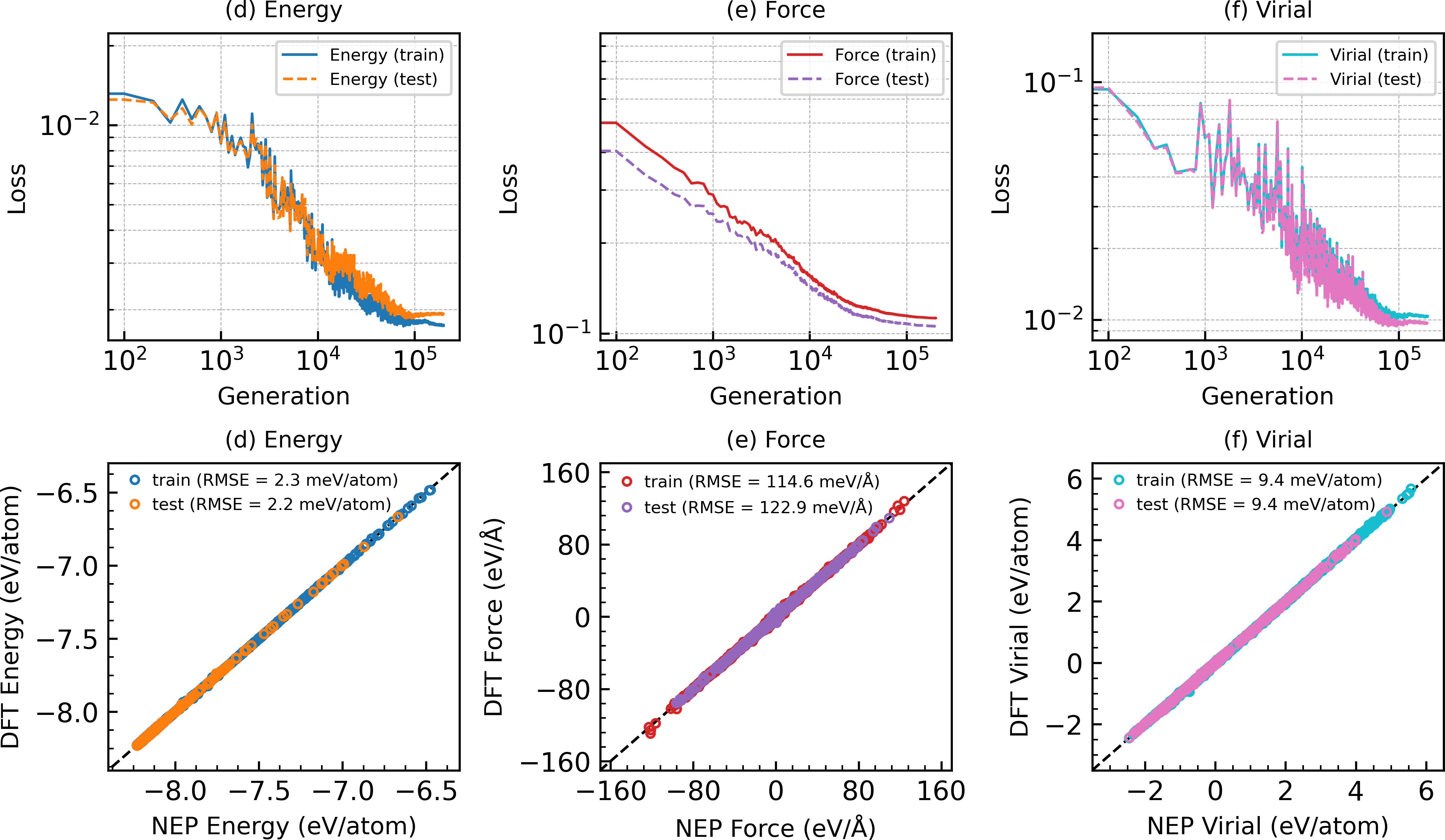}
    \caption{Evolution of the NEP training process and model accuracy. (a–c) Log‑loss curves for energy, force, and virial, respectively, plotted against generation for both training (solid lines) and testing (dashed lines) data sets. (d–f) Correlation between NEP‑predicted and DFT reference values for energy, force, and virial on the training and testing sets; the dashed diagonal indicates perfect agreement, and RMSE values are reported in each legend.}
    \label{fig:nep}
\end{figure*}

\subsection*{7. Bonded $\rightarrow$ Non-Bonded Transition Coordinate with Cu}
We constructed a one–dimensional configurational coordinate connecting interlayer–bonded and non–bonded states in the presence of Cu adatoms (see Fig. S4):
21 geometries along this path (plus their rattled/deformed counterparts; total 42) were added.
These enable evaluation of transition energetics and structural relaxation pathways.

\subsection*{8. Iterative Cu/Defect Co-Sampling (MLMD + Curation)}

Successive MLMD simulations ($NVT$, where $T=300$ K) were initiated from pristine rectangular supercells containing targeted vacancy patterns and one or more Cu adatoms.
During these runs, Cu atoms frequently anchored at vacancy sites and, in some cases, promoted additional interlayer bond formation.
The raw numbers of configurations collected for each compositional motif are:

\begin{center}
\begin{tabular}{l r}
\toprule
System & Raw Structures \\
\midrule
\(1V_\mathrm{B} + 1\)Cu & 108 \\
\(1V_\mathrm{B} + 2\)Cu & 128 \\
\(1V_\mathrm{B} + 3\)Cu & 106 \\
\(2V_\mathrm{B} + 1\)Cu & 106 \\
\(2V_\mathrm{B} + 2\)Cu & 75 \\
\(1V_\mathrm{B} + 4\)Cu & 109 \\
\(1V_\mathrm{B} + 5\)Cu & 92 \\
\(1V_\mathrm{B} + 6\)Cu & 66 \\
\(1V_\mathrm{B} + 7\)Cu & 17 \\
\(4V_\mathrm{B} + 9\)Cu & 9 \\
\midrule
\textbf{Total} & 816 \\
\bottomrule
\end{tabular}
\end{center}

The final curated dataset comprises 3953 structures, partitioned into 3361 training (85\%) and 592 test (15\%) configurations.
A final NEP model employing radial and angular cutoffs of \SI{5.0}{\angstrom} and \SI{3.5}{\angstrom}, respectively, and a single hidden layer of 50 neurons, achieved high accuracy as shown in Fig~\ref{fig:nep}.
Long-timescale MLMD trajectories (Movies~S1–S6) demonstrate the model's dynamical stability across diverse thermodynamic conditions and defect/adatom configurations.
All curated structures (3953 entries), split into training and test subsets, along with the final \texttt{nep.txt} model file, will be deposited at \href{https://github.com/sahashemip/cu-hbn-sic-ml-potential}{https://github.com/sahashemip/cu-hbn-sic-ml-potential}.

%%%%%%%%%%%%%%%%%%%%%%%%%%%%%%%%%%%%%%%%%%%%%%%%%%%%%%%%
\section{Fundamental properties of pristine hBN and SiC}

Table \ref{tab:pritine-infor} highlights various properties of monolayers.
The in-plane lattice constant of hBN ($2.51$ \AA) is within $\approx1.5$\% of graphene’s ($2.46$ \AA), rationalising its routine use as an atomically flat encapsulant, whereas the $\approx 25$\% mismatch with SiC ($3.09$ \AA) explains why graphene/SiC stacks are typically realised by Si-sublimation growth rather than by mechanical stacking.
Interlayer spacings cluster around $3.3$ \AA, yet the slightly shorter value for SiC hints at a marginally stronger out-of-plane coupling that could raise the exfoliation energy.
In-plane stiffness follows the expected order, graphene > hBN > SiC, with 2D Young's moduli of $\approx 1.29$, $1.11$ and $0.65$ TPa, respectively, while the corresponding shear moduli ($C_{66}$) show that SiC's lattice resists shape distortion less than either hBN or graphene, making it more prone to wrinkle formation.
SiC also exhibits the largest Poisson ratio ($0.32$), indicating greater lateral dilation under uniaxial strain and hence a higher likelihood of brittle failure in flexible architectures.  Electrically, hBN's wide $5.46$ eV gap cements its role as an ultra-thin dielectric, SiC's moderate $3.27$ eV gap positions it for high-power or UV optoelectronic layers, and graphene remains the prototypical zero-gap conductor.
These contrasts underscore why hBN, SiC and graphene occupy complementary, rather than competing, niches in emerging two-dimensional heterostructures.
\begin{table}[H]
\centering
\caption{Calculated data: $a$(\AA), $d_{zz}$(\AA), $l_{\rm b}$(\AA) denote the lattice constants, interlayer distance, and bond length (B--N/Si--C//C--C), respectively, for the AA' stacking configuration (where B(C) and Si(N) are positioned over each other). The elastic constant tensor elements are represented by $C_{ij}$ (GPa): Longitudinal compression ($C_{11}$): the applied force and atomic displacement are aligned along the same axis.  
Transverse expansion ($C_{12}$): the response when the applied force and atomic displacement are perpendicular to each other.  
Shear modulus ($C_{66}$): the deformation of the lattice structure from a triangular to a rectangular configuration under the influence of an applied force, as an illustrative example.
Poisson ratio ($\nu^{\rm 2D} = C_{12}/C_{11}$) and Young modulus $Y^{\rm 2D} =({C_{11}^{2} - C_{12}^{2}})/{C_{11}}$ (GPa) are also reported.
$\Delta E_{\rm g}$(eV), calculated using HSE, indicates electronic structure band gap.}
\vspace{0.1cm}
\label{tab:pritine-infor}
\begin{tabular}{c|ccccccccc}
 material   &$a$   & $d_{zz}$  & $l_{b}$ & $C_{11}$ & $C_{12}$ & $C_{66}$ & $\nu^{\rm 2D}$ & $Y^{\rm 2D}$ & $\Delta E_{\rm g}$\\\hline 
 hBN       & 2.51 & 3.26 & 1.45 & 1171.5 & 270.0 & 450.7 & 0.23 & 1109.3 & 5.46\\
 SiC      & 3.09 & 3.23 & 1.78 & 725.2  & 228.6 & 248.3 & 0.32 & 653.1  & 3.27\\
 graph.   & 2.46 & 3.35 & 1.42 & 1353.4 & 287.3 & 533.1 & 0.21 & 1292.4 & ---
\end{tabular}
\end{table}

%%%%%%%%%%%%%%%%%%%%%%%%%%%%%%%%%%%%%%%%%%%%%%%%%%
\section{Interlayer distance of pristine hBN/SiC}

To quantify the interlayer spacing between the hBN monolayer and the SiC monolayer during MD simulations, we computed spatially resolved interlayer distances. For each selected frame in the trajectory, atoms are mapped onto a $25 \times 25$ lateral grid based on their fractional $x$–$y$ positions.
Within each grid tile, the mean $z$-coordinate of B/N atoms and Si/C atoms is calculated separately, and their difference yields the local interlayer spacing $d_{zz}$. 

The MLMD simulations were carried out at different temperatures ($25$, $50$, $100$, $200$, $300$, $400$, $500$, $600$, $700$, $800$, and $900$ K). For each temperature, the system was first equilibrated in the $NpT$ ensemble for $10^{6}$ integration steps with a $0.5$ fs time step. Subsequently, a production run of $10^{6}$ steps was performed in the $NVT$ ensemble, and atomic configurations were saved every $2$ ps for analysis.

\begin{figure}
    \centering
    \includegraphics[width=0.98\linewidth]{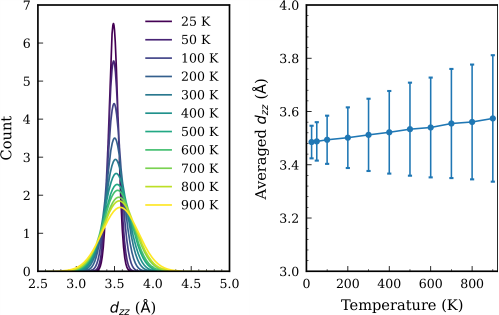}
    \caption{(Left panel): Distribution of interlayer distance ($d_{zz}$) as a function of temperature, obtained from molecular dynamics simulations. (Right panel): Temperature dependence of the averaged $d_{zz}$. Error bars represent the standard deviation.}
    \label{fig:sm_intlayerdist}
\end{figure}
Figure~\ref{fig:sm_intlayerdist} summarizes how the \ch{hBN-SiC} separation responds to temperature in our MLMD runs.
The distributions of the local interlayer distance $d_{zz}$ are plotted for temperatures from 25 K to 900 K (left panel).
As the temperature rises, the peak progressively shifts to larger spacings and broadens.
By $900$ K the mode has moved to $\approx 3.57$ \AA~with a standard deviation ($\sigma$) of $0.24$ \AA.
The monotonic right-shift reflects anharmonic out-of-plane vibrations, while the growing width captures the increasing amplitude of thermally driven ripples.

The time-averaged $d_{zz}$ for each simulation; vertical bars show one standard deviation of the underlying histogram.
The mean separation increases almost linearly with temperature, by about $0.11$ \AA~between $25$ K and $900$ K ($\approx3$ \% relative dilation).
The modest slope underscores that the hBN layer remains weakly bound to the SiC surface.
Error bars grow with temperature because larger out-of-plane excursions make the instantaneous spacing more dispersed, consistent with the left-panel broadening.

Taken together, our results demonstrate that the hBN/SiC interface is mechanically stable yet compliant: the average interlayer distance increases only slightly under heating, but thermal fluctuations introduce substantial local corrugation. At room temperature, $d_{zz}$ equals $3.51 \pm  0.14 $ \AA~with a minimum and maximum  of  $2.97$ and $4.09$ \AA, respectively.

%%%%%%%%%%%%%%%%%%%%%%%%%%%%%%%%%%%%%%%%%%%%%%%%%%%%%
\section{Pristine hBN/SiC heterostructure}
\begin{table}[ht!]
\centering
\caption{The calculated values include the averaged interlayer distance (\(d\)), the layer buckling for the hBN (\(\Delta z_{\rm{BN}}\)) and SiC (\(\Delta z_{\rm{SiC}}\)) monolayers, and the lattice mismatch for various vdW interactions implemented in VASP (IVDW).}
\label{tab:infohbnsic2}
\begin{tabular}{lcccc}
    \hline   
 IVDW & $d_{zz}$ (\AA) & $\Delta z_{\rm{BN}}$ (pm) & $\Delta z_{\rm{SiC}}$ (pm) & $\Delta a$ (\%) \\ \hline
 0    & 4.25  &  9.6 & 0.3 & 1.47 \\
 3    & 3.47  &  42.6& 8.0 & 1.50 \\ 
 4    & 3.55  &  38.9& 5.3 & 1.44 \\ 
 10   & 3.33  &  49.9& 13.2& 1.52 \\ 
 11   & 3.62  &  33.9& 5.5 & 1.43 \\
 12   & 3.49  &  41.2& 9.0 & 1.48 \\
 20   & 3.54  &  33.7& 10.8& 1.33 \\
 21   & 3.80  &  27.4& 2.1 & 1.44 \\ \hline
\end{tabular}
\end{table}

The average interlayer distance ($d$) is defined as:
\begin{equation}
    d_{zz} = |\overline{z}_{\rm{BN}} - \overline{z}_{\rm{SiC}}|,
\end{equation}
where $\overline{z}_{\rm{BN}}$ and $\overline{z}_{\rm{SiC}}$ represent the average heights of the hBN and SiC layers, respectively.

Each layer's fluctuation range is defined as follows:
\begin{equation}
    \Delta z = z_{\rm{max}} - z_{\rm{min}}.
\end{equation}

The lattice mismatch is calculated as:
\begin{equation}
    \Delta a = (\frac{a_{\rm{hBN}}^{\rm{HS}}- a_{\rm{SiC}}^{\rm{HS}}}{a_{\rm{hBN}}}) \times 100,
\end{equation}
where $a_{\rm{hBN}}^{\rm{SC}}$ and $a_{\rm{SiC}}^{\rm{SC}}$ denote the lattice constants of the $10\times10\times1$ hBN monolayer supercell and the $8\times8\times1$ SiC supercell, respectively.

Table \ref{tab:infohbnsic2} highlights how sensitive the relaxed hBN/SiC heterostructure is to the particular vdW correction implemented in VASP (IVDW tag).
In the reference calculation without any dispersion term (IVDW = 0) the two sheets sit far apart, $d_{zz}=4.25$ \AA, and both layers remain essentially flat ($\Delta z_{\text{BN}}=9.6\;\text{pm}$, $\Delta z_{\text{SiC}}=0.3\;\text{pm}$).
Once a vdW functional is switched on, the interlayer spacing contracts by almost $1$ \AA, converging to the $3.3-3.8$ {\AA} range that is expected for weakly bound vdW interfaces.
Accompanying this vertical contraction, the hBN sheet develops a pronounced buckling of $34-50$ pm, whereas the SiC distortion never exceeds $13$ pm, underscoring that the mechanically softer hBN layer absorbs most of the out-of-plane strain.
Despite these sizeable changes in $d_{zz}$ and buckling, the in-plane lattice mismatch, $\Delta a$, remains remarkably stable around $1.4-1.5$ \%, with IVDW = $20$ giving the only noticeably lower value ($1.33$ \%).
Overall, the data show that accurate vdW forces are essential for capturing both the equilibrium spacing and the corrugation amplitude of this heterobilayer, while the lattice mismatch is largely insensitive to the specific vdW scheme.

%%%%%%%%%%%%%%%%%%%%%%%%%%%%%%%%%%%%%%%%%%%%%%%%
\section{Boron vacancy vs interlayer distance}
\begin{figure}
    \centering
    \includegraphics[width=0.92\linewidth]{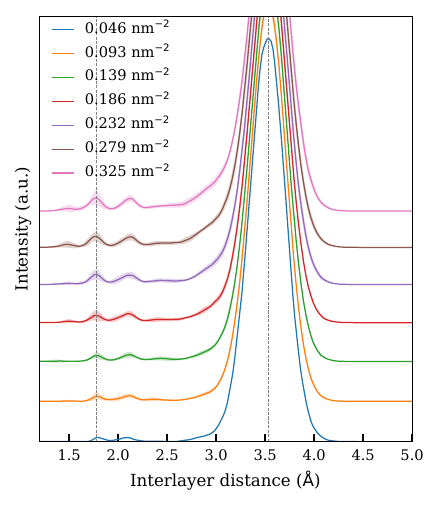}
    \caption{Interlayer spacing in \ch{hBN–SiC} monolayers as a function of $V_\mathrm{B}$ defect concentration at $T=300 \rm{K}$.
    Vertical dotted lines are drawn at $1.8$ \AA~and $3.52$ \AA, and each plot features a shaded band indicating the error bars.}
    \label{fig:sm_intlayerdist_vb}
\end{figure}
Figure \ref{fig:sm_intlayerdist_vb} traces the distribution of interlayer distances in hBN/SiC heterostructure for a series of boron vacancy ($V_\mathrm{B}$) areal concentrations ($0.046-0.325$ nm$^{-2}$).
As the vacancy density rises, the dominant peak at $3.52$ \AA \ slightly weakens and broadens while a low-distance peak near $2$ \AA \ indicates that defect-mediated rumpling locally collapses the interface and widens the overall spacing distribution.
This systematic evolution confirms that out-of-plane disorder is strongly coupled to the in-plane point-defect landscape in the \ch{hBN-SiC} monolayer system.

The MLMD simulations were carried out at $300$ K using $NVT$ ensemble for $2\times10^{6}$ steps.
Time step is $0.5$ fs.
Initial structure (before defect creation) was equilibrated at room temperature.
Ten various defective structures were generated at each concentration except for the single defect case, where only one configuration was used.

\section{Transition Metal Adsorption Energy Profile}
\label{subsec:tmadsorption}

\begin{figure*}[ht!]
    \centering
    \includegraphics[scale=0.95]{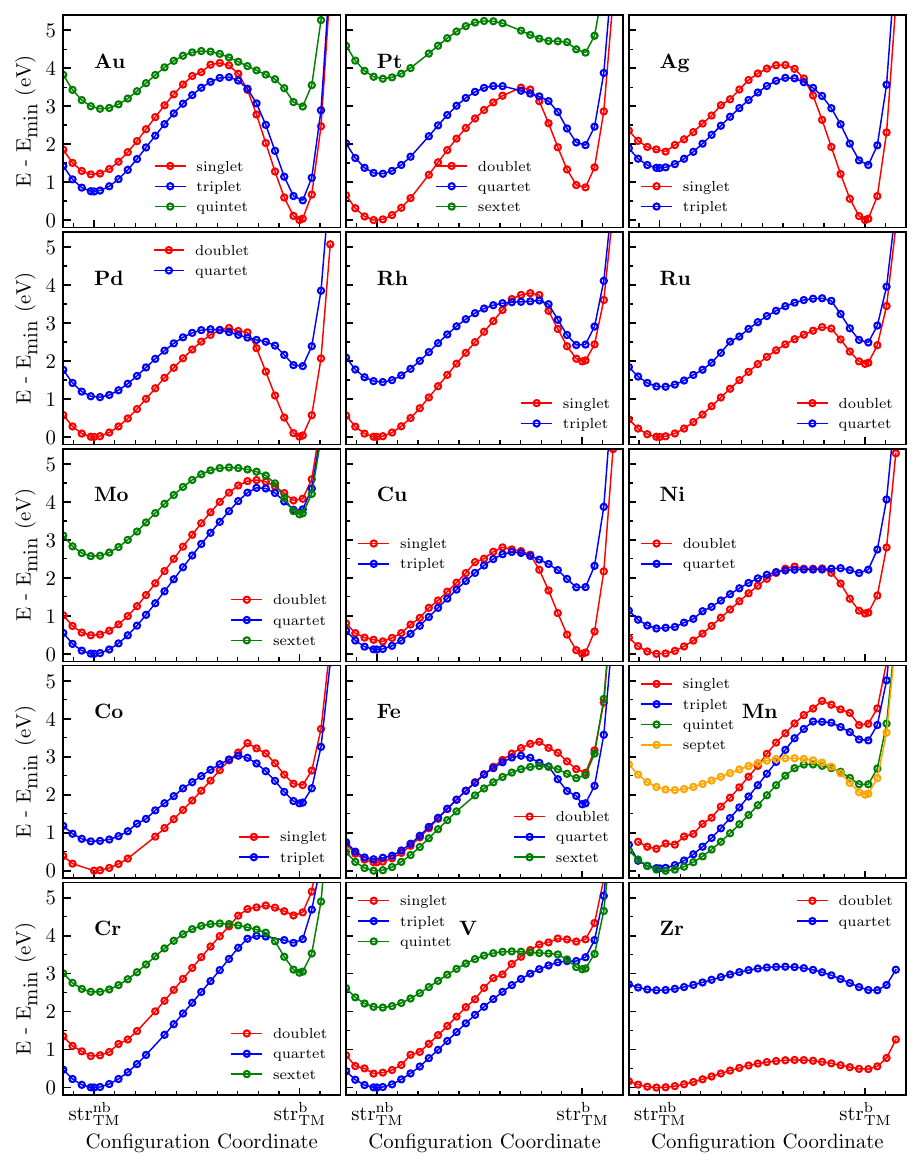}
    \caption{Energy profile of the structural transition from a non-bonded \ch{hBN-SiC} configuration in the presence of various transition metals (TMs), denoted as \({\rm str^{nb}_{\rm{TM}}}\), to a bonded structure, \({\rm str^{b}_{\rm{TM}}}\) in different spin states.}
    \label{fig:TMatab}
\end{figure*}
Figure \ref{fig:TMatab} compiles configuration-coordinate diagrams ($E-E_{\rm min}$ versus reaction coordinate) for the structural conversion of a non-bonded hBN/SiC interface "${\rm str_{\rm{TM}}^{\rm nb}}$" to a chemisorbed "${\rm str_{\rm{TM}}^{\rm b}}$" complex in the presence of 15 different transition metals.
Each panel corresponds to one metal and overlays the potential-energy surfaces for all spin multiplicities that converge in our calculations (singlet through sextet).
Based on the shape of the potential-energy surface, we sort the metals as follows:
(i) For the coinage metals Cu, Ag, and Au (group IB) the lowest‐energy state is reached when the interlayer \ch{Si-N} bonds have been established.
Pd behaves almost similarly, displaying a dual-well potential-energy surface with two nearly degenerate minima ($\Delta E \approx 0.01$ eV).
In all of these cases, the transition-state energy separating the wells is high enough ($2.68-3.76$ eV) to kinetically lock the system in the bonded configuration once the barrier has been crossed, thereby stabilizing the chemisorbed state. The bonded hBN/SiC in the presence of these metals is stable in \emph{singlet} spin state.
(ii) For Pt, the bonded \ch{Pt–hBN/SiC} system is metastable: it sits $0.84$ eV higher in energy than the non-bonded minimum, yet it is insulated by a substantial $2.66$ eV dissociation barrier.
This large kinetic barrier effectively traps the system in the chemisorbed state, allowing the bonded configuration to persist under typical experimental conditions despite its thermodynamic handicap.
(iii) For the remaining transition metals, a shallow local minimum does appear at the bonded geometry (${\rm str_{\rm{TM}}^{\rm b}}$).
However, the barrier is low and the non-bonded configuration (${\rm str_{\rm{TM}}^{\rm nb}}$) lies markedly lower in energy, so interlayer bond cleavage remains highly probable.
These metals also exhibit a preference for high-spin manifolds, and spin-state crossings frequently accompany the structural interconversion.

\begin{table}[!htb]
\centering
\begin{tabular}{lrccccccr}
   &\multicolumn{5}{c}{hBN/SiC} & & \multicolumn{2}{c}{hBN} \\ \cline{2-6} \cline{8-9}
TM & $\Delta E^{0}$ & $\Delta E^{*}$ & $q_{\rm{TM}}^{\rm nb}$ & $q_{\rm{TM}}^{\rm b}$ & $E_{\rm{bind}}^{\rm b}$ & &$q_{\rm{TM}}$ & $E_{\rm{bind}}$ \\ \hline
 V  & $3.12$ & $0.22$ & $1.38$ & $0.86$ &$-4.06$ && $1.31$ & $-10.89$\\
  Cr & $3.03$ & $0.95$ & $1.28$ & $0.85$ &$-2.72$ && $1.28$ & $-9.45$ \\
   Mn & $1.99$ & $0.79$ & $1.25$ & $0.69$ &$-2.75$ && $1.26$ & $-8.61$ \\
    Fe & $1.76$ & $1.01$ & $1.25$ & $0.80$ &$-3.54$ && $1.06$ & $-8.91$ \\
     Co & $1.44$ & $1.26$ & $0.91$ & $0.67$ &$-3.86$ && $0.93$ & $-9.32$ \\
      Ni & $1.07$ & $1.19$ & $0.91$ & $0.59$ &$-3.95$ && $0.91$ & $-8.67$ \\
  \bf{Cu} & $-0.11$& $2.68$ & $0.92$ & $0.60$ &$-2.73$ && $0.92$ & $-6.35$ \\
        Zr & $0.48$ & $0.15$ & $1.61$ & $1.74$ &$-8.28$ && $1.61$ & $-12.45$\\
         Mo & $3.68$ & $0.70$ & $1.23$ & $0.73$ &$-6.28$ && $1.24$ & $-13.67$\\
          Ru & $1.92$ & $0.97$ & $0.91$ & $0.48$ &$-4.11$ && $0.91$ & $-9.71$ \\
           Rh & $1.99$ & $1.61$ & $0.79$ & $0.33$ &$-3.60$ && $0.79$ & $-9.29$ \\
       \bf{Pd} & $0.01$ & $2.75$ & $0.73$ & $0.32$ &$-2.32$ && $0.73$ & $-3.91$ \\
        \bf{Ag} & $-1.37$& $3.75$ & $0.73$ & $0.53$ &$-1.56$ && $0.73$ & $-3.91$\\
         \bf{Pt} & $0.84$ & $2.66$ & $0.71$ & $0.20$ &$-2.93$ && $0.70$ & $-7.38$\\
          \bf{Au} & $-0.85$& $3.76$ & $0.67$ & $0.24$ &$-1.19$ && $0.66$ & $-4.09$\\ \hline
\end{tabular}
\caption{The energy difference between the bonded and non-bonded states ($\Delta E^{0}$) and the transition energy barrier ($\Delta E^{*}$), Bader net charge on TMs in the bonded and non-bonded \ch{hBN-SiC} heterostructure, represented by $q_{\rm{TM}}^{\rm b}$ and $q_{\rm{TM}}^{\rm nb}$, and in the monolayer hBN represented by $q_{\rm{TM}}$.
The TM binding energy to the hBN/SiC with interlayer chemical bonds formed, as well as to the monolayer hBN, are denoted by $E_{\rm{bind}}^{\rm b}$ and $E_{\rm{bind}}$, respectively.
The energy and charge values are expressed in units of eV and e, respectively.}
\label{tab:tm-info4}
\end{table}
Table \ref{tab:tm-info4} summarizes the information related to Fig. \ref{fig:TMatab}. Indeed, it compares relevant values for monolayer hBN with those for the hBN/SiC heterostructure.
The binding energy, $E_{\mathrm{bind}}$, quantifies how firmly a TM is anchored at the interface and, by extension, its propensity to desorb.
On this basis, the adatoms can be separated into three categories - weak, moderate, and strong binders.
Au and Ag reside in the weak-binding class, consistent with limited orbital overlap with the substrate.
Comparing the bonded-state values, $E_{\mathrm{bind}}^{\rm b}$, with those obtained for TMs at the $V_\mathrm{B}$ site of an isolated hBN sheet shows that the monolayer retains adatoms roughly $1.5 - 3.5\times$ more strongly than the hBN/SiC heterostructure.
This additional strength arises from greater electron sharing between the TM $d$-orbitals and the \ch{B–N} framework, yielding a more covalent bond in the monolayer.
In the bilayer system, newly formed interlayer \ch{Si–N} bonds already alleviate the electron deficiency of the $V_\mathrm{B}$ site, reducing the amount of charge the defect must draw from the adatom and, consequently, weakening the TM–substrate interaction.

\section{hBN on 4\ch{H‑SiC}(0001)}
\label{sec:4hsic-hbn}

To investigate pristine and defective hBN on a SiC substrate, we employ a slab of 4\ch{H-SiC} cleaved along the (0001) direction.
A (0001)-oriented 4\ch{H‑SiC} slab preserves the in-plane hexagonal lattice of monolayer SiC; however, unlike an isolated monolayer (which is strictly planar), the bulk-derived slab contains buckled \ch{Si-C} bilayers in which the Si and C sub-lattices occupy alternating planes along the surface normal. This structural buckling must be retained to capture realistic interfacial registry and electrostatics.

The substrate consists of nine SiC layers (i.e., nine repetitions of \ch{Si-C} double layers; total thickness $\simeq 20.12$ \AA).
The bottom surface (carbon‑terminated) is passivated by hydrogen atoms to saturate dangling bonds and suppress spurious surface states.
This passivation also minimizes artificial electrostatic interactions between the two periodic slab surfaces.
The substrate supercell contains 144 Si, 144 C, and 16 H atoms (Si$_{144}$C$_{144}$H$_{16}$).
A vacuum region of $18$ \AA \ is introduced along the surface normal to avoid interactions between periodic images.

A $5\times5\times1$ hBN monolayer is placed on the top (unpassivated) surface to form the heterostructure.
This supercell size accommodates the 4\ch{H‑SiC}(0001) surface with a residual lattice mismatch of approximately 1.2\% (defined as $|a_{\text{hBN}}-a_{\text{SiC}}|/a_{\text{SiC}}$);
geometry optimizations were performed at fixed cell volume (ISIF = $4$), allowing the cell shape and all internal atomic positions to relax until the maximum residual force on any atom was below $10^{-2}\ \mathrm{eV},\mathrm{\AA^{-1}}$.

\begin{figure}[ht!]
    \centering
    \includegraphics[scale=0.45]{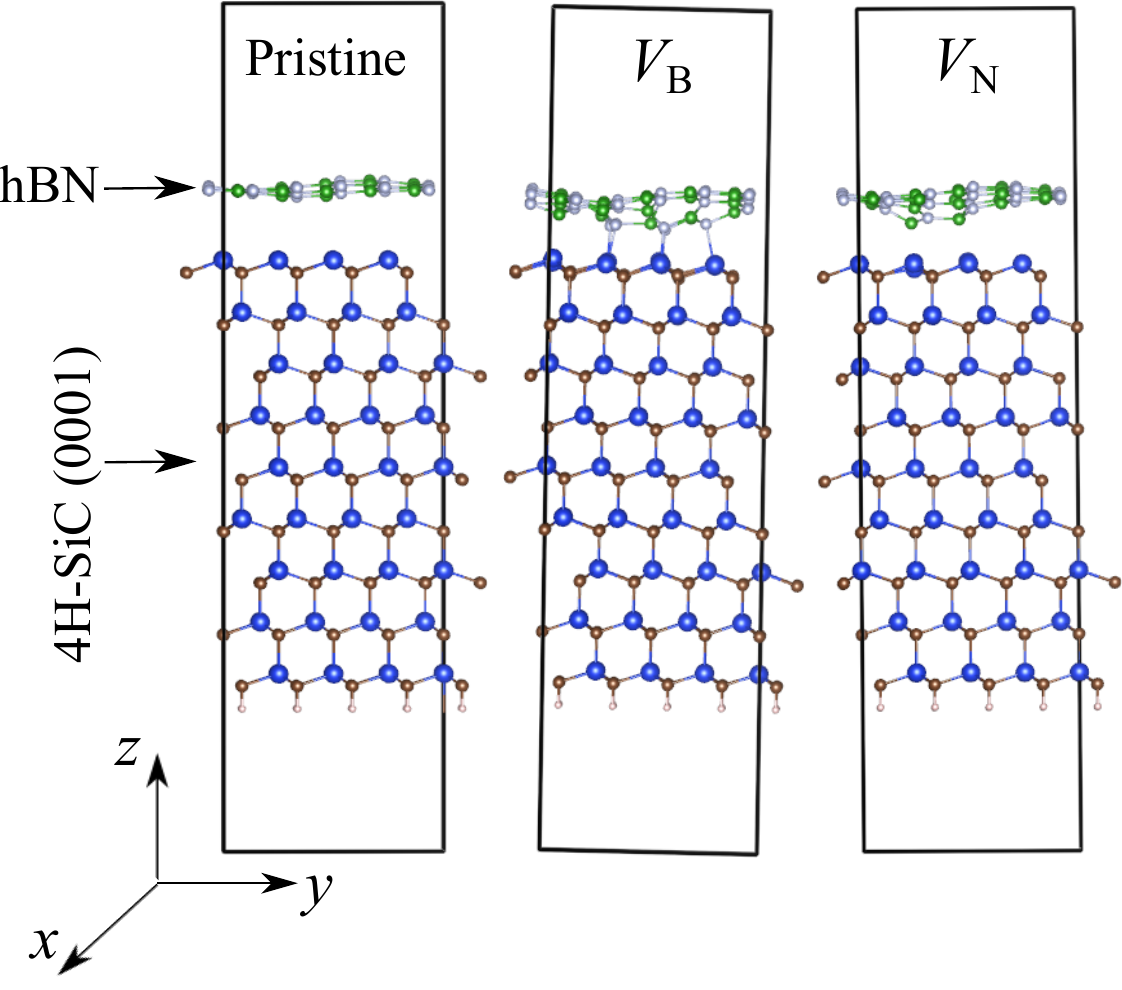}
    \caption{Optimized atomic structures of pristine hBN/4H-SiC(0001), hBN with a boron vacancy \vb , and hBN with a nitrogen vacancy \vn. Si: (blue), C: (brown), B: (green), N: (gray), H: (white).
    Side views are shown. Slight corrugation of the hBN layer and local relaxations around the vacancies are evident.}
    \label{fig:hsichbn}
\end{figure}
Point defects in the hBN layer were modeled by removing a single B or N atom from the $5\times5$ sheet, generating boron and nitrogen vacancies, \vb and \vn, respectively.
The chosen supercell size yields a defect concentration of $1/25 = 4$\%.
It is four times larger than what we had earlier in the hBN/SiC heterostructure.
Figure~\ref{fig:hsichbn} shows the fully relaxed pristine and defective interface models.

In the pristine system, the hBN layer exhibits a slight corrugation (maximum height variation $\approx 0.68$ \AA) induced by the underlying buckled SiC surface.
We obtained a minimum interlayer distance of $3.15$ \AA, which is $0.06$ \AA~lower than the reported value for hBN/SiC heterostructure.
The hBN/4\ch{H‑SiC}(0001) binding energy is about $-20$ meV/\AA$^{2}$, higher than the SiC bilayer but lower than the hBN bilayer.

Local structural relaxations around \vb and \vn follow the same trends reported previously for hBN/SiC, and the \ch{Si-N} interfacial covalent bond length of $\approx 3.75\ \text{\AA}$ confirms that our 4\ch{H-SiC}(0001) slab (representative of a bulk SiC interface) adequately captures the relevant interfacial effects.

\section{Descriptions of Supplemental Movies}
\begin{itemize}
    \item[]{{\bf{Movie S1:}} Molecular-dynamics trajectory of a pristine hBN/SiC heterostructure showing structural dynamics and stability at room temperature ($300$ K) and at $900$ K.}
    \item[]{{\bf{Movie S2:}} Molecular-dynamics trajectory of a defective hBN/SiC heterostructure containing a single \vb, showing the formation of interlayer chemical bonds at $300$ K (room temperature).}
    \item[]{{\bf{Movie S3:}} Molecular-dynamics trajectory of a defective hBN/SiC heterostructure containing four and seven \vb, showing the formation of interlayer chemical bonds at $300$ K (room temperature).}
    \item[]{{\bf{Movie S4:}} Molecular‑dynamics simulation of an hBN/SiC heterostructure containing multiple vacancies and hosting Cu atoms at $300$ K (Example 1).}
    \item[]{{\bf{Movie S5:}} Molecular‑dynamics simulation of an hBN/SiC heterostructure containing multiple vacancies and hosting Cu atoms at $300$ K (Example 2).}
    \item[]{{\bf{Movie S6:}} Molecular‑dynamics simulation of an hBN/SiC heterostructure containing multiple vacancies and hosting Cu atoms at $300$ K (Example 3).}
\end{itemize}
%refs
\bibliography{ars.bib}